\documentclass[12pt,twoside]{toptesi}
\usepackage[latin1]{inputenc}
\usepackage{graphicx}
\usepackage{epsfig}
\usepackage{amsfonts}
\usepackage{pstricks}
\usepackage{bm}
\usepackage{axodraw}
\usepackage{changepage}
\usepackage{calc}
\newcommand{\htm}{\frac{1}{m}}
\newcommand{\mth}{m}
\newcommand{\bbox}[1]{%
{\mbox{\boldmath {$#1$}}}
}
\newcommand{\qt}{{\widetilde q}}
\newcommand{\mod}[1]{|{\bf #1}|}
\def\be{\begin{equation}}
\def\ee{\end{equation}}
\def\g{g_{ij}}

\def\met{\frac{1}{2}}
\def\coppie{\frac{N(N-1)}{2}}
\def\nanb{n_\alpha+n_h}
\def\rmo{{\bf r}}
\def\hm{{\bf h}}
\def\del{(2\pi)^3\delta^{(3)}(\ppm-\qm-\hm)}
\def\due{\bbox{\tau}_1\bbox{\tau}_2}
\def\tre{\bbox{\sigma}_1\bbox{\sigma}_2}
\def\quattro{\due\tre}

\def\suma{\sum_{\sigma_\alpha,\tau_\alpha}}
\def\J{{\bf J}}
\def\gam{\bbox{\gamma}}
\def\mq{\mod{q}}

\def\ub{\overline{u}}

\def\lsim{\buildrel < \over {_{\sim}}}
\def\gsim{\buildrel > \over {_{\sim}}}
\def\qm{{\bf q}}
\def\ppm{{\bf p}}

\newcommand{\primo}{^{\prime}}
\newcommand{\magq}{|{\bf q}|}

\newcommand{\beq}{\begin{equation}}
\newcommand{\eeq}{\end{equation}}
\newcommand{\bea}{\begin{eqnarray}}
\newcommand{\eea}{\end{eqnarray}}
\newcommand{\Lmunu}{L^{\mu\nu}}
\newcommand{\WmunuA}{W_{\mu\nu}^{A}}

\def\q24m{\frac{q^2}{4m^2}}

\def\lsim{\mathrel{\rlap{\lower4pt\hbox{\hskip1pt$\sim$}}
    \raise1pt\hbox{$<$}}}         
\def\gsim{\mathrel{\rlap{\lower4pt\hbox{\hskip1pt$\sim$}}
    \raise1pt\hbox{$>$}}}         
\newcommand{\sigsig}{({\bm \sigma}_1 \cdot {\bm \sigma}_2)}
\newcommand{\tautau}{({\bm \tau}_1 \cdot {\bm \tau}_2)}
\def\openone{\leavevmode\hbox{\small1\kern-3.8pt\normalsize1}}
\def\<{\langle}
\def\>{\rangle}
\begin{document}
\english
\begin{titlepage}

\newlength{\logolength}
\settowidth{\logolength}{\textsc{Scuola di Dottorato ``Vito Volterra''}}
\newlength{\advisorlength}
\settowidth{\advisorlength}{\textbf{Program Coordinator}}
\newlength{\addmargin}
\setlength{\addmargin}{((\paperwidth -\textwidth) / 2 ) -\oddsidemargin -\hoffset -1in}
\begin{adjustwidth*}{\addmargin}{-\addmargin}
\begin{center}
	\vspace*{-2cm}
    \includegraphics[width=\logolength]{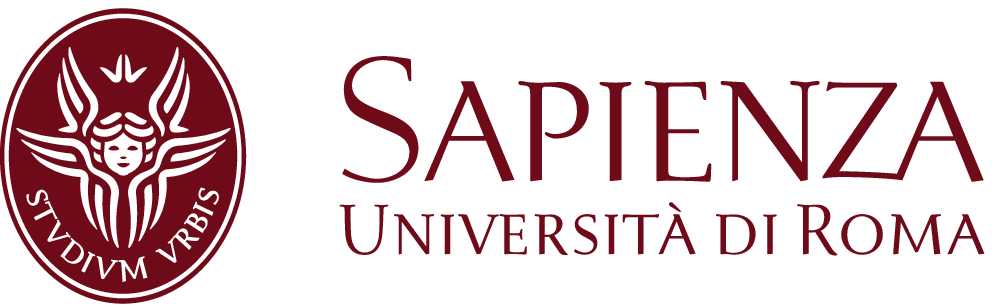}\\
    \vspace{0.4cm}
    {\textsc{Scuola di Dottorato ``Vito Volterra''}}\\
    {\textsc{Dottorato di Ricerca in Fisica}}\\
    \vspace{3.8cm}
	\begin{huge}
    \textbf{Weak Response of Nuclear Matter}
    \end{huge}\\
    \vspace{3.6cm}
    {\textsc{Thesis submitted to obtain the degree of}}\\
    {\textit{``Dottore di Ricerca'' -- Philosophi}\ae{} \textit{Doctor}}\\
    {\textsc{PhD in Physics -- XXI cycle -- October 2008}}\\
    \vspace{1.3cm}
    {\textsc{by}}\\
    \vspace{0.4cm}
    {\Large \textbf{Nicola Farina}}\\
    \vspace{4cm}
	\makebox[\linewidth][c]{ 
		\large
		\begin{tabular}{cp{5cm}c} 
		\textbf{Program Coordinator} & & \textbf{Thesis Advisor}\\
		\rule{0pt}{0.5cm}Prof. Enzo Marinari & &  Dr. Omar Benhar\\
		\rule{\advisorlength}{0pt}& & \rule{\advisorlength}{0pt}
		\end{tabular}
	}
\end{center}
\end{adjustwidth*}

\end{titlepage}
\thispagestyle{empty}
\vfil\null
\begin{center}
\bigskip \bigskip \bigskip
\vspace{1.0in}
\large
\bigskip \bigskip
\end{center} 
\normalsize
\vfil\null
\clearpage
\indici
\addcontentsline{toc}{chapter}{Introduction}
\pagenumbering{arabic}
\chapter*{Introduction}
\markboth{Introduction}{}
\markright{Introduction}{}
The description of neutrino interactions with nuclei, and nuclear matter in general, 
is relevant to the study of many different problems, from supernov$\ae$ explosions 
to neutron star cooling, as well as to the determination of the properties of neutrino 
itself, most notably its mass.

The appearance of a supernova is the last stage of the evolution of stars with 
initial mass bigger than $\sim$ 4 $M_\odot$, where $M_\odot \approx$ 2 
$\times 10^{33}$ g denotes the solar mass \cite{shapiro,weigert,arnett}. 
Although the first attempts to simulate a supernova explosion 
date more than 30 years back \cite{colgate}, the problem is not solved yet. In fact, the 
results of numerical calculations predict that, due to loss of energy carried away by 
neutrinos, produced in the dissociation of atomic nuclei in the core, 
the explosion does not occur. 

The systematic uncertainty associated to simulations depends heavily on the values 
of the neutrino-nucleon and neutrino-nucleus reaction rates used as inputs. 
As many existing programs use values obtained from models based on somewhat oversimplified 
nuclear dynamics \cite{bs1,bs2,horo1}, such uncertainty may be significantly reduced 
adopting more realistic models, which have proved very successfull in the description 
of electro-magnetic interaction of nuclei (see, e.g., Refs. \cite{rocco_joe,RMP}).

A similar problem arises in the field of neutrino physics, 
which is rapidly developing after the discovery of atmospheric and solar neutrino 
oscillation \cite{Kajita,Solar,KamLAND}. 
The experimental results point to two very distinct mass
differences \footnote{A third mass difference, $\Delta m^2_{LSND} \sim 1$ eV$^2$, 
suggested by the LSND experiment \cite{lsnd}, has not been confirmed yet \cite{boone}.},
$\Delta m^2_{sol} \approx 8.2 \times 10^{-5}$ eV$^2$ and
$|\Delta m^2_{atm}| \approx 2.5 \times 10^{-3}$ eV$^2$.
Only two out of the four parameters of the three-family leptonic mixing 
matrix $U_{PMNS}$ are known: $\theta_{12} \approx 34^\circ$ and 
$\theta_{23}\approx 45^\circ$.
The other two parameters, $\theta_{13}$ and $\delta$, are still unknown: 
for the mixing angle $\theta_{13}$ direct searches at reactors \cite{chooz} 
and three-family global analysis of the
experimental data \cite{globalfit,Gonzalez-Garcia:2004jd} give the 
upper bound $\theta_{13}
\leq 11.5^\circ$, whereas for the leptonic CP-violating phase $\delta$ 
we have no information whatsoever. Two additional discrete unknowns are the sign 
of the atmospheric mass difference and
the $\theta_{23}$-octant.

Neutrino oscillation experiments measure energy and emission angle of the charged 
leptons produced in neutrino-nucleus interactions, and use the obtained results 
to reconstruct the incoming neutrino energy. Hence, the quantitative understanding 
of the neutrino-nucleus cross section, as well as of the energy spectra and angular 
distribution of the final state leptons, is critical to reduce the systematic
uncertainty of data analysis. A number of theoretical studies aimed at providing accurate
predictions of neutrino-nucleus scattering observables are discussed in 
Refs. \cite{NUINT01,NUINT04,NUINT05}.

It is important to realize that, while neutrinos interacting in stellar matter typically 
have energies of the order of few MeV, the energies involved in long baseline 
oscillations experiments 
are much larger. For example, K2K takes data in the region $E_{\nu} =0.5-3$ GeV. 

The huge difference in kinematical conditions is reflected by different reaction mechanisms. 
For neutrinos of energy  $\lsim 10$ MeV, since the spatial resolution of incoming particle 
is much bigger than the average distance between nucleons, the nuclear response is largely 
determined by many-body effects. 

On the other hand, at energies of the order of 1 GeV, it is reasonable to expect that 
the scattering process on a nucleus reduce to the incoherent sum of 
elementary processes involving individual nucleons. Furthermore, in this kinematical regime 
one needs to take into account the fact that elementary neutrino-nucleon interactions can 
give rise to inelastic processes, leading to the appearance of hadrons other than protons and
neutrons.

The results of electron- and hadron-induced nucleon knock-out experiments have provided
overwhelming evidence of the inadequacy of the independent particle model to describe the
full complexity of nuclear dynamics \cite{book,specfact}. While the peaks corresponding 
to knock-out from shell model orbits can be clearly identified in the measured energy spectra, 
the corresponding strengths turn out to be consistently and sizably lower than expected, 
independent of the nuclear mass number.
This discrepancy is mainly due to the effect of dynamical correlations induced
by the nucleon-nucleon force, whose effect is not taken into account in the independent
particle model. 

Nuclear many body theory provides a scheme allowing for a consistent treatment of 
neutrino-nucleus interactions at both high and low energies. Within this approach, nuclear 
dynamics is described by a phenomenological hamiltonian, whose structure is completely determined 
by the available data on two- and three-nucleon systems, and dynamical correlations are taken 
into account. 

Over the past decade, the formalism based on correlated wave functions, originally proposed to 
describe quantum liquids \cite{feenberg}, has been employed to carry out highly accurate 
calculations of the binding energies of both nuclei and nuclear matter, using either 
the Monte Carlo method \cite{WP,GFMC,AFDMC1} or the cluster expansion formalism and 
the Fermi Hypernetted Chain integral equations \cite{APR1,gpc1,gpc2}.

A different approach, recently proposed in Refs. \cite{shannon2,valli} exploits the 
correlated wave functions to construct an effective interaction suitable for use
in standard perturbation theory. This scheme has been employed to obtain a variety 
of nuclear matter properties, including the neutrino mean free path \cite{shannon2} and the 
transport coefficients \cite{valli,transport}.

In this work we describe the application of the formalism based on correlated wave functions
and the effective interaction to the calculation of the weak response of atomic nuclei and 
uniform nuclear matter.

The Thesis is structured as follows.

In Chapter \ref{nuclear:physics}, we briefly describe the main features of nuclear matter and  
nuclear forces.

In Chapter \ref{CBF}, we focus on the theory of nuclear matter, introducing the correlated states and
the cluster expension formalism, needed to define the effective interaction. The applicability 
of perturbation theory within this framework is also discussed.

Chapter \ref{response} is devoted to an overview of the many body theory of the nuclear matter response, 
and its connection to the spectral function.

In Chapter \ref{SF} we discuss the assumptions underlying the impulse approximation, and its 
applicability in the high energy regime. After a short description of the main features of the 
spectral function obtained from nuclear many-body theory, we compare the 
calculated electron-nucleus cross sections to data, and show the predictions of our approach 
for charged current neutrino-nucleus interactions.

Finally, in Chapter \ref{RESP} we focus on the low energy regime. We develop a formalism based
on the effective interaction of Ref. \cite{valli} and an effective Fermi transition operator, 
obtained at the same order of the cluster expansion. The effects of both short and long range
correlatios, described within the correlated Hartree Fock and Tamm-Dancoff approximations, are
discussed.

Note that in this Thesis we use a system of units in which $\hbar = h/2\pi = c = 1$, 
where $h$ is the Planck constant and $c$ is the speed of light in the vacuum.

\chapter{Nuclear matter and nuclear forces}
\label{nuclear:physics}
Nuclear matter can be thought of as a giant nucleus, with given numbers of protons
and neutrons interacting through nuclear forces only. As the typical thermal energies 
are negligible compared to the nucleon Fermi energies, such a system can be safely
considered to be at zero temperature. 

A quantitative understanding of the properties 
of nuclear matter, whose calculation is greatly simplified by translational
invariance, is needed both as an intermediate step towards the description of real 
nuclei and for the development of realistic models of matter in the neutron star core.

The large body of data on nuclear masses can be used to extract empirical 
information on the equilibrium properties of symmetric nuclear matter (SNM), 
consisting of equal numbers of protons and neutrons.

\begin{figure}[hbt]
\begin{center}
{\epsfig{figure=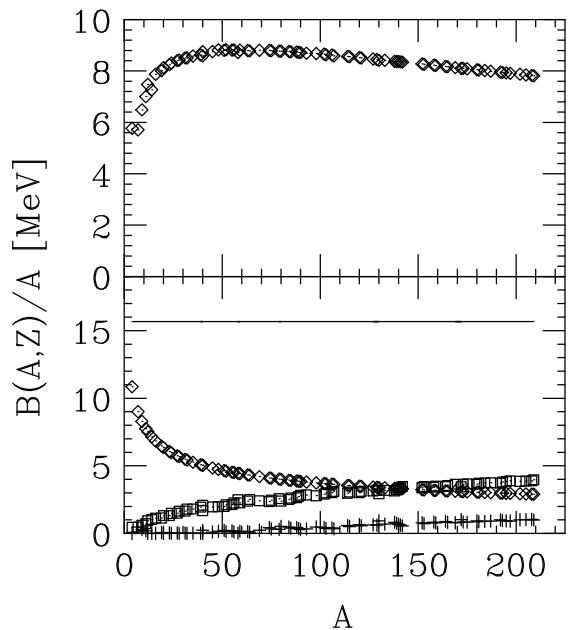,width=7.5cm }}
\vspace{-0.1in}
\caption{ \small
Upper panel: A-dependence of the binding energy per nucleon of
stable nuclei, evaluated according to Eq. (\protect{\ref{mass:formula}}) with
$a_{\rm V}$~=~15.67~MeV, $a_{\rm s}$~=~17.23~MeV, $a_{\rm c}$~=~.714 MeV,
$a_{\rm A}$~=~93.15~MeV and $a_{\rm p}$~=~11.2~MeV. Lower panel:
the solid line shows the magnitude of the volume contribution to the
binding energy per nucleon, whereas the A-dependence of the surface, coulomb
and symmetry contributions are represented by diamonds, squares and crosses,
respectively.
}
\label{binding:energy}
\end{center}
\end{figure}

The (positive) binding energy of nuclei of mass number $A$ and electric 
charge $Z$, defined as
\beq
\frac{B({\rm Z},{\rm A})}{A} =
\frac{1}{A} \left[{\rm Z}m_p + ({\rm A} - {\rm Z})m_n
+ {\rm Z}m_e -  M({\rm Z},{\rm A}) \right]\ ,
\eeq
where $M({\rm Z},{\rm A})$ is the measured nuclear mass and $m_p$, $m_n$ and $m_e$ denote
the proton, neutron and electron mass, respectively, is almost constant for 
A$\ge$ 12 , its value being $\sim$ 8.5 MeV (see Fig.~\ref{binding:energy}).
The $A$-dependence is well described by the semi-empirical mass formula 
\beq
\frac{B}{A} = \frac{1}{A} \left[ \right. a_{\rm V} A -  a_{\rm s} A^{2/3} -
 a_{\rm c} \frac{Z^2}{A^{1/3}} - 
a_A \frac{ (A-2Z)^2 }{ 4 A } + \lambda \  a_{\rm p}
\frac{1}{A^{1/2}} \left. \right]\ .
\label{mass:formula}
\eeq
The first term in square brackets, proportional to A, is called the {\it volume term}
and describes
the bulk energy of nuclear matter. The second term, proportional to the nuclear
radius squared, is associated with the surface energy, while the third one accounts
for the Coulomb repulsion between Z protons uniformly distributed within a sphere of
radius R. The fourth term, that goes under the name of {\it symmetry energy} is required
to describe the experimental observation that stable nuclei tend to have the same number of
neutrons and protons. Moreover, even-even nuclei (i.e. nuclei having even Z and even
A $-$ Z) tend to be more stable than even-odd or odd-odd nuclei. This property is
accounted for by the last term in the above equation, where $\lambda$ $-$1, 0  and +1
for even-even, even-odd and odd-odd nuclei, respectively. Fig. \ref{binding:energy} shows
the different contributions to $B({\rm Z},{\rm A})/{\rm A}$, evaluated using
Eq.~(\ref{mass:formula}).

In the
 $A \rightarrow \infty$ limit, and neglecting the effect of Coulomb repulsion 
between protons, the
only contribution surviving in the case $Z=A/2$ is the term linear in $A$. Hence, the
coefficient $a_V$ can be identified with the binding energy per particle of SNM.

\begin{figure}[htb]
\begin{center}
{\epsfig{figure=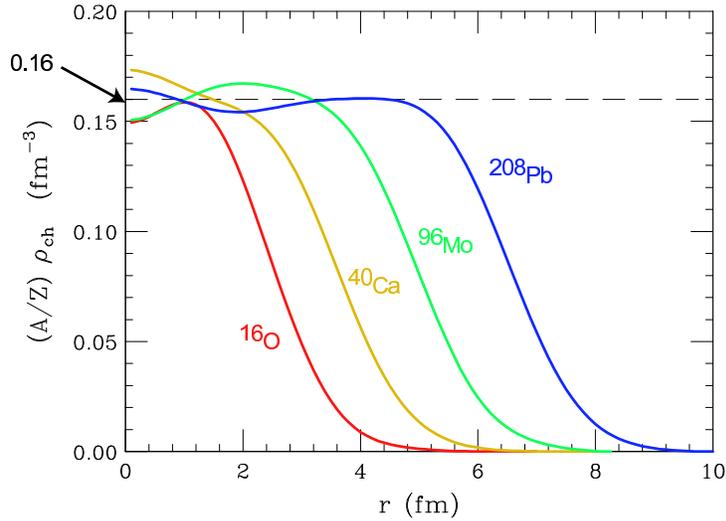,width=9.5cm }}
\vspace*{-.1in}
\caption{ \small
Saturation of central nuclear densities measured by elastic electron-nucleus
scattering.
}
\label{satdens}
\end{center}
\end{figure}

The equilibrium density of SNM, $\rho_0$, can be
inferred exploiting the saturation of nuclear densities, i.e. the experimental 
observation that the central charge density of atomic nuclei, measured by elastic
electron-nucleus scattering, does not depend upon $A$ for
large $A$. This property is illustrated in Fig. \ref{satdens}.

The empirical values of the binding energy and equilibrium density of SNM
are
\beq
\rho_0 = 0.16\, \textrm{fm}^{-3} \ \ , \ \ E = - 15.7\, \textrm{MeV} \ .
\label{eq:prop}
\eeq
In principle, additional information can be obtained from measurements of the 
excitation energies of nuclear vibrational states, yielding the (in)-compressibility 
module $K$. However, the data analysis of these experiments is non trivial, and the 
resulting values of $K$ range from $K \sim 200$ MeV (corresponding to more compressible
nuclear matter, i.e. to a {\it soft} equation of state (EOS)) to $K \sim 300$ MeV (corresponding
to a {\it stiff} EOS) \cite{compress}.

The main goal of nuclear matter theory is deriving a EOS  at 
zero temperature (i.e. the density dependence of the binding energy per 
particle $E=E(\rho)$) capable to explain the above data starting from the elementary 
nucleon-nucleon (NN) interaction. However, many important applications 
of nuclear matter theory require
that its formalism be also flexible enough to describe the properties of 
matter at finite temperature. 

Unfortunately, due to the complexity of the fundamental theory of strong interactions, 
 the quantum chromo-dynamics (QCD), an {\em ab initio} description of nuclear matter 
at finite density and zero temperature is out of reach of the present computational 
techniques. As a consequence, one has to rely on dynamical models in which nucleons and 
mesons play the role of effective degrees of freedom.

In this work we adopt the approach based on nonrelativistic quantum mechanics and
phe\-no\-me\-no\-lo\-gi\-cal nuclear hamiltonians, that allows for a quantitative
description of both the two-nucleon bound state and nucleon-nucleon scattering data. 

In this Chapter we outline the main features of nuclear interactions and briefly describe 
the structure of the NN potential models employed in many-body calculations.

\section{Nuclear forces}
\label{NM1}

The main features of the NN interaction, inferred from
the analysis of nuclear systematics, may be summarized as follows.

\begin{itemize}

\item The {\it saturation} of nuclear density (see Fig. \ref{satdens}), i.e. the 
fact that density in
the interior of atomic nuclei is nearly constant and independent of the mass number $A$,
tells us that nucleons cannot be packed together too tightly. Hence, at short distance the
NN force must be repulsive. Assuming that the interaction can be
described by a nonrelativistic potential $v$ depending on
the inter-particle distance, ${\bf r}$, we can then write:
\beq
v({\bf r}) > 0\ \ \ ,\ \ |{\bf r}|<r_c\ ,
\eeq
$r_c$ being the radius of the repulsive core.

\item The fact that the nuclear binding energy per nucleon
 is roughly the same for all nuclei with $A\ge$ 12 suggests that the NN interaction 
has a finite range $r_0$, i.e. that
\beq
v({\bf r}) = 0\ \ \ ,\ \ |{\bf r}|>r_0\ .
\eeq

\item The spectra of the so called {\it mirror nuclei}, i.e. pairs of nuclei having the
same $A$ and charges differing by one unit (implying that the number of protons in a nucleus
is the same as the number of neutrons in its mirror companion), e.g. $^{15}_{\ 7}$N
($A$~=~15, $Z$~=~7) and $^{15}_{\ 8}$O ($A$~=~15, $Z$~=~8), exhibit striking
similarities. The energies of the levels with the same parity and angular momentum are
the same up to small electromagnetic corrections, showing that protons and neutrons have
similar nuclear interactions, i.e. that nuclear forces are {\it charge symmetric}.

\end{itemize}

Charge symmetry is a manifestation of a more general property  of the NN interaction,
called {\em isotopic invariance}. Neglecting the small mass difference, proton and 
neutron can be viewed as two states of the same particle, the nucleon (N), described by the 
Dirac equation obtained from the Lagrangian density
\beq
{\cal L} =  \bar{\psi}_N \left( i \gamma^\mu \partial_\mu - m \right) \psi_N
\label{lagsu2}
\eeq
where
\beq
\psi_N = \left(
\begin{array}{c} p \\  n \end{array}
\right)\ ,
\label{Nspinor}
\eeq
$p$ and $n$ being the four-spinors associated with the proton and the neutron, 
respectively. The lagrangian density (\ref{lagsu2}) is invariant under the $SU(2)$ 
global phase transformation
\beq
U = {\rm e}^{i \alpha_j \tau_j} \ ,
\eeq
where ${\bm \alpha}$ is a constant (i.e. independent of $x$) vector and the $\tau_j$ ($j=1,2,3$) 
are Pauli matrices (whose properties are briefly collected in Appendix
\ref{pauli}). The above equations show that 
the nucleon can be described as a doublet in isospin space. 
Proton and neutron correspond to isospin projections $+$1/2 and $-$1/2,
respectively. Proton-proton and neutron-neutron pairs always have total isospin T=1 whereas
a proton-neutron pair may have either $T=0$ or $T=1$. The two-nucleon isospin states 
$|T,M_T\rangle$ can be summarized as follows (see also Appendix \ref{pauli})
\bea
\nonumber
| 1,1 \rangle & = & | pp \rangle \\ 
\nonumber
| 1,0 \rangle & = & 
\frac{1}{\sqrt{2}}\left( | pn \rangle + | np \rangle \right)  \\
\nonumber
| 1,-1 \rangle & = & | nn \rangle  \\
\nonumber
| 0,0 \rangle & = &
\frac{1}{\sqrt{2}}\left( | pn \rangle - | np \rangle \right) \ . 
\eea
Isospin invariance implies that the interaction between two nucleons separated by 
a distance $r = |{\bf r}_1 - {\bf r}_2|$ and having total spin $S$ depends on 
their total isospin $T$ but not on its projection $M_T$. For example, the potential $v({\bf r})$ 
acting 
between two protons with spins coupled to $S=0$ is the same as the potential acting between a
proton and a neutron with spins and isospins coupled to $S=0$ and $T=1$.

\subsection{The two-nucleon system}
\label{twobody}

The details of the NN interaction can be best understood in the two-nucleon system.
There is {\em only one} NN bound state, the nucleus of deuterium, or
deuteron ($^2$H), consisting of a proton
and a neutron coupled to total spin and isospin $S=1$ and $T=0$, respectively. 
This is a clear manifestation of the {\em spin dependence} of nuclear forces.

Another important piece of information can be inferred from the
observation that the deuteron exhibits a non vanishing electric quadrupole moment,
implying that its charge distribution is not spherically symmetric. Hence, 
the NN interaction is {\em non-central}.

Besides the properties of the two-nucleon bound state, the large data base of phase
shifts measured in NN scattering experiments (the Nijmegen data base \cite{Nij} includes 
$\sim$ 4000 data points, corresponding
to energies up to 350 MeV in the lab frame) provides valuable additional information
on the nature of NN forces.

The theoretical description of the NN interaction was first attempted by Yukawa
in 1935. He made the hypothesis that nucleons interact through the exchange of a
particle, whose mass $\mu$ can be related to the interaction range $r_0$ according to
\beq
r_0 \sim \frac{1}{\mu}\ . 
\eeq
Using $r_0 \sim$ 1 fm, the above relation yields $\mu \sim 200$ MeV (1 fm$^{-1}$ = 197.3 MeV).

\begin{figure}[htb]
\begin{center}
{\epsfig{figure=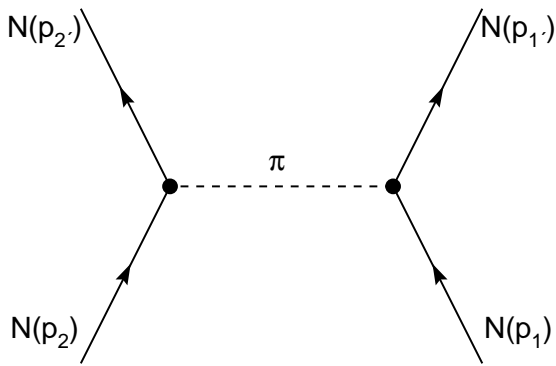,width=7.5cm }}
\vspace*{-.1in}
\caption{ \small
Feynman diagram describing the one-pion-exchange process between two nucleons.
The corresponding amplitude is given by Eq. (\protect{\ref{OPEP:amplitude}}).
}
\label{pion:exchange}
\end{center}
\end{figure}

Yukawa's idea has been successfully implemented identifying the
exchanged particle with the $\pi$ meson (or {\em pion}), discovered in 1947, whose
mass is $m_\pi \sim$ 140 MeV.
Experiments show that the pion is a spin zero pseudo-scalar particle\footnote{The pion spin has been deduced from the balance of the reaction
$\pi^+ + ^2\!{\rm H} \leftrightarrow p + p$, while its intrinsic parity was determined
observing the $\pi^-$ capture from the K shell of the deuterium atom,
leading to the appearance of two neutrons: \mbox{$\pi^- + d \rightarrow n + n$}.}
 (i.e. it has spin-parity 0$^-$) that comes in three charge states, denoted $\pi^+$,
$\pi^-$ and $\pi^0$. Hence, it can be regarded as an
isospin T=1 triplet, the charge states being associated with isospin projections
$M_T$=$+$~1, 0 and $-$1, respectively.

The simplest $\pi$-nucleon coupling compatible with the observation that nuclear
interactions conserve parity has the pseudo-scalar form $i g \gamma^5 {\bm\tau}$, where 
$g$ is a coupling constant and ${\bm \tau}$ describes
the isospin of the nucleon. With this choice for the interaction vertex, the amplitude
of the process depicted in Fig. \ref{pion:exchange} can readily be written, using standard
Feynman's diagram techniques, as
\beq
\langle f | M | i \rangle = -i g^2 \
\frac{{\bar u}(p_2\primo,s_2\primo) \gamma_5 u(p_2,s_2)
{\bar u}(p_1\primo,s_1\primo) \gamma_5 u(p_1,s_1)}{k^2 - m_\pi^2}\
\langle {\bm \tau_1}\cdot{\bm \tau_2} \rangle \ ,
\label{OPEP:amplitude}
\eeq
where $m_\pi$ is the pion mass, $k = p_1\primo - p_1 = p_2 - p_2\primo$, 
$k^2 = k_\mu k^\mu = k_0^2 - |{\bf k}|^2$,  
$u(p,s)$ is the Dirac spinor associated with a nucleon of four momentum
$p \equiv ({\bf p},E)$ (E=$\sqrt{{\bf p}^2 + m^2}$) and spin projection $s$ and
\beq
\langle{\bm\tau}_1\cdot{\bm\tau}_2\rangle =
\eta_{2^\prime}^\dagger{\bm\tau}\eta_2 \ \eta_{1^\prime}^\dagger{\bm\tau}\eta_1 \ ,
\eeq 
$\eta_i$ being the two-component Pauli spinor describing the isospin state of particle $i$. 

In the nonrelativistic limit, Yukawa's theory leads to define a NN interaction potential 
that can be written in coordinate space as
\begin{eqnarray}
\nonumber
v_{\pi} & = &  \frac{g^2}{4 m^2}\ ({\bm \tau}_1 \cdot {\bm \tau}_2)
({\bm \sigma}_1 \cdot {\bm \nabla})({\bm \sigma}_2 \cdot {\bm \nabla})\
\frac{e^{-m_\pi r}}{r} \\
\nonumber
                   & = & \frac{g^2}{(4\pi)^2}\ \frac{m_\pi^3}{4m^2}\ \frac{1}{3}
({\bm \tau}_1 \cdot {\bm \tau}_2)
\left\{ \left[ ({\bm \sigma}_1 \cdot {\bm \sigma}_2) +
S_{12} \left(1+\frac{3}{x}+\frac{3}{x^2}\right)\right]\frac{e^{-x}}{x} \right.\\
& & \ \ \ \ \ \ \ \ \ \ \ \ 
 -  \left. \frac{4\pi}{m_\pi^3}({\bm \sigma}_1 \cdot {\bm \sigma}_2)\delta^{(3)}({\bf r})
  \right\}\ ,
\label{opep:1}
\end{eqnarray}
where $x = m_\pi |{\bf r}|$ and 
\beq
S_{12} = \frac{3}{r^2}({\bm \sigma}_1 \cdot {\bf r})
({\bm \sigma}_2 \cdot {\bf r}) - ({\bm \sigma}_1 \cdot {\bm \sigma}_2)\ ,
\label{opep:2}
\eeq
reminiscent of the operator describing the non-central interaction between two magnetic 
dipoles, is called the tensor operator. The properties of $S_{12}$ are
summarized in Appendix \ref{pauli}

For $g^2/(4\pi) \sim 14$, the above potential provides an accurate description of the long
range part
($|{\bf r}|~>$~1.5 fm) of the NN interaction, as shown by the very good fit of the
NN scattering phase
shifts in states of high angular momentum. In these states, due to the
strong centrifugal barrier, the probability of finding the
two nucleons at small relative distances becomes in fact negligibly small.

At medium- and short-range other more complicated processes, involving
the exchange of two or more pions (possibly interacting among themselves) or
 heavier particles (like the $\rho$ and the $\omega$ mesons, whose masses are
$m_\rho$ = 770 MeV and $m_\omega$ = 782 MeV, respectively), have to be taken
into account. Moreover, when their relative distance becomes very small
($|{\bf r}| \lsim 0.5$ fm) nucleons, being composite and finite in size,
are expected to overlap. In this regime, NN interactions should in principle
be described in terms of interactions between nucleon
constituents, i.e. quarks and gluons, as dictated by QCD.

Phenomenological potentials describing the {\it full} NN interaction are generally
written as
\beq
v = {\widetilde v}_{\pi} + v_{R}
\label{phenv:1}
\eeq
where ${\widetilde v}_{\pi}$ is the one-pion-exchange potential, defined by Eqs. (\ref{opep:1}) 
and (\ref{opep:2}), stripped of the $\delta$-function contribution, whereas $v_{R}$ 
describes the interaction at medium and short range. The spin-isospin dependence
and the non-central nature of the NN interactions can be properly described 
rewriting Eq. (\ref{phenv:1}) in the form
\beq
v(ij) = \sum_{ST} \left[ v_{TS}(r_{ij}) + \delta_{S1} v_{tT}(r_{ij}) S_{12} \right]
P_{2S+1} \Pi_{2T+1} \ ,
\label{pot:TS}
\eeq
$S$ and $T$ being the total spin and isospin of the interacting pair, respectively. In 
the above equation $P_{2S+1}$ ($S=0,1$) and $\Pi_{2T+1}$ ($T=0,1$) are the spin and isospin 
projection operators, whose definition and properties are given in Appendix \ref{pauli}.  

The functions $v_{TS}(r_{ij})$ and $v_{tT}(r_{ij})$ describe the radial dependence of the 
interaction in the different spin-isospin channels and reduce to the corresponding components 
of the one-pion-exchange potential at large $r_{ij}$. Their shapes are chosen in such a way as 
to reproduce the available NN data (deuteron binding energy, charge radius and quadrupole moment
and the NN scattering data).

An alternative representation of the NN potential, based on the set of six operators
(see Appendix \ref{pauli})
\beq
O^{n \leq 6}_{ij} = [1, (\bm{\sigma}_{i}\cdot\bm{\sigma}_{j}), S_{ij}]
\otimes[1,(\bm{\tau}_{i}\cdot\bm{\tau}_{j})] \ ,
\label{pot2}
\eeq
is given by
\beq
v(ij) = \sum_{n=1}^{6} v^{(n)}(r_{ij})O^{(n)}_{ij}\ .
\label{pot1}
\eeq
While the static potential of Eq.(\ref{pot1}) provides a reasonable account of deuteron 
properties, in order to describe NN scattering in S and P wave, one has to include 
the two additional momentum dependent operators
\beq
O^{n = 7, 8}_{ij} = {{\bf L}\cdot {\bf S}}\otimes[1,(\bm{\tau}_{i}\cdot\bm{\tau}_{j})] \ ,
\eeq
${\bf L}$ being the orbital angular momentum.

The potentials yielding the best available fits of NN scattering data, with a $\chi^2$/datum $\sim$ 1, 
are written in terms of eighteen operators, with
\bea
O^{n=9,\ldots,14}_{ij}  & = & [{\bf L}^{2},{\bf L}^{2}(\bm{\sigma}_{i}\cdot\bm{\sigma}_{j}),
({\bf L\cdot S})^{2}]\otimes[1,\bm{\tau}_{i}\cdot\bm{\tau}_{j}] \ ,\\ 
O^{n=15,\ldots,18}_{ij}  & = & [1, \bm{\sigma}_{i}\cdot\bm{\sigma}_{j}, S_{ij}]\otimes
T_{ij} \ , \  (\tau_{zi}+\tau_{zj})
\eea
where
\beq
T_{ij} = \frac{3}{r^2} ({\bm \tau}_i \cdot {\bf r})({\bm \tau}_j \cdot {\bf r}) 
 - ({\bm \tau}_i \cdot {\bm \tau}_j)  \ .
\eeq
The $O^{n=15,\ldots,18}_{ij}$ take care of small charge symmetry breaking effects,
due to the different masses and coupling constants of the charged and neutral pions.

\begin{figure}[hbt]
\begin{center}
{\epsfig{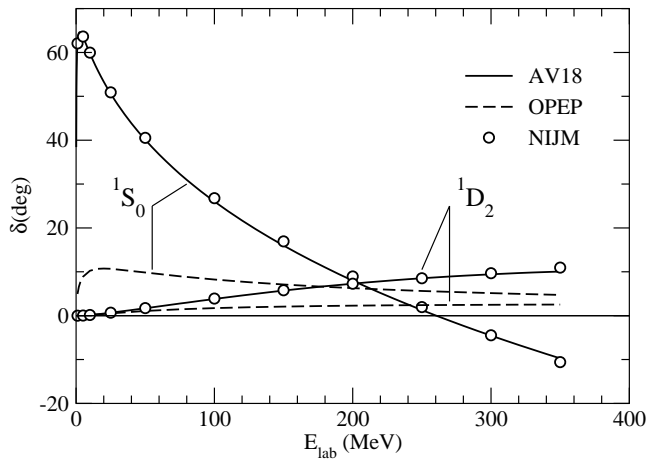}}
\vspace{-.1in}
\caption{ \small Comparison between the $^1$S$_0$ and  $^1$D$_2$ phase shifts resulting from
 the
Nijmegen analysis \cite{Nij} (open circles) and the predictions of the Argonne $v_{18}$ (AV1
8) and
one-pion-exchange (OPEP) potentials.
}
\label{phase}
\end{center}
\end{figure}

The calculations discussed in this Thesis are based on a
widely employed potential model, obtained within the phenomenological approach outlined 
in this Section, generally referred to as Argonne $v_{18}$ potential \cite{ANL18}. It
is written in the form
\beq
v(ij)=\sum_{n=1}^{18} v^{n}(r_{ij}) O^{n}_{ij} \ .
\eeq

As an example of the quality of the phase shifts obtained from the Argonne $v_{18}$ potential,
in Fig. \ref{phase} we show the results for the $^1$S$_0$ and  $^1$D$_2$ partial waves, 
compared with the predictions of the one-pion-exchange model (OPEP).

\begin{figure}[hbt]
\begin{center}
{\epsfig{figure=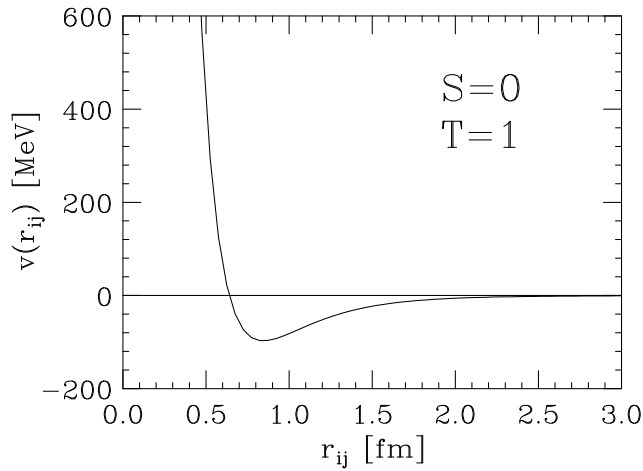,width=8.5cm }}
\vspace{-.1in}
\caption{ \small Radial dependence of the NN potential describing the interaction
between two nucleons in the state of relative angular momentum $\ell = 0$,
and total spin and isospin $S = 0$ and $T = 1$.
}
\label{NN:pot}
\end{center}
\end{figure}

We have used a simplified version of the above potential, obtained including the 
operators $O^{n \leq 8}_{ij}$, originally proposed in Ref.\cite{V8P}. It reproduces the scalar 
part of the full interaction in all S and P waves, as well as in the $^3$D$_1$ wave and its 
coupling to the $^3$S$_1$.

The typical shape of the NN potential in the state of relative angular momentum $\ell = 0$ 
and total spin and isospin $S = 0$ and $T = 1$ is shown in Fig. \ref{NN:pot}. 
The short-range repulsive core, to be ascribed to heavy-meson exchange or to
more complicated mechanisms involving nucleon constituents, is followed by an
intermediate-range attractive region, largely due to two-pion-exchange processes. 
Finally, at large interparticle distance the one-pion-exchange mechanism dominates.

\subsection{Three-nucleon interactions}
\label{fewbody}

\begin{figure}[hbt]
\begin{center}
{\epsfig{figure=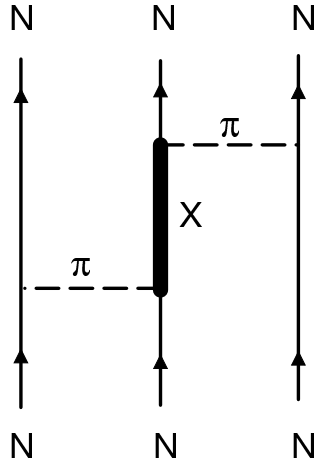,width=4.0cm }}
\vspace{-.1in}
\caption{ \small Diagrammatic representation of the process providing the 
main contribution to the three-nucleon interaction. The thick solid line 
corresponds to an excited state of the nucleon. 
}
\label{TBF}
\end{center}
\end{figure}

The NN potential determined from the properties of the two-nucleon system can be 
used to solve the many-body nonrelativistic Schr\"odinger equation for $A>2$. 
In the case $A=3$ the problem can be
still solved exactly, but the resulting ground state energy, $E_0$, turns out to 
be slightly different from the experimental value. For example, for $^3$He one typically 
finds $E_0=7.6$ MeV, to be compared to $E_{exp} = 8.48$~MeV. In order to exactly 
reproduce $E_{exp}$ one has to add to the nuclear hamiltonian a term containing 
three-nucleon interactions described by a potential $V_{ijk}$. The most important 
process leading to three-nucleon interactions is two-pion exchange associated with 
the excitation of a nucleon resonance in the intermediate state, depicted in 
Fig. \ref{TBF}.

The three-nucleon potential is usually written in the form
\beq
V_{ijk} = V_{ijk}^{2\pi} + V_{ijk}^{N} \ ,
\label{TBV}
\eeq
where the first contribution takes into account the process of Fig. \ref{TBF} 
while $V_{ijk}^{N}$ is purely phenomenological. The two parameters entering the 
definition of the three-body potential are adjusted in such a way as to reproduce 
the properties of $^3$H and $^3$He \cite{TBF}. Note that the inclusion of $V_{ijk}$ leads to
a very small change of the total potential energy, the ratio 
$\langle v_{ij} \rangle / \langle V_{ijk} \rangle$ being $\sim 2$ \%.

For $A>3$ the Scr\"odinger equation is no longer exactly solvable.  
However, very accurate solutions can be obtained using stochastic techniques, such as
variational Monte Carlo (VMC) Green function Monte Carlo (GFMC) \cite{WP}.

\begin{figure}[hbt]
\vspace{.1cm}
\begin{center}
\includegraphics[width=2.2in,angle=270]{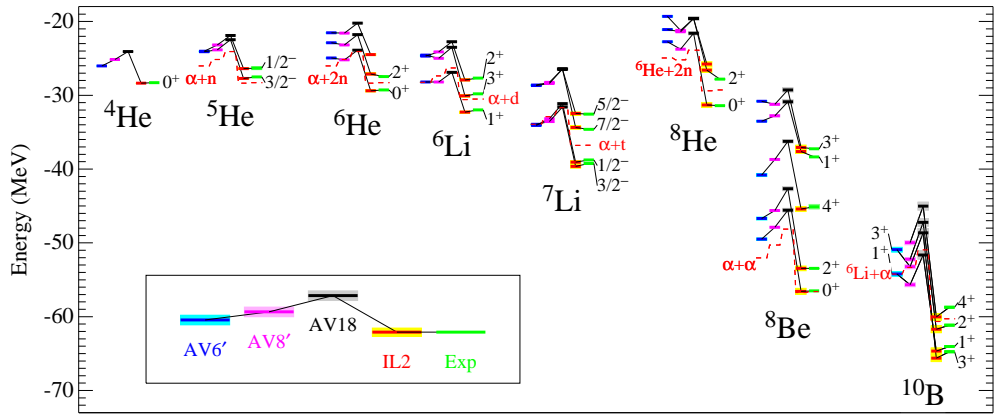}
\vspace{-.0in}
\caption{ \small
VMC and GFMC energies of nuclei with $A\leq10$ compared to experiment (from Ref.\cite{WP2}).
}
\label{GFMC:results}
\end{center}
\end{figure}

The GFMC approach has been successfully employed to describe the ground state and
the low lying excited states of nuclei having $A$ up to 10. The results of VMC and GFMC
calculations carried out using realistic nuclear hamiltonian, summarized in 
Fig. \ref{GFMC:results},
show that the nonrelativistic
approach, based on a dynamics modeled to reproduce the properties of two- and
three-nucleon systems, has a remarkable predictive power.

\chapter{Nuclear matter theory}
\label{CBF}
Understanding the properties of matter at densities comparable to the central density of
atomic nuclei is made difficult by {\it both} the complexity of the interactions, 
discussed in the previous Chapter, {\it and} the approximations implied in any theoretical 
description of quantum mechanical many-particle 
systems.

The main problem associated with the use of the nuclear potential models described in 
Chapter \ref{nuclear:physics} in a many-body calculation lies in the strong repulsive 
core of the NN force, which cannot be handled within standard perturbation theory. 

In non-relativistic many-body theory (NMBT), a nuclear system is seen as a collection of
point-like protons and neutrons whose dynamics are described by the hamiltonian
\beq
H = \sum_{i} t(i) + \sum_{j>i} v(ij) + \ldots \ ,
\label{NMBT:H}
\eeq
where $t(i)$ and $v(ij)$ denote the kinetic energy operator and the {\it bare} NN potential,
respectively, while the ellipses refer to the presence of additional many-body
interactions (see Chapter \ref{nuclear:physics}).

Carrying out perturbation theory in the basis provided by the eigenstates of 
the noninteracting system requires a renormalization of the NN potential. This is 
the foundation of the widely employed approach developed by Br\"uckner, Bethe and Goldstone,
in which $v(ij)$ is replaced by the well-behaved G-matrix, describing NN scattering in the 
nuclear medium (see, e.g. Ref.\cite{Marcello}). 
Alternatively, the many-body Schr\"odinger equation, with the hamiltonian 
of Eq.(\ref{NMBT:H}), can be solved using either the variational method or 
stochastic techniques. These approaches have been successfully applied to the study of 
both light nuclei \cite{WP} and uniform neutron and
nuclear matter \cite{APR1,GFMC,AFDMC1,AFDMC2}.

Our work has been carried out using a scheme, formally similar to standard
perturbation theory, in which nonperturbative effects due to the short-range repulsion 
are embodied in the basis functions. The details of this approach will be discussed
in the following Sections. 

It has to be emphasized that within NMBT the interaction is
completely determined by the analysis of the {\em exactly solvable} two- and three-nucleon systems.
As a consequence, the uncertainties associated with the dynamical model and the 
many-body calculations are decoupled, and the properties of nuclear systems ranging from 
deuteron to neutron stars can be obtained in a fully consistent fashion, without including 
any adjustable parameters. 

\section{Correlated basis function theory}

The {\em correlated} states of nuclear matter are obtained from the Fermi gas
(FG) states $\vert n_{FG}\rangle$
through the transformation \cite{CBF0,CBF1}
\begin{equation}
| n ) = \frac{ F |n_{{\rm FG}} \rangle }
{ \langle n_{{\rm FG}} | F^\dagger F | n_{{\rm FG}} \rangle^{1/2} } \ .
\label{def:corr_states}
\end{equation}
In the above equation, $|n_{{\rm FG}} \rangle$ is a determinant of single particle states 
describing $N$ noninteracting nucleons. The 
operator $F$, embodying the correlation structure induced by the NN
interaction, is written in the form
\begin{equation}
F(1,\ldots,N)=\mathcal{S}\prod_{j>i=1}^N f_{ij} \  ,
\label{def:F}
\end{equation}
where $\mathcal{S}$ is the symmetrization operator which takes care of the fact that, in general,
\beq
\left[ f_{ij} , f_{ik} \right] \neq 0 \ .
\eeq
The structure of the two-body {\em correlation functions} $f_{ij}$
must reflect the complexity of the NN potential. Hence, it is generally cast in the form
(compare to Eq.(\ref{pot1}))
\beq
f_{ij}=\sum_{n=1}^6 f^{n}(r_{ij}) O^{n}_{ij} \ ,
\label{def:corrf}
\eeq
with the $O^n_{ij}$ defined by Eq.(\ref{pot2}). Note that the operators included in the 
above definition provide a fairly accurate description of the correlation structure of the 
two-nucleon bound state. The shape of the radial functions $f^{n}(r_{ij})$ is determined
through functional minimization of the expectation value of the nuclear hamiltonian in 
the correlated ground state
\beq
E^V_0 = ( 0 | H | 0 )  \ .
\label{def:E0V}
\eeq
The correlated states defined in Eq.(\ref{def:corr_states}) are not orthogonal to one another.
However, they can be orthogonalized using an approach, based on standard techniques of 
many-body theory, that preserves diagonal matrix elements of the hamiltonian \cite{CBF2}. 
Denoting the orthogonalized states by $\vert n \rangle$, the procedure of Ref. \cite{CBF2} amounts 
to defining a transformation ${\widehat T}$ such that 
\beq
| n ) \rightarrow | n \rangle = {\widehat T} | n ) \ ,
\eeq
with
\beq
( n | H | n ) = \langle n | H | n \rangle \ .
\eeq

Correlated basis function (CBF) perturbation theory is based on the decomposition 
of the nuclear hamiltonian
\beq
H = H_0 + H_I \ ,
\label{CBF:H1}
\eeq
where $H_0$ and $H_I$ denote the diagonal and off-diagonal components of $H$, respectively, 
 defined by the equations
\bea
\label{CBF:H2}
\langle m | H_0 | n \rangle & = & \delta_{mn} \langle m | H | n \rangle  \ , \\ 
\langle m | H_I | n \rangle & = & (1 - \delta_{mn}) \langle m | H | n \rangle \ .
\eea
The above definitions obviously imply that,  
if the correlated states have large overlaps with the eigenstates of $H$, the matrix elements 
of $H_I$ are small, and the perturbative expansions in powers of $H_I$ is rapidly convergent.

Let us consider, for example, the Green function describing the propagation of a nucleon in a 
hole state \cite{FetterWalecka} 
\beq
\label{GF:1}
G({\bf k},\omega) = \langle \widetilde{0} \vert a^\dagger_{{\bf k}} \
\frac{1}{H-E_0-\omega-i\eta} \ a_{{\bf k}} \vert \widetilde{0} \rangle / 
\langle \widetilde{0} \vert \widetilde{0} \rangle \ .
\eeq
In the above equation, $\eta=0^+$, $a^\dagger_{{\bf k}}$ and $a_{{\bf k}}$ are creation
and annihilation operators and the {\it exact} ground state $\vert \widetilde{0} \rangle$, 
satisfying the
Schr\"odinger equation $H \vert \widetilde{0} \rangle = E_0 \vert \widetilde{0} \rangle$
can be obtained from the expansion \cite{BFF1,BFF2}
\beq
\vert \widetilde{0} \rangle = 
\sum_n(-)^n \left(
\frac { H_I-\Delta E_0}{ H_0-E_0^V } \right)^n\vert 0 \rangle \ ,
\label{GF:2}
\eeq
where $\Delta E_0 = E_0 - E^V_0$, with $E^V_0$ defined by Eq.(\ref{def:E0V}).

In principle, using Eq.(\ref{GF:2}) and the similar expansion \cite{BFF1,BFF2}
\beq
\frac{1}{H-E_0-\omega-i\eta} =
\frac{1}{H_0-E^V_0-\omega-i\eta} \
\sum_n (-)^n \left( \frac{H_I - \Delta E_0}{H_0-E^V_0-\omega-i\eta} \right)^n \ ,
\label{GF:3}
\eeq
the Green function can be consistently computed at any order in $H_I$. 
However, the calculation of the matrix elements of the hamiltonian 
appearing in Eqs.(\ref{GF:1})-(\ref{GF:3}) involves prohibitive difficulties and 
requires the development of a suitable approximation scheme, to be discussed in the 
following Section.

\section{Cluster expansion formalism}

The correlation operator of Eq.(\ref{def:F}) is defined such that, 
if any subset of the particles, say $i_1, \ldots i_p$, is removed far from the 
remaining $i_{p+1}, \ldots i_N$, it factorizes according to
\beq
F(1,\ldots,N) \rightarrow F_p(i_1, \ldots i_p) F_{N-p}(i_{p+1}, \ldots i_N) \ .
\eeq
The above property is the basis of the cluster expansion formalism, that allows 
one to write the matrix element of a many-body operator between correlated states as a sum, 
whose terms correspond to contributions arising from isolated subsystems 
({\it clusters}) involving an increasing number of particles. 

Note that in this Section we will use {\em non normalized} correlated states, defined as
(compare to Eq.(\ref{def:corr_states}))
\beq
| n \rangle = F | n_{FG} \rangle \ .
\label{nonnorm}
\eeq

\subsection{Ground state expectation value of the hamiltonian}

Let us consider, as an example, the expectation value of the hamiltonian $H$ in the correlated
state $\vert 0 \rangle$, defined in Eq.(\ref{nonnorm}). We will closely follow the 
derivation of the corresponding cluster expansion given in Ref. \cite{CBF1} and neglect, 
for the sake of simplicity, the three body potential $V_{ijk}$. Under this assumption, 
we can write the hamiltonian as in Eq.(\ref{NMBT:H}).

The starting point is the definition of the generalized normalization integral 
\beq
\label{CE:1}
I(\beta) = \langle 0 \vert {\rm exp} [ \beta(H-T_0) ] \vert 0 \rangle \ ,
\eeq
where 
\beq
T_0 = \sum_{|{\bf p}|<p_F} \frac{ {\bf p}^2 }{2m}  \ ,
\eeq 
$p_F$ being the Fermi momentum, is the ground state energy of the noninteracting 
Fermi gas at density $\rho = 2p_F^3/3\pi^2$. Using the definition of Eq.(\ref{CE:1}) we 
can rewrite the expectation value of the hamiltonian in the form
\beq
\langle H \rangle = \frac{\langle 0 \vert H \vert 0 \rangle}{\langle 0 \vert 0 \rangle}  
= T_0 + \left. \frac{\partial}{\partial \beta} \ln I(\beta) \right|_{\beta=0}\ . 
\eeq
The cluster property of $F$ can be exploited to define a set of $N!/(N-p)!p!$ 
sub-normalization integrals, associated with each $p$-particle subsystem ($p=1,\ldots,N$)
\begin{eqnarray}
\nonumber
I_i(\beta) & = & \langle i \vert {\rm exp} [ \beta(t(1)-\epsilon^0_i)] \vert i \rangle \ ,\\
\nonumber
I_{ij}(\beta) & = & \langle i j \vert F^\dagger_2(12) 
 {\rm exp} [ \beta(t(1)+t(2)+v(12)-\epsilon^0_i-\epsilon^0_j)] F_2(12) \vert i j \rangle_a \ ,\\
\nonumber
 & \vdots &  \\
I_{1\ldots N}(\beta) & = & I(\beta) \ ,
\label{GPC0}
\end{eqnarray}
where the indices $i,j,\ldots$ label states belonging to the Fermi sea, 
 the ket $|i_1 \ldots i_n \rangle$ describes $n$ non interacting particles in the 
states $i_1 \ldots i_n$, $\epsilon^0_i=|{\bf p}|_i^2/2m$ is the 
kinetic energy eigenvalue associated with the state $\vert i \rangle$ and the subscript $a$ 
refers to the fact that the corresponding state is antisymmetrized. For example, in the case of
two particles
\beq
\vert i j \rangle_a = \frac{1}{\sqrt{2}} \ \left( \vert i j \rangle - 
\vert j i \rangle \right) \ .
\eeq
To express $\ln I(\beta)$ in terms of the $\ln I_{i_1 \ldots i_p}(\beta)$, we start noting that 
$I_{ij}$ is close to the product of $I_i$ and $I_j$. It would be exactly equal if we could neglect 
the interaction, described by the potential $v(12)$, and the correlations induced by {\em both}
 $F_2(12)$ {\em and} Pauli exclusion principle. This observation suggests that $I_{ij}$ can be 
written as
\beq
\label{FIY1}
I_{ij} = I_i I_j Y_{ij} \ ,
\eeq
with $Y_{ij} \sim 1$. Extending the same argument to the $I$'s with more than two indices, we obtain
\bea
\nonumber
I_i & = & Y_i \\
\nonumber
I_{ij} & = & Y_i Y_j Y_{ij} \\
\nonumber
 & \vdots &  \\
I_{1\ldots N} & = & I = \prod_i Y_i \ \prod_{j>i} Y_{ij} \ldots Y_{1 \ldots N} \ , 
\label{FIY2}
\eea
implying 
\beq
\label{CE:2}
\ln I(\beta) = \sum_i \ln Y_i + \sum_{j>i} \ln Y_{ij} + \ldots + \ln Y_{1 \ldots N} \ .
\eeq
It can be shown \cite{CBF1} that each term in the rhs of Eq.(\ref{CE:2}) goes like $N$ in the 
thermodynamic limit. In addition, the $p$-th term collects all 
contributions to 
the cluster development of $\ln I(\beta)$ involving, in a connected manner, exactly $p$ 
Fermi sea orbitals. Therefore, the $p$-th term can be referred to as the $p$-body cluster 
contribution to $\ln I(\beta)$.

The decomposition (\ref{FIY2}) allows one to rewrite the expectation value of the hamiltonian
in the form
\beq
\langle H \rangle = T_0 + (\Delta E)_2 + (\Delta E)_3 + \ldots  + (\Delta E)_N
\label{expansion:H}
\eeq
with 
\beq
\label{def:deltaE}
(\Delta E)_p = \sum_{i_1 < i_2 < \ldots < i_p}  \left. \frac{\partial}{\partial \beta} 
\ln Y_{i_1 i_2 \ldots i_p} \right|_{\beta=0} \ .
\eeq

Note that $(\Delta E)_1 = 0$, as the above definitions imply
\beq
I_i = Y_i = 1 \ ,
\eeq
and
\beq
\frac{\partial I_i}{\partial \beta} = 0 \ .
\eeq

To make the last step we have to use Eq.(\ref{FIY2}) and express $(\Delta E)_p$ in 
terms of the $I_{i_1 \ldots i_p}$. Substitution of the resulting expressions 
\bea
\nonumber
Y_i & = & I_i \\
Y_{ij} & = & I_{ij} (I_i I_j)^{-1}  \ ,\\ 
& \vdots &
\eea
into Eq.(\ref{def:deltaE}) with $p=2$ yields 
\beq
\label{def:deltaE2}
(\Delta E)_2 = \sum_{i<j} \left[ \frac{1}{I_{ij}} \frac{\partial I_{ij}}{\partial \beta} 
    - \left( \frac{\partial I_{i}}{\partial \beta} + \frac{\partial I_{j}}{\partial \beta} 
 \right) \right]_{\beta=0}  \ .
\eeq
The ``normalizations'' $I_{ij}|_{\beta=0}$ appearing in the denominator differ from unity 
by terms $O(1/N)$ at most, that can be disregarded in the $N \rightarrow \infty$ limit.
As a result, we obtain
\beq
(\Delta E)_2 = \sum_{i<j} \ w_{ij} \ ,
\eeq
where (see Eq.(\ref{GPC0}))
\beq
\label{def:w}
w_{ij} = \langle i j \vert \ \frac{1}{2} 
[ F_2(12) ,\ [ t(1) + t(2) ,\  F_2(12) ] \ ]+  F_2(12) v(12) F_2(12)  \ 
\vert i j \rangle_a \ , 
\eeq
Note that in the above equation we have assumed that the correlation operator be hermitian, 
i.e. that $F_2(12)=F^\dagger_2(12)=f_{12}$ (see Eq.(\ref{def:F})). The explicit expression of
$(\Delta E)_2$, in the case of six component potential and correlation operator, 
is given in Appendix \ref{2body}.

Each term of the expansion (\ref{expansion:H}) can be represented by a diagram featuring 
$p$ vertices, representing the nucleons in the cluster, connected by lines corresponding to 
dynamical and statistical correlations. The terms in the resulting diagrammatic expansion
can be classified according to their topological structure, and selected classes of diagrams 
can be summed up to all orders solving a set of coupled integral equations, called
Fermi hyper-netted chain (FHNC) equations \cite{FA:RO,PW:RMP}. 

\subsection{Transition matrix elements}

The nuclear matter response to an external probe delivering energy $\omega$ 
and momentum ${\bf q}$ can be written in the form
\beq
S({\bf q}, \omega) = \sum_f |\langle f | \mathcal{O}({\bf q}) | 0 \rangle |^2 
\delta(\omega + E_0 - E_f) \ ,
\label{def:densresp}
\eeq
where $\mathcal{O}$ is the operator inducing a transition from the ground state 
$| 0 \rangle$, carrying 
energy $E_0$, to a final state $| f \rangle$, carrying energy $E_f$. In the simple case
of interaction with a scalar probe, resulting in a density fluctuation 
\beq
\mathcal{O}({\bf q}) = \rho({\bf q}) = \sum_{{\bf k}} a^\dagger_{{\bf k}+{\bf q}} a_{{\bf k}} \ ,
\label{def:rho}
\eeq
$a^\dagger_{{\bf k}}$ and $a_{{\bf k}}$ being nucleon creation and annihilation operators, respectively.

In order to obtain the response within the CBF approach, the cluster expansion formalism 
discussed in the previous section must be extended to the case of transition matrix elements.

Consider a (non normalized) correlated one particle-one hole final state
\beq
| f \rangle = | {\bf p} {\bf h} \rangle = 
F a^\dagger_{{\bf p}} a_{{\bf h}}| 0_{FG}\rangle \ .
\eeq
To obtain the response we need to calculate the matrix elements
\beq
\frac{\langle 0 |\rho_{{\bf q}}^\dagger | {\bf p} {\bf h} \rangle }{ \< 0 | 0 \>^{1/2} 
\langle {\bf p} {\bf h} | {\bf p} {\bf h} \rangle^{1/2} }
\eeq
The cluster expansion of the above quantity can be carried out using a formalism somewhat
different from the one described in the previous Section, originally developed in 
Ref. \cite{CW}. The starting point is again the generalized normalization integral, that in 
this case is written in the form
\beq
I_{0,ph}(\beta,\alpha) = \sqrt{N!} \int \prod_{i=1}^N dx_i 
 \ \Phi_0^\dagger(1,\ldots,N) F^\dagger \ {\rm e}^{\beta \rho_{{\bf q}}^\dagger } F \  
 {\rm e}^{\alpha W_{ph}} \ \phi_{m_1}(1)\ldots \phi_{m_N}(N) \ ,
\label{GNI}
\eeq
In the above equation $x_1, \ldots , \ x_N$ denote the nucleon degrees of freedom, 
and $\Phi_0^\dagger(1,\ldots,N) = \langle x_1 \ldots  x_N | 0_{FG} \rangle$ is the FG ground 
state wave function, i.e. the antisymmetrized product of the single particle orbitals
$\phi_{m_1}(1)\ldots \phi_{m_N}N(N)$. Note that the calculation of matrix elements involving
states describing Fermi systems, only requires the antisymmetrization of either the initial or
the final state. The coordinate space expressions of the 
operators $\rho_{{\bf q}}$ and $W_{ph}$ are
\beq
\rho_{{\bf q}} = \sum_{i=1}^N \ {\rm e}^{i{\bf q}\cdot{\bf r}_i} \ = 
\sum_{i=1}^N \ \rho_{{\bf q}}(i) \ ,
\eeq
\beq
W_{ph} = \sum_{i=1}^N \ \frac{\phi_p(i)}{\phi_{m_i}(i)} \delta_{h m_i} = 
\sum_{i=1}^N \ W_{ph}(i) \ .
\eeq
Acting on the product $\phi_{m_1} \ldots \phi_{m_N}$, $W_{ph}$ replaces the hole state 
orbital $\phi_{m_i}$ with the particle state orbital $\phi_p$.
In terms of generalized normalization integrals we can write
\beq
\langle {\bf p}{\bf h} |\rho_{{\bf q}}^\dagger| 0 \rangle = \int \prod_{i=1}^N dx_i 
\ \Phi_0^\dagger F^\dagger \rho_{{\bf q}}^\dagger \ F \Phi_0 = \left.
\frac{\partial}{\partial \beta} \frac{\partial I_{0,ph}(\beta,\alpha)}{\partial \alpha} 
\right|_{\alpha=\beta=0} \ ,
\eeq
\beq
\langle {\bf p}{\bf h} | {\bf p}{\bf h} \rangle = \int \prod_{i=1}^N dx_i 
 \Phi_{ph}^\dagger F^\dagger \ F \Phi_{ph} = I_{ph,ph}(0,0)
\eeq
where $I^{ph,ph}(0,\alpha)$ is obtained from Eq.(\ref{GNI}) replacing the FG ground state 
with the one particle-one hole state $\Phi_{ph}$, and 
\beq
\langle 0 | 0 \rangle = \int \prod_{i=1}^N dx_i
 \Phi_0^\dagger F^\dagger \ F \Phi_0 =I_{0,0}(0,0) \ .
\eeq

To carry out the cluster expansion we need to define subnormalization integrals, involving
an increasing number of orbitals
\bea
\nonumber
I_i & = & \int dx_1 \phi_{m_i}(1) {\rm e}^{\beta \rho_{{\bf q}}^\dagger(1)} 
{\rm e}^{\alpha W_{ph}(1)} \phi_{m_i}(1) = X_i\\
\nonumber
I_{ij} & = & \sqrt{2} \int dx_1 dx_2 \ \varphi_{m_i m_j}^\dagger(1,2) F_2^\dagger(12)  
{\rm e}^{\beta [\rho_{{\bf q}}^\dagger(1)+  \rho_{{\bf q}}^\dagger(2)]} F_2(12) \\
\nonumber
& \times & {\rm e}^{\alpha [W_{ph}(1) + W_{ph}(2)]} \phi_{m_i}(1) \phi_{m_j}(2) 
 = X_i X_j  + X_{ij} \\
\nonumber
 & \vdots &  \\
I_{1\ldots N} & = & I^{0,ph}(\beta,\alpha) = 
\sum \ X_{i_1 \ldots a} \ldots X_{j_1 \ldots b} \ ,
\label{IY}
\eea
where $\varphi_{m_i m_j}(1,2) = [\phi_{m_i}(1) \phi_{m_j}(2) - \phi_{m_i}(2) \phi_{m_j}(1)]/\sqrt{2})$, 
and the sum in the last line is extended to all partitions such that $a~+~\ldots~+~b~=~N$.
Note that in this case the (small) deviation between $I_{ij}$ and the product $I_i I_j$ is 
characterized through their difference, rather than the ratio (compare to Eq.(\ref{FIY2})). 

The thermodynamic limit (i.e. the limit $A,V \rightarrow \infty$, with $A/V = {\rm const}$) 
of $I_{0,ph}(\beta, \alpha)$ can be best identified rewriting it in the form originally obtained 
in Ref. \cite{CW}:
\beq
I_{0,ph}(\beta,\alpha) = \prod_{i=1}^N X_i(\beta,\alpha) {\rm e}^{G_{0,ph}(\beta,\alpha)} \ ,
\eeq
with
\beq
G_{0,ph} = \sum_{j>i} \xi_{ij} + O(N^{-1})\ ,
\eeq
where 
\beq
\xi_{ij} = \frac{X_{ij}}{X_iX_j} \ .
\eeq
Defining 
\beq
\mathcal{P}_{0,ph}(\beta) = 
\left[ \frac{\partial \ {\rm e}^{G_{0,ph}(\beta,\alpha)}}{\partial \alpha} \right]_{\alpha=0}
 \ {\rm e}^{-[ G_{ph,ph}(0,0) + G_{0,0}(0,0)]/2} 
\eeq
we finally obtain \cite{CW}
\beq
\frac{\langle 0 |\rho_{{\bf q}}^\dagger | {\bf p} {\bf h} \rangle }{ \< 0 | 0 \>^{1/2}
\langle {\bf p} {\bf h} | {\bf p} {\bf h} \rangle^{1/2} } = 
\langle ({\bf p} {\bf h})_{FG} | \rho_{{\bf q}}^\dagger | 0_{FG} \rangle \ 
\mathcal{P}_{0,ph}(0) + \left. \frac{ \partial}{\partial \beta} 
\mathcal{P}_{0,ph}(\beta) \right|_{\beta=0} \ .
\eeq
From the above equations it follows that, at two-body cluster level \cite{CW},
\beq
\frac{\langle 0 |\rho_{{\bf q}}^\dagger | {\bf p} {\bf h} \rangle }{ \< 0 | 0 \>^{1/2}
\langle {\bf p} {\bf h} | {\bf p} {\bf h} \rangle^{1/2} } =
 \sum_i \left. \frac{\partial}{\partial \beta} 
\frac{\partial X_i}{\partial \alpha} \right|_{\beta=\alpha=0}+ \sum_{j>i} 
\left. \frac{\partial}{\partial \beta} 
 \frac{\partial X_{ij}}{\partial \alpha} \right|_{\beta=\alpha=0} \ .
\label{matel:final}
\eeq
Note that the derivation of Eq.(\ref{matel:final}) has been carried out consistently with 
that of Eq.(\ref{def:deltaE2}), i.e. neglecting all contributions $O(N^{-1})$ to the 
normalization of the correlated two-nucleon states.

\section{Effective interaction}

At lowest order of CBF, the effective interaction $V_{{\rm eff}}$ is {\em defined} by the equation
\beq
\label{veff:1}
\langle H \rangle = \langle 0_{FG} | T_0 + V_{{\rm eff}}| 0_{FG} \rangle  \ .
\eeq
As the above equation suggests, the approach based on the effective interaction allows one 
to obtain any nuclear matter observables using perturbation theory in the 
FG basis. However, as discussed in the previous Section, the calculation of the hamiltonian 
expectation value in the correlated ground state, needed to extract $V_{{\rm eff}}$ from
Eq.(\ref{veff:1}), involves severe difficulties. 

In this Thesis we follow the procedure developed in Refs. \cite{shannon1,shannon2}, whose 
authors derived the expectation value of the effective interaction by carrying out a 
cluster expansion of the rhs of Eq.(\ref{veff:1}), and 
keeping only the two-body cluster contribution. The resulting expression, that can be 
obtained from Eqs.(\ref{def:deltaE})-(\ref{def:w}) through a simple rearrangement of the 
kinetic energy contributions, reads
\begin{eqnarray}
\nonumber
& & \langle 0_{FG} | V_{{\rm eff}} | 0_{FG} \rangle = 
\sum_{i<j} \langle ij | v_{{\rm eff}}(12) | ij \rangle_a  \\ 
& & \ \ \ \ = \sum_{i < j} \langle ij | f_{12} \left[ -\frac{1}{m} (\nabla^2 f_{12}) 
 - \frac{2}{m} (\bm{\nabla} f_{22}) \cdot \bm{\nabla} + 
v(12) f_{12} \right] | ij \rangle_a \ ,
\label{veff:2}
\end{eqnarray}
where the laplacian and the gradient operate on the relative coordinate. 
Note that $v_{{\rm eff}}$ defined by the above equation exhibits a momentum dependence 
due to the operator $(\bm{\nabla} f_{ij}) \cdot \bm{\nabla}$, yielding contributions
to nuclear matter energy through the exchange terms \footnote{The direct contribution
is vanishing, as it involves the integration of an odd function of 
${\bf k} = {\bf k}_i - {\bf k}_j$.}. However, 
our numerical calculations show that these contributions are small, compared to 
the ones associated with the momentum independent terms. As a consequence, the results 
presented in this Thesis have been obtained using only the static part of the effective 
interaction (\ref{veff:2}), i.e. setting 
\beq
\label{veff:3}
v_{{\rm eff}}(ij) =  f_{ij} \left( -\frac{1}{m} \nabla^2 + v(ij) \right) f_{ij}  
 = \sum_n v_{{\rm eff}}^{n}(r_{ij}) O^n_{ij} \ ,
\eeq
The properties of the operators $O^n$ with $n = 1, \ldots,\  6$, leading to the above result,  
are given in Appendix \ref{pauli}.

The definition of $v_{{\rm eff}}$ given by Eqs.(\ref{veff:2}) and (\ref{veff:3}) 
obviously neglects the 
effect of three-nucleon interactions, whose inclusion in the hamiltonian is known to be
needed in order to explain the binding energies of the few-nucleon systems, as well as
the saturation properties of nuclear matter. To circumvent this problem, we have used 
the approach originally proposed by Lagaris and Pandharipande \cite{LagPan}, in which the 
main effect of the three-body force is simulated through a density dependent modification 
of the two-nucleon potential at intermediate range, where two-pion exchange is believed 
to be the dominant interaction mechanism. Neglecting, for simplicity, the charge-symmetry
breaking components of the interaction, the resulting potential can be written in the form
\beq
\label{TNR}
\widetilde{v}(ij) = \sum_{n=1,14} \left[ v^n_\pi(r_{ij})
+ v^n_I(r_{ij}){\rm e}^{-\gamma_1 \rho} + v^n_S(r_{ij}) \right] O^n_{ij} \ ,
\eeq
where $v^n_\pi$, $v^n_I$ and $v^n_S$ denote the long- (one-pion-exchange), intermediate-
and short-range part of the potential, respectively. The above modification results in a 
repulsive contribution to the binding energy of nuclear matter.
The authors of Ref.\cite{LagPan} also include the small additional attractive contribution 
\beq
\label{TNA}
\Delta E_{TNA} = \gamma_2 \rho^2 (3-2\beta^2){\rm e}^{-\gamma_3 \rho} \ , 
\eeq
with $\beta = (\rho_p - \rho_n)/(\rho_p + \rho_n)$, where $\rho_p$ and $\rho_n$ denote the 
proton and neutron density, respectively. The values of the parameters 
$\gamma_1$, $\gamma_2$ and $\gamma_3$ appearing in 
Eqs.(\ref{TNR}) and (\ref{TNA}) have been determined in such a way as to reproduce the 
binding energy and equilibrium density of nuclear matter \cite{LagPan}.

Besides the bare two- and three body-potentials, the effective interaction is determined by 
the correlation operators $f_{ij}$ defined by Eq.(\ref{def:corrf}).
The shapes of the radial functions $f^n(r_{ij})$ are obtained from the functional 
minimization of the energy at the two-body cluster level, yielding a set of coupled
differential equations to be solved with the boundary conditions
\beq
\label{bound0:f}
f^n(r_{ij} \geq d) = \left\{
\begin{array}{ll}
1  \ , & n = 1 \\
0  \ ,& n = 2,3,4 \\
\end{array}
\right. \ , 
\eeq
\beq
\label{bound1:f}
f^n(r_{ij} \geq d_t) = 0 \ , \ n = 5,6
\eeq
and 
\bea
\nonumber
\left. \frac{df^n}{dr_{ij}} \right|_{r_{ij}=d} & = & 0  \ , \ n = 1,2,3,4 \\
\left. \frac{df^n}{dr_{ij}} \right|_{r_{ij}=d_t} & = & 0 \ , \ n = 5,6 \ ,
\label{bound2:f}
\eea
$d$ and $d_t>d$ being variational parameters. The above conditions simply express the
requirements that i) for relative distances larger than the interaction range
the two-nucleon wave function reduces to the one describing non interacting particles 
and ii) tensor interactions have longer range. 

For any given value of nuclear matter density, we have solved the Euler-Lagrange 
equations resulting from the minimization of the binding energy at two-body cluster level, 
whose derivation and explicit form is given in Appendix \ref{corrfcn},
using the values of $d$ and $d_t$ reported in Ref.\cite{AKMAL}.
\begin{figure}[hbt]
\vspace{0.1cm}
\begin{center}
{\epsfig{figure=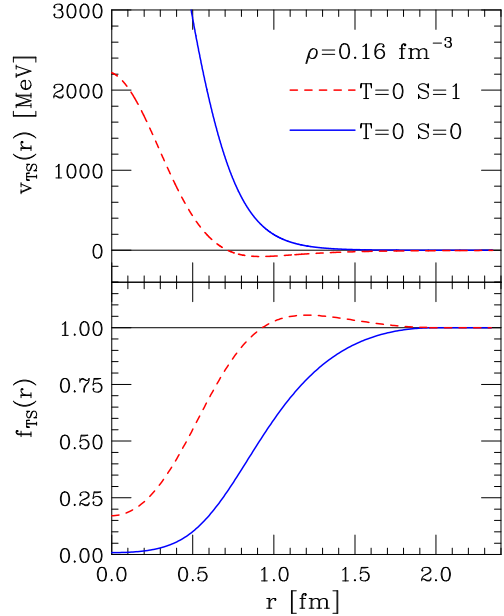,width=6.5cm }}
\vspace{0.1cm}
\caption{{\small
Interaction potentials (upper panel) and correlation functions (lower panel)
acting in the spin-isospin channels $S=0$ and $T=0$ (solid lines)
and $S=0$ and $T=1$ (dashed lines). The potential is the Argonne $v^\prime_8$ and
the correlation functions correspond to nuclear matter at equilibrium density. }
\label{corrf} }
\end{center}
\end{figure}

As an example, the results corresponding to nuclear matter at equilibrium density 
are illustrated in Fig.\ref{corrf}, showing the central component of the correlation 
functions acting between 
a pair of nucleon carrying total spin and isospin $S$ and $T$, respectively.
The relations between the $f_{TS}$ of Fig.\ref{corrf} and the $f^n$ of Eq.(\ref{def:corrf})
are given in Appendix \ref{pauli}. The shapes of the $f_{TS}$ clearly reflect the nature 
of the interaction. In the $T=0$ $S=0$ channel, in which the potential exhibits a strong 
repulsive core, the correlation function is very small at $r \lsim 0.5$ fm. On the other 
hand, in the $T=0$ $S=1$ channel, the spin-isospin state corresponding to the deuteron, 
the repulsive core is much weaker and the potential becomes attractive 
at $r \gsim 0.7$ fm. As a consequence, the correlation function does not approach zero 
as $r \rightarrow 0$ and exceeds unity at intermediate range.

\begin{figure}[h!]
\vspace{0.1cm}
\begin{center}
\psfig{file=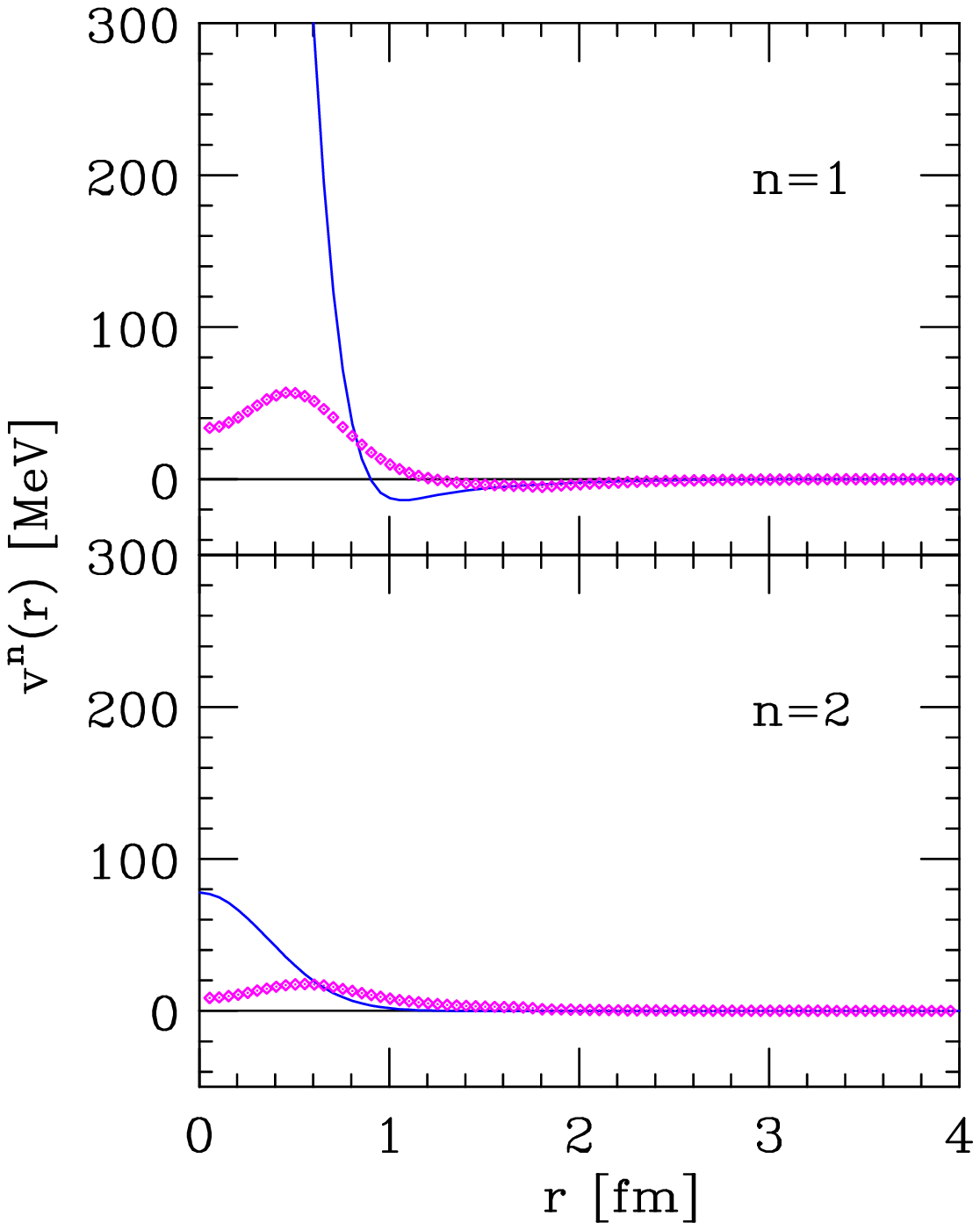,width=6.4cm,angle=0}\quad\quad\quad
\psfig{file=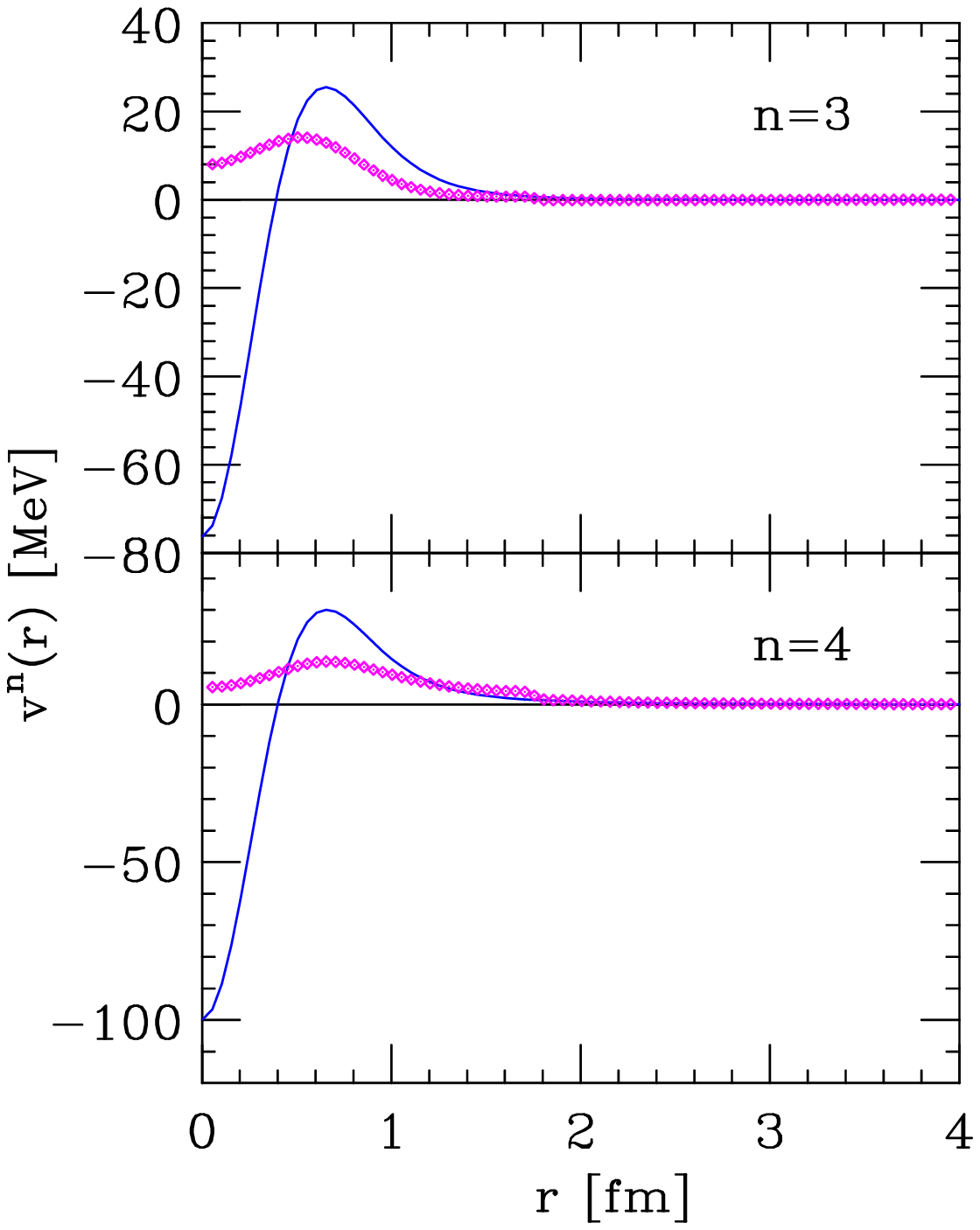,width=6.4cm,angle=0} 
\end{center}
\vspace{0.1cm}
\begin{center}
\psfig{file=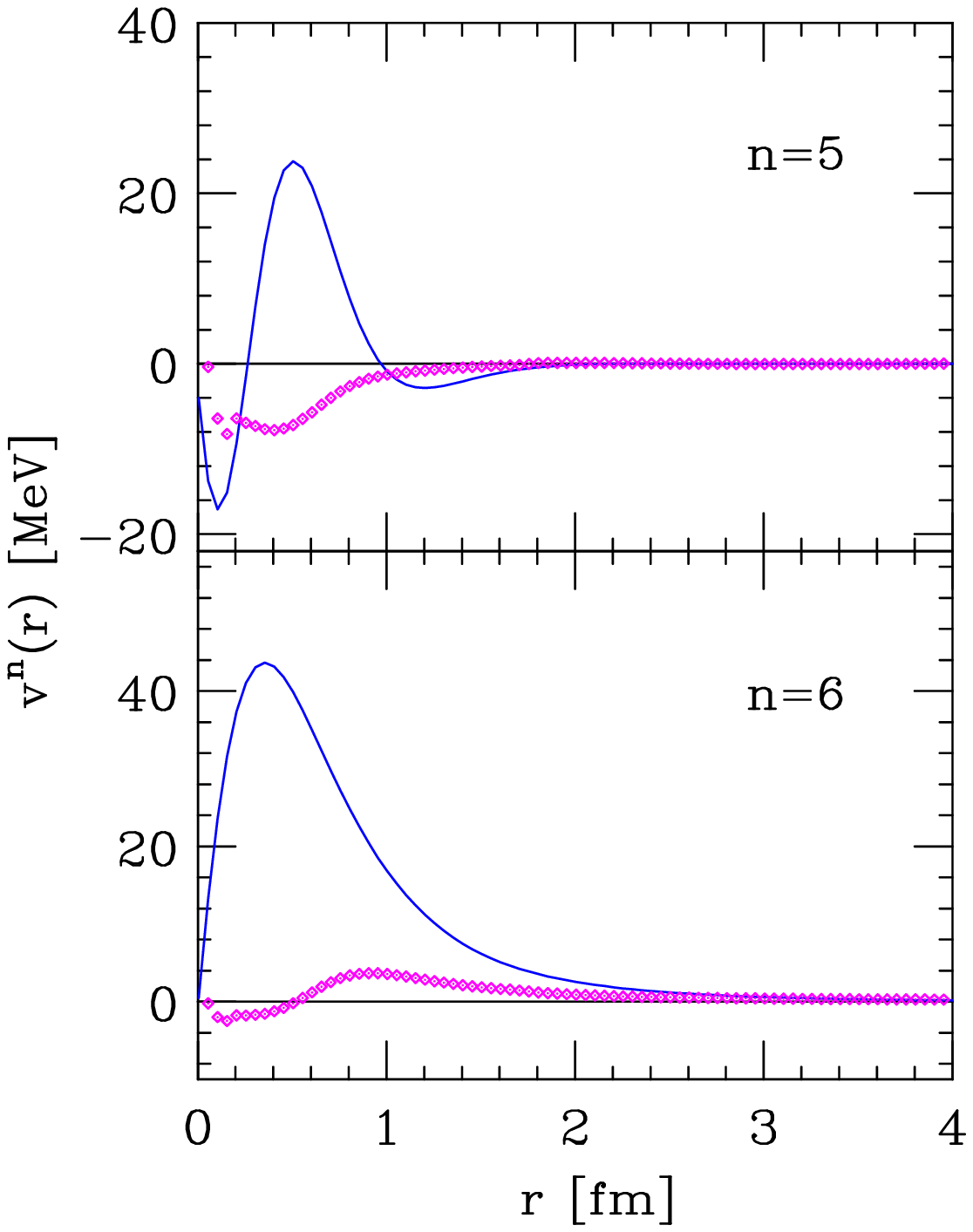,width=6.4cm,angle=0}
\vspace{0.1cm}
\caption{{\small Comparison between the components of the bare Argonne $v^\prime_8$ potential 
(dashed lines) and the effective potential defined by Eq.(\ref{veff:3}) (solid lines), 
calculated at nuclear matter equilibrium density.} \label{veff1} }
\end{center}
\end{figure}

In Fig.\ref{veff1} the components of the effective interaction at equilibrium density
are compared to the corresponding components of the truncated $v_8^\prime$ potential.
It clearly appears that screening effects due to NN correlations lead to a significant
quenching of the interaction.

\begin{figure}[hbt]
\vspace{0.2cm}
\begin{center}
\psfig{file=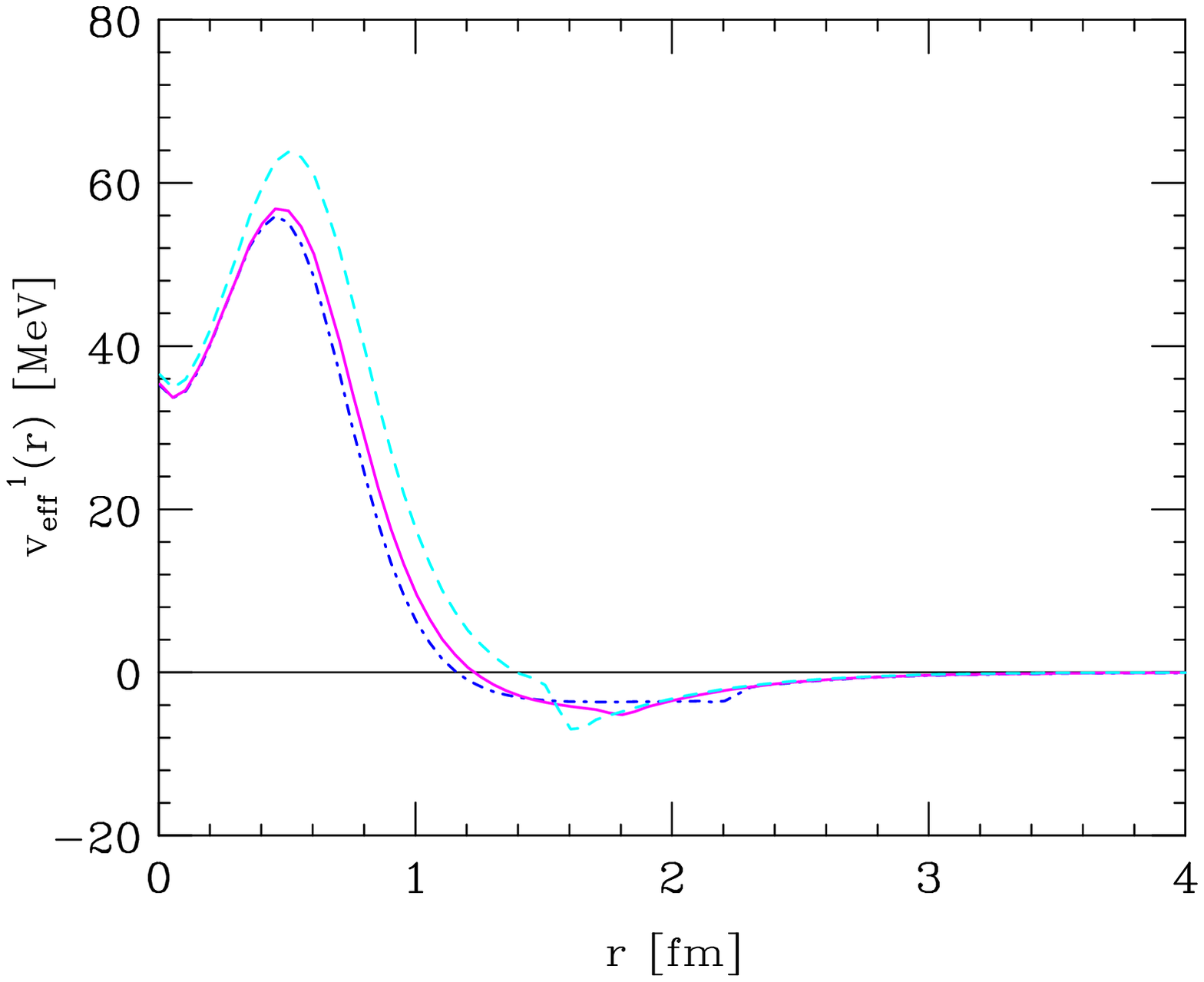,width=6.8cm,angle=0}\quad
\psfig{file=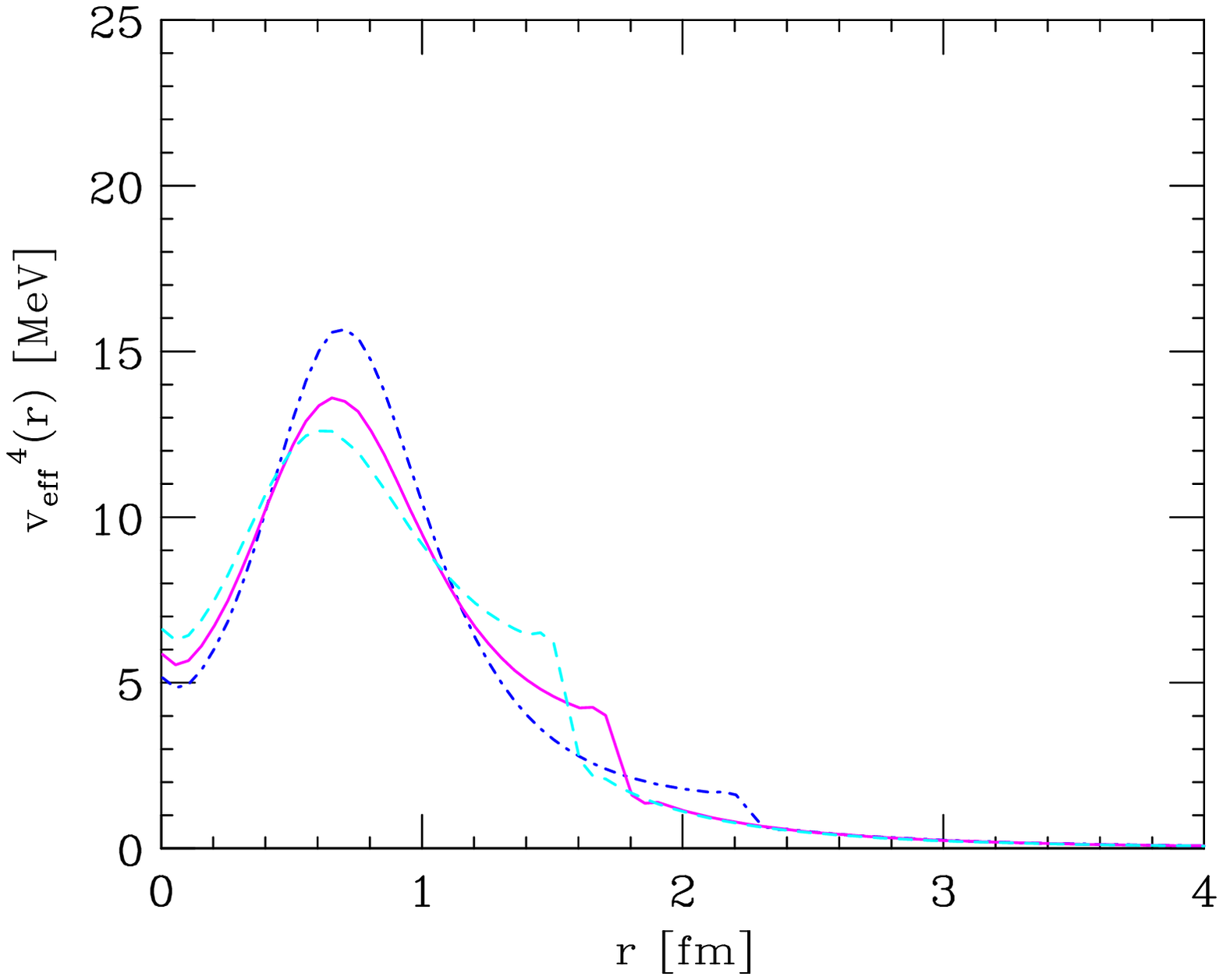,width=6.8cm,angle=0}
\vspace{0.1cm}
\caption{{\small Density dependence of the central ($n=1$, left panel) and spin-isospin 
($n=4$, right panel) components of the effective interaction of Eq.(\ref{veff:3}),
calculated using the Argonne $v^\prime_8$ potential. }
The dot-dash, dashed and solid lines correspond to $\rho =$ 0.04, 0.16 and 0.32 fm$^{-3}$, 
respectively.  \label{veff2} }
\end{center}
\end{figure}

Figure \ref{veff2} shows a comparison between the central ($n=1$, left panel) and spin-isospin
($n=4$, right panel) components of the effective interaction of Eq.(\ref{veff:3}),
calculated at different densities using the Argonne $v^\prime_8$ potential. The density
dependence is associated with the correlation functions, which depend on $\rho$ through 
the correlation ranges, $d$ and $d_t$, and the Fermi distributions. 
The jumps in the radial behavior of the effective interactions, clearly visible in 
Figs. \ref{veff1} and \ref{veff2} are due to the discontinuity in the second derivative 
of the correlation functions. 

\subsection{Energy per particle of neutron and nuclear matter}

The effective interaction described in the previous Section was tested by computing the energy
per particle of symmetric nuclear matter and pure neutron matter in first order perturbation
theory using the FG basis. 

Let us consider nuclear matter at density
\beq
\rho = \sum_{\lambda=1}^4 \ \rho_\lambda  \ ,
\eeq
where $\lambda = 1,2,3,4$ labels spin-up protons, spin-down protons, spin-up neutrons and
spin-down neutrons, respectively, the corresponding densities being
$\rho_\lambda = x_\lambda \rho$, with $\sum_\lambda x_\lambda = 1$.
For example, for symmetric nuclear matter $x_1=x_2=x_3=x_4=1/4$, while for pure 
neutron matter $x_1=x_2=0$ and $x_3=x_4=1/2$.
Within our approach, the energy of such a system can be obtained from
\beq
\frac{E}{N} = \frac{3}{5} \ \sum_\lambda \frac{ p_{F,\lambda}^2}{2m} 
 + \frac{\rho}{2} \sum_{\lambda \mu} \sum_n \ x_\lambda x_\mu \int d^3r \ v^n_{{\rm eff}} 
   \left[  A^n_{\lambda \mu} - B^n_{\lambda \mu} \ell( p_{F,\lambda} r) \ell( p_{F,\mu} r)  \right] \ .
\label{energy:nm}
\eeq
In the above equation, $p_{F,\lambda}=(6 \pi^2 \rho_\lambda)^{1/3}$ and the Slater function $\ell$ is
defined as
\beq
\ell(p_{F,\lambda} r) = \sum_{\bf k} \ {\rm e}^{i {\bf k}\cdot{\bf r}}  \ 
  \theta(p_{F,\lambda} - |{\bf k}|)\ .
\label{slater:l}
\eeq
The explicit expression of the matrices
\beq
A^n_{\lambda \mu} = \langle \lambda \mu \vert O^n \vert \lambda \mu \rangle \ \ , \ \ 
B^n_{\lambda \mu} = \langle \lambda \mu \vert O^n \vert \mu \lambda \rangle \ ,
\eeq
where $\vert \lambda \mu \rangle$ denotes the two-nucleon spin-isospin state, 
is given in Appendix \ref{pauli}.

\begin{figure}[htb]
\vspace*{.25cm}
\centerline
{\epsfig{figure=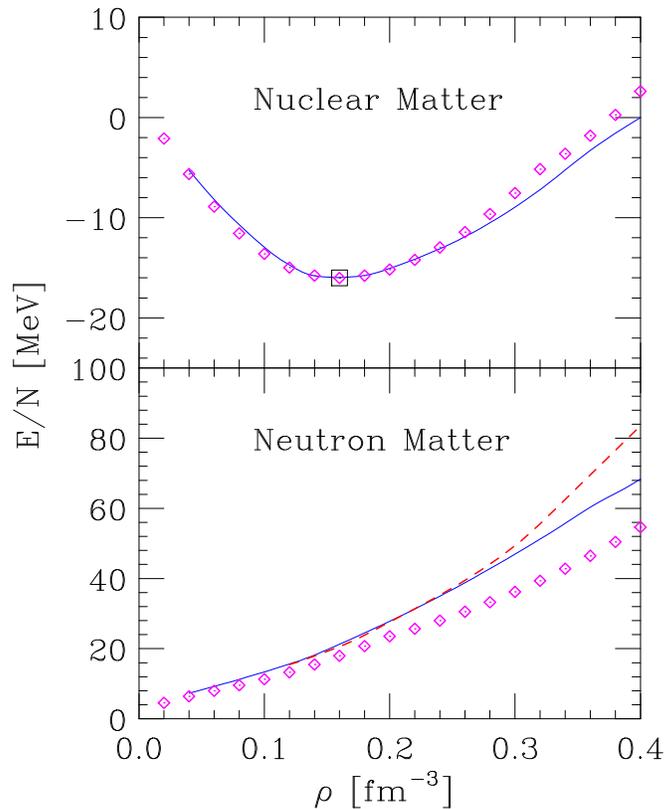,angle=000,width=8.8cm}}
\vspace*{.1cm}
\caption{{\small 
Energy per particle of symmetric nuclear matter (upper panel) and pure neutron matter
(lower panel). The solid lines represent the results obtained using Eq.(\ref{energy:nm}),
whereas the diamonds correspond to the results of Akmal, Pandharipande and
Ravenhall \cite{APR2}. The dashed line of the lower panel represents the results
of the AFDMC approach or Ref. \cite{AFDMC1}. The square in the upper panel
shows the empirical saturation point of symmetric nuclear matter.}\label{energy_new} }
\end{figure}

In Fig. \ref{energy_new} our results are compared to those of
Refs. \cite{APR2} and \cite{AFDMC1}.
The calculations of Ref. \cite{APR2} (diamonds, results given in the sixth column of Table VI) 
have been carried out using a variational
approach based on the FHNC-SOC formalism, with a hamiltonian including the Argonne $v_{18}$ NN
potential and the Urbana IX three-body potential \cite{TBF}. The results of Ref. \cite{AFDMC1}
(dashed line of the lower panel) have been obtained using the $v_8^\prime$ model and the same
three-body potential, within the framework of the Auxiliary field diffusion Monte Carlo (AFDMC)
technique. It appears that the effective interaction approach 
provides a fairly reasonable description of the EOS over a broad density range.

The comparison between the results of our calculation, based on the two-body cluster approximation, 
and those obtained taking into account higher order many-body effects deserves a comment.
In view of the fact that the contribution of clusters involving more than two nucleons is known to be
sizable, our approach has to be regarded as an {\em effective theory}, {\em designed}
to provide {\em lowest order} results in agreement with the available ``data''.
Effective theories are widely employed in many areas of Physics, including nuclear matter 
theory. For example, the Walecka model \cite{walecka_model} is designed to reproduce the 
nuclear matter empirical saturation properties in the mean field approximation, i.e. at tree level, 
although the corresponding loop corrections are known to be large.

It is worth noting that the 
empirical equilibrium properties of symmetric nuclear matter are accounted for without
including the somewhat {\em ad hoc} density dependent correction of Ref. \cite{APR2}.
The authors of Ref. \cite{valli} argued that this may be ascribed to the different
description of the three-body force. It should also be emphasized that, using $v_{{\rm eff}}$ of
Eq.(\ref{veff:3}) and the three-nucleon interaction (TNI) model of Ref. \cite{LagPan}, one effectively 
includes the contribution of clusters involving more than two nucleons. 

In addition to the correct binding energy per nucleon 
and equilibrium density ($E/N=15.96$ MeV at $\rho=0.16$ fm$^{-3}$), our calculation also yields
a quite reasonable value of the compressibility module, $K =$ 230 MeV.

It has to be kept in mind that our approach does not involve adjustable parameters. 
The correlation ranges $d$ and $d_t$ have been taken from Ref. \cite{AKMAL}, while the 
parameters entering the definition of the TNI have been determined 
by the authors of Ref. \cite{LagPan} through a fit of nuclear matter equilibrium properties.

\subsection{Single particle spectrum and effective mass}
\label{deflambda}

\begin{figure}[tb]
\centerline
{\epsfig{figure=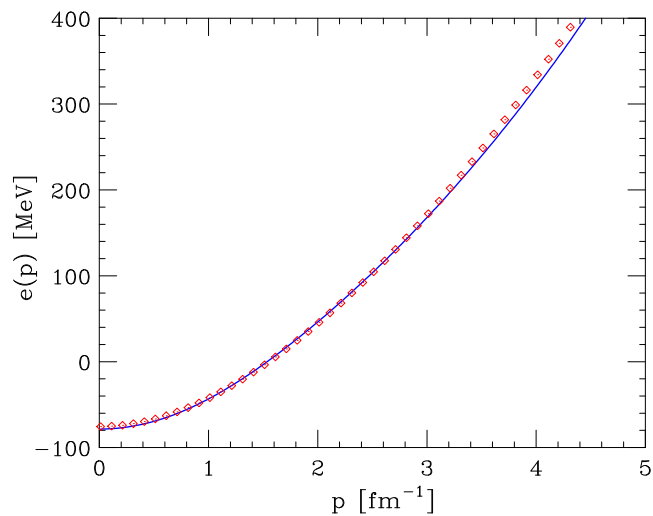,angle=000,width=8.6cm}}
\vspace*{.1cm}
\caption{ {\small Solid Line:
momentum dependence of the single particle energies obtained from the CBF effective 
interaction in the Hartree-Fock approximation (see Eq.(\ref{HF})).
The diamonds show the results of Ref.\cite{bob:ep}. }\label{sp:spectrum} }
\end{figure}

Using the effective interaction, the effective mass can be obtained from the 
single-particle energies $e_\lambda({\bf p})$, that can be easily computed in Hartree-Fock 
approximation \cite{FetterWalecka}. The resulting expression is (compare to Eq.(\ref{energy:nm})):
\beq
e_\lambda(p) = \frac{p^2}{2m} + \rho \ 
 \sum_\mu \sum_n x_\mu \int d^3r  v^n_{\rm eff}(r) \left[ A^n_{\lambda\mu} 
 - B^n_{\lambda\mu} j_0(p r) \ell(p_{F,\mu} r)  \right] \ ,
\label{HF}
\eeq
where $p = |{\bf p}|$ and $j_0$ is the spherical Bessel function: $j_0(x) = \sin(x)/x$.
Figure \ref{sp:spectrum} shows $e(p) = \sum_\lambda x_\lambda e_\lambda(p)$ for symmetric
nuclear matter at equilibrium density. For comparison the corresponding results from
 Ref. \cite{bob:ep} are also displayed. They have been obtained using the FHNC-SOC 
approach and the Urbana $v_{14}$ two-nucleon potential, modified according 
to the TNI model of Lagaris and Pandharipande \cite{LagPan}.

\begin{figure}[h]
\vspace*{.25in}
\centerline
{\epsfig{figure=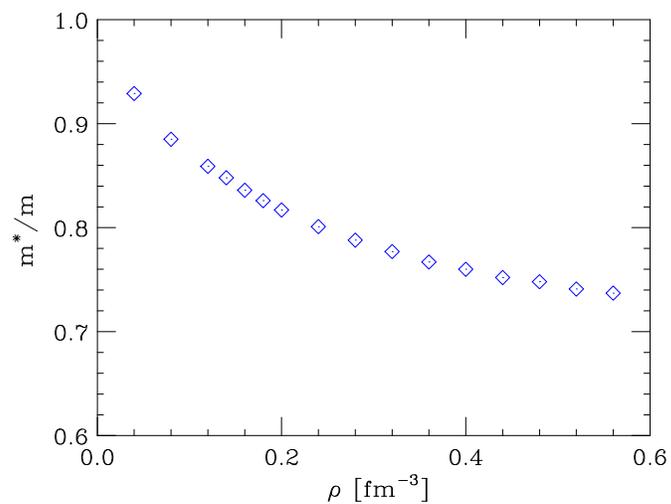,angle=000,width=8.8cm}}
\vspace*{.1cm}
\caption{ {\small
Density dependence of the ratio $m^\star(p_F)/m$ of PNM, obtained from Eqs.(\ref{HF}) and (\ref{mstar})
using the effective interaction described in the text. }\label{eff:mass} }
\end{figure}

The nucleon effective mass, $m^\star$, is related to the single-particle energy through
\beq
\frac{1}{m^\star} = \frac{1}{p} \ \frac{de}{dp} \ .
\label{mstar}
\eeq
The density dependence of the ratio $m^\star(p_F)/m$ of PNM, obtained from the $v_{\rm eff}$ 
discussed in this Chapter, is shown in Fig. \ref{eff:mass}. 
It is worth mentioning that for SNM at equilibrium, we find $m^\star(p_F)/m = 0.65$, in close 
agreement with the lowest order CBF result of Ref. \cite{FFP}.
The results of CBF calculations at second order show a $\sim$20\% enhancement of the effective 
mass at the Fermi surface, due to medium polarization effects \cite{FFP}. We do not find this 
enhancement, as these effects are not taken into account in the Hartree-Fock approximation. 

\subsection{Spin susceptibility of neutron matter}

The results of numerical calculations show that 
the energy per particle of nuclear matter can be accurately approximated using the expression 
\beq
\frac{1}{N} \ E(\alpha,\beta,\gamma) = E_0 + E_\sigma \alpha^2 + E_\tau \beta^2 + 
 E_{\sigma\tau} \gamma^2 \ ,
\
\eeq
with
\bea
\nonumber
\alpha & = & (x_3-x_4) + (x_1-x_2) \\ 
\beta & = & (x_3+x_4) - (x_1+x_2)  \\
\nonumber
\gamma & = & (x_3-x_4) - (x_1-x_2) \ .
\eea
In symmetric nuclear matter $x_\lambda = 1/4$ for all $\lambda$ 
(see the definition in Section \ref{deflambda}), yielding $E/N=E_{{\rm SNM}}=E_0$, 
while in pure neutron matter, corresponding to $x_1=x_2=0$ and $x_3=x_4=1/2$,  
$E/N=E_{{\rm PNM}}=E_0 + E_\tau$, implying that $E_\tau$ can be identified with the symmetry energy.

Let us consider fully spin-polarized neutron matter. The two degenerate states corresponding to $x_3=1$ 
and $x_4=0$ ($\alpha=1$, spin-up) and $x_3=0$ and $x_4=1$ ($\alpha= -1$, spin-down) have energy, 
\beq
E^\uparrow = E^\downarrow = E_{{\rm PNM}} + \widetilde{E}_\sigma \ ,
\label{Epol}
\eeq
with $\widetilde{E}_\sigma = E_\sigma + E_{\sigma\tau}$. For arbitrary polarization $\alpha$, the 
energy can be obtained from the expansion 
\beq
E(\alpha) = E(0) + \left. \frac{\partial E}{\partial \alpha} \right|_{\alpha=0} \alpha 
 + \frac{1}{2}  \left. \frac{\partial^2 E}{\partial \alpha^2} \right|_{\alpha=0} \alpha^2 
 + \ldots  \ .
\label{alpha:exp} 
\eeq
As $E$ must be an even function of $\alpha$ (see Eq.(\ref{Epol})), the linear term in the above series 
must be vanishing and, neglecting terms of order $\alpha^3$, we can write
\beq
\Delta E = E(\alpha) - E(0) = \frac{1}{2}  
\left. \frac{\partial^2 E}{\partial \alpha^2} \right|_{\alpha=0} \alpha^2 \ .
\eeq

\begin{figure}[htb]
\vspace*{.25cm}
\centerline
{\epsfig{figure=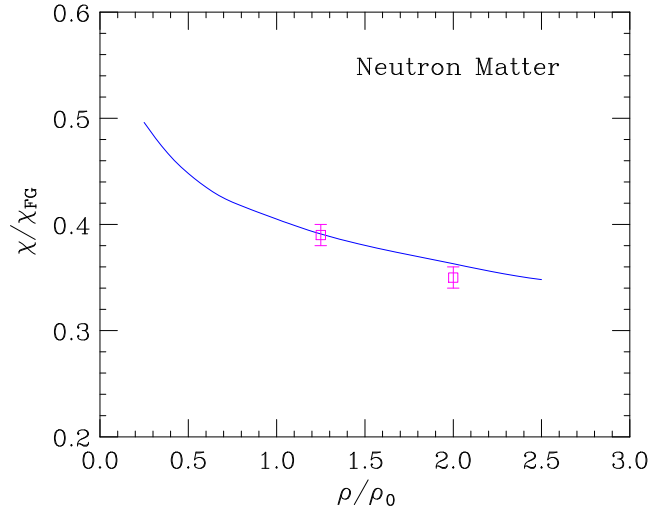,angle=000,width=8.6cm}}
\vspace*{.1cm}
\caption{ {\small
Ratio between the spin susceptibility obtained from Eqs.(\ref{def:chi}) and (\ref{energy:nm})
and the FG model result. The points with error bars show the AFDMC results of
Ref.\cite{AFDMC:chi}. }\label{spin} }
\end{figure}

In the presence of a uniform magnetic field ${\bf B}$ the energy of the system becomes
\beq
E_B(\alpha) = E(\alpha) - \alpha \mu B , 
\eeq
where $B$ denotes the magnitude of the external field, whose direction is chosen as spin 
quantization axis, and $\mu$ is the neutron magnetic moment.

Assuming that equilibrium is achieved at $\alpha=\alpha_0$, i.e. that 
\beq
\left. \frac{\partial E}{\partial \alpha} \right|_{\alpha=\alpha_0} - \mu B = 0 \ ,
\eeq
we obtain
\beq
\alpha_0 = \mu B \left( \frac{\partial^2 E}{\partial \alpha^2} \right)^{-1}_{\alpha=0} \ .
\eeq
From the definitions of the total magnetization 
\beq
M = \mu (\rho_3 - \rho_4) = \mu \alpha_0 \rho = \mu^2 
\left( \frac{\partial^2 E}{\partial \alpha^2} \right)^{-1}_{\alpha=0} B \rho \ ,
\eeq
and the spin susceptibility $\chi$
\beq
M = \chi B \ ,
\eeq
we finally obtain
\beq
\chi = \mu^2 \left( \frac{\partial^2 E}{\partial \alpha^2} \right)^{-1}_{\alpha=0} \rho = 
\mu^2 \frac{1}{2(E^\uparrow - E_{{\rm PNM}})} \ \rho \ .
\label{def:chi}
\eeq
The above equation shows that, within our approach, the spin susceptibility of neutron matter 
can be easily calculated from Eq.(\ref{energy:nm})

Figure \ref{spin} shows the density dependence of the ratio between the susceptibility of 
neutron matter obtained from the effective interaction and that corresponding to the FG model. 
For comparison, the results of Ref.\cite{AFDMC:chi}, obtained within the AFDMC approach using
the Argonne $v^\prime_8$ NN potential and the Urbana IX three-body force, are also displayed.
It appears that the inclusion of interactions leads to a substantial decrease of the susceptibility
over the whole density range, and that the agreement between the two theoretical calculations
is remarkably good.

\chapter{Nuclear matter response}
\label{response}
In this Chapter, we will discuss the response of nuclear matter to an external probe, 
defined in Eq.(\ref{def:densresp}) of Chapter \ref{CBF}.
Extensive experimental studies of the nuclear response have been 
carried out mostly through inclusive electron scattering experiments (for a recent review see, e.g., 
\cite{RMP}). 
The wealth of available data, corresponding to a variety of targets, ranging from the few nucleon systems, 
having $A \leq 4$, to nuclei as heavy as Gold ($A=197$) and Lead ($A=208$), can be reliably extrapolated
to the $A \rightarrow \infty$ limit to obtain quantitative empirical information on the nuclear matter 
response \cite{nmextr}.

Electron scattering experiments have exposed the deficiencies of the independent
particle model of nuclear dynamics. On the other hand, many body approaches
explicitly including dynamical correlation effects provide a quantitative
account of the measured cross sections in a broad kinematical domain \cite{RMP}.

As a pedagogical example, we will first consider the response to 
a scalar probe. The generalization to the case of electromagnetic and charged current weak 
interactions will be discussed in Chapters \ref{SF} and \ref{RESP}.

\section{Many-body theory of the nuclear response}

Within NMBT, the nuclear response to a scalar probe delivering
momentum {\bf q} and energy $\omega$, defined in Eq.(\ref{def:densresp}), can be written 
in terms 
of the the imaginary part of the polarization propagator $\Pi({\bf q},\omega)$ according
to \cite{FetterWalecka,BFF2}
\beq
\label{def:resp}
S({\bf q},\omega) = \frac{1}{\pi}\ {\rm Im}\  \Pi({\bf q},\omega) = \frac{1}{\pi}\
{\rm Im}\ \langle 0 | \rho^\dagger_{{\bf q}}  \
\frac{1}{H-E_0-\omega-i\eta} \ \rho_{{\bf q}} | 0 \rangle \ ,
\eeq
where $\eta=0^+$ denotes an infinitesimal positive quantity and the operator $\rho_{{\bf q}}$, 
describing the density fluctuation induced 
by the probe, is given in Eq(\ref{def:rho}).

The above definition is best suited to establish the relation between $S({\bf q},\omega)$ and
the nucleon Green function, leading to the popular expression of the response in
terms of spectral functions \cite{BFF2,BFF1}. 

Equation (\ref{def:resp}) clearly shows that the interaction with the probe leads to
a transition of the struck nucleon from a {\em hole state} of momentum ${\bf k}$,
with $|{\bf k}|<p_F$, to a {\em particle state} of momentum ${\bf k}+{\bf q}$,
with $|{\bf k}+{\bf q}|>p_F$. Hence, the calculation of $S({\bf q},\omega)$ amounts to 
describing the propagation of a particle-hole pair through the nuclear medium.

The Green function is the quantum mechanical amplitude associated with the propagation
of a particle from $x\equiv(t,{\bf x})$ to $x^\prime\equiv(t^\prime,{\bf x}^\prime)$ \cite{FetterWalecka}.
In uniform matter, due to translation invariance,
it only depends on the difference $x-x^\prime$, and after Fourier
transformation to the conjugate variable $k~\equiv~(~{\bf k}~,~E~)$ can be
written in the form
\begin{eqnarray}
\nonumber
G({\bf k},E) & = & \langle 0 | a^\dagger_{{\bf k}} \
\frac{1}{H-E_0-E-i\eta} \ a_{{\bf k}} | 0 \rangle
 - \langle 0 | a_{{\bf k}} \ \frac{1}{H-E_0+E-i\eta} \ a^\dagger_{{\bf k}} | 0 \rangle \\ & = &  G_h({\bf k},E) + G_p({\bf k},E) \ ,
\label{green:1}
\end{eqnarray}
where $G_h$ and $G_p$ correspond to propagation of nucleons in hole and particle states, respectively.

The connection between Green function and spectral functions is established
through the Lehman representation\cite{FetterWalecka}
\beq
G({\bf k},E) = \int dE^\prime \left[ \frac{P_h({\bf k},E^\prime)}{E^\prime - E - i\eta}
    -  \frac{P_p({\bf k},E^\prime)}{E - E^\prime - i\eta} \right] \ ,
\eeq
implying
\beq
P_h({\bf k},E) = \sum_n
\vert \langle n_{(N-1)}(-{\bf k}) \vert a_{{\bf k}} \vert 0_N \rangle \vert^2
\delta(E-E^{(-)}_n+E_0) = \frac{1}{\pi}\ {\rm Im}\ G_h({\bf k},E) \ , 
\label{def:Ph}
\eeq
\beq
P_p({\bf k},E) = \sum_n
\vert \langle n_{(N+1)}({\bf k}) \vert a^\dagger_{{\bf k}} \vert 0_N \rangle \vert^2
\delta(E+E^{(+)}_n-E_0) = \frac{1}{\pi} {\rm Im}\ G_p({\bf p},E) \ ,
\label{def:Pp}
\eeq
where $\vert \langle n_{(N \pm 1)}(\pm {\bf k}) \rangle$ denotes an eigenstate of the
$(A \pm 1)$-nucleon system, carrying momentum $\pm{\bf k}$ and energy $E^{(\pm)}_n$.

Within the FG model the matrix elements of the creation and annihilation operators
reduce to step functions, and the Green function takes a very simple form.
For example, for hole states we find\footnote{Note that, according to our
definitions, the hole spectral function is defined for $E \geq -\mu$, $\mu$ being
the Fermi energy.}
\beq
G_{FG,h}({\bf k},E) = \frac{ \theta(p_F-|{\bf k}|) }{ E+\epsilon^0_k-i\eta }  \ ,
\label{green:FG}
\eeq
with $\epsilon^0_k = |{\bf k}^2| /2m$, implying
\beq
P_{FG,h}({\bf k},E) = \theta(p_F-|{\bf k}|) \delta(E+\epsilon^0_k) \ .
\eeq
Strong interactions modify the energy of a nucleon carrying momentum ${\bf k}$ according to
$\epsilon^0_k \longrightarrow \epsilon^0_k + \Sigma({\bf k},E)$, where $\Sigma({\bf k},E)$
is the {\em complex} nucleon self-energy, describing the effect of nuclear dynamics.
As a consequence, the Green function for hole states becomes
\beq
G_h({\bf k},E) = \frac{1}{ E + \epsilon^0_k + \Sigma({\bf k},E)} \ .
\label{greenh:2}
\eeq

A very convenient decomposition of $G_h({\bf k},E)$
can be obtained inserting a complete set of $(A-1)$-nucleon states
(see Eqs.(\ref{green:1})-(\ref{def:Ph})) and isolating the contributions of
one-hole {\em bound} states, whose weight is given by\cite{BFF:1990}
\beq
Z_k = | \langle -{\bf k} | a_{{\bf k}} | 0 \rangle |^2 = \theta(p_F-|{\bf k}|)
\Phi_k \ .
\label{def:Z}
\eeq
Note that in the FG model these are the only non-vanishing terms,
and $\Phi_k \equiv 1$, while in the presence of interactions
$\Phi_k < 1$.
The resulting contribution to the Green function exhibits a pole at
$-\epsilon_k$, the {\em quasi-particle} energy $\epsilon_k$ being defined
by the equation
\beq
\epsilon_k = \epsilon^0_k + {\rm Re}\ \Sigma({\bf k},\epsilon_k) \ .
\label{QP:energy}
\eeq
The full Green function can be rewritten
\beq
G_h({\bf k},E) = \frac{Z_k}{E+\epsilon_k+i Z_k\ {\rm Im}\ \Sigma({\bf k},e_k)}
 + G^B_h({\bf k},E) \ ,
\eeq
where $G^B_h$ is a smooth contribution, associated with $(A-1)$-nucleon states
having at least one nucleon excited to the continuum (two hole-one particle,
three hole-two particles \ldots) due to virtual scattering processes
induced by nucleon-nucleon (NN) interactions. The corresponding spectral function is
\beq
P_h({\bf k},E) =  \frac{1}{\pi}\
\frac{ Z_k^2 \ {\rm Im}\ \Sigma({\bf k},\epsilon_k) }
{ [E + \epsilon^0_k + {\rm Re}\ \Sigma({\bf k},\epsilon_k)]^2 +
          [Z_k {\rm Im}\ \Sigma({\bf k},\epsilon_k)]^2 }
 + P^B_h({\bf k},E) \ .
\eeq
The first term in the right hand side of the above equation yields the spectrum of
a system of independent quasi-particles, carrying momenta $|{\bf k}|<p_F$, moving in
a complex mean field whose real and imaginary parts determine the quasi-particle
effective mass and lifetime, respectively. The presence of the second term is
a consequence of nucleon-nucleon correlations, not taken into account in the mean
field picture. Being the only one surviving at $|{\bf k}|>p_F$, in the FG model
this correlation term vanishes.

\begin{figure}[th]
\centerline{\psfig{file=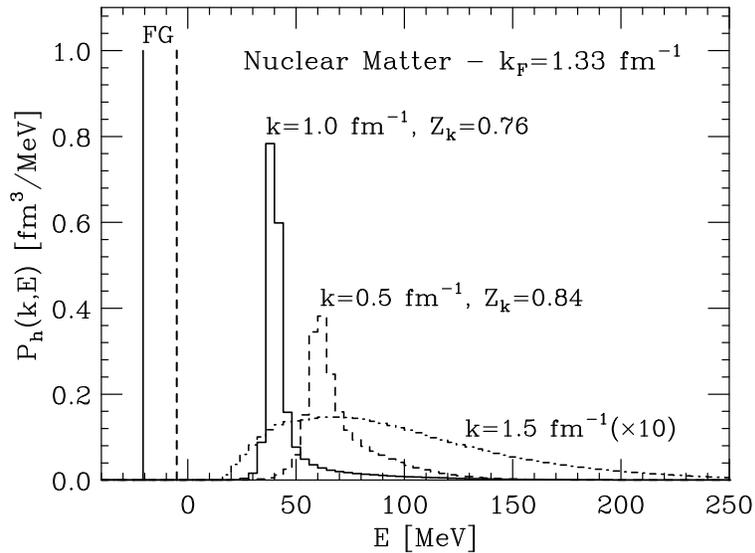,width=10cm}}
\vspace*{8pt}
\caption{ Energy dependence of the hole spectral function of nuclear
matter.\protect\cite{BFF1} The solid, dashed and dot-dash lines
correspond to $|{\bf k}|=$ 1, 0.5 and 1.5 fm$^{-1}$,
respectively. The FG spectral function at $|{\bf k}|=$ 1 and 0.5 fm$^{-1}$ is shown
for comparison.
The quasi-particle strengths of Eq.(\ref{def:Z}), are also reported. \label{fig1} }
\end{figure}

Figure \ref{fig1} illustrates the energy dependence of the hole spectral function
of nuclear matter, calculated in Ref.\cite{BFF1} using CBF perturbation theory and
a realistic nuclear hamiltonian, including the Urbana $v_{14}$ potential and the TNI 
discussed in the previous Chapter. Comparison with the FG model clearly shows
that the effects of nuclear dynamics and NN correlations are large, resulting
in a shift of the quasi-particle peaks, whose finite width becomes
large for deeply-bound states with $|{\bf k}| \ll p_F$. In addition, NN
correlations are responsible
for the appearance of strength at $|{\bf k}|>p_F$. The energy integral
\beq
\label{def:nk}
n(k) = \int dE\  P_h({\bf k},E)
\eeq
yields the occupation probability of the state of momentum ${\bf k}$. The results
of Fig. \ref{fig1} clearly show that in presence of correlations
$n(|{\bf k}|>p_F)\neq0$.

\section{Nuclear response and spectral functions}

In general, the calculation of the response requires the knowledge of $P_h$ and $P_p$, as
well as of the particle-hole effective interaction.\cite{BFF2,WIM:2004}
The spectral functions are mostly affected by short range NN correlations
(see Fig. \ref{fig1}), while the inclusion of the effective interaction, e.g. within the
framework of the Tamm-Dancoff approximation (TD) or the Random Phase Approximation (RPA),
\cite{WIM:2004} is needed 
to account for collective excitations induced by long range correlations, involving more than
two nucleons.

At large momentum transfer, as the space resolution of the probe becomes small compared
to the average NN separation distance, $S({\bf q},\omega)$ is no longer significantly
affected by long range correlations. The authors of Ref. \cite{gpc3} found that for 
$\magq \gsim$ 500 MeV RPA corrections are negligibly small, if computed using finite
size interactions.

In this kinematical regime the zero-th order
approximation in the effective interaction, according to which hole and particle propagate
independent of one another, is expected to be applicable. 
The corresponding response can be written in the simple form
\beq
\label{L0}
S({\bf q},\omega) = \int d^3k dE\ P_h({\bf k},E) P_p({\bf k}+{\bf q},\omega-E) \ .
\eeq
The widely employed {\em plane wave} impulse approximation (IA) \cite{RMP} can be readily 
obtained from
the above definition replacing $P_p$ with the FG result, which amounts to
disregarding final state interactions (FSI) between the struck nucleon and the spectator
particles. The resulting expression reads
\beq
\label{IA}
S_{IA}({\bf q},\omega) = \int d^3k dE\ P_h({\bf k},E) \theta(|{\bf k}+{\bf q}|-p_F)
\delta(\omega-E-\epsilon^0_{|{\bf k}+{\bf q}|}) \ .
\eeq

\begin{figure}[th]
\centerline{\psfig{file=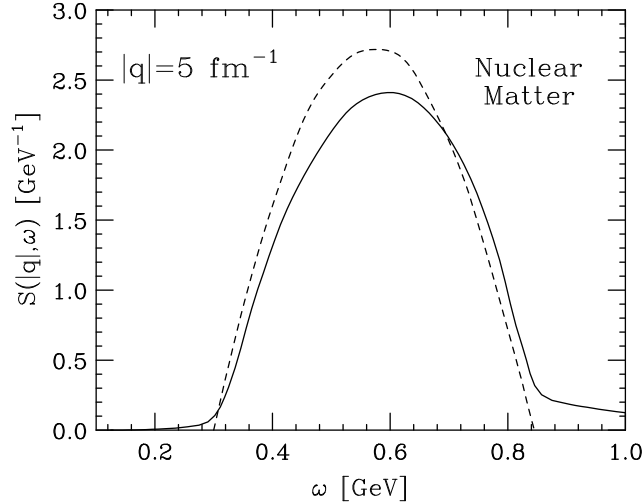,width=8.5cm}}
\vspace*{8pt}
\caption{ Nuclear matter $S_{IA}({\bf q},\omega)$ (see Eq.(\ref{IA})), as a function
of $\omega$ at $|{\bf q}|=5$ fm$^{-1}$.  The solid and dashed lines correspond to
the spectral function of Ref.\protect\cite{BFF1} and to the FG model (shifted in such
a way as to account for nuclear matter binding energy), respectively (taken from Ref.\cite{IAPAP}). 
\label{iscorr} }
\end{figure}

Figure \ref{iscorr}, showing the $\omega$ dependence of the nuclear matter structure
function at $|{\bf q}|=5$ fm$^{-1}$, illustrates the role of correlations
in the target ground state. The solid and dashed lines have been obtained from
Eq.(\ref{IA}) using the spectral function of Ref.\cite{BFF1} and that resulting
from the FG model (shifted in such a way as to account for nuclear matter binding energy),
respectively. It clearly appears that the inclusion of correlations produces a
significant shift of the strength towards larger values of energy transfer.

At moderate momentum transfer, both the full response and the particle and hole
spectral functions can be obtained using non relativistic many-body theory.
The results of Ref.\cite{BFF1} suggest that the zero-th order approximations
of Eqs.(\ref{L0}) and (\ref{IA}) are fairly accurate at $|{\bf q}|~\gsim~500$~MeV.
However, in this kinematical
regime the motion of the struck nucleon in the final state can no longer
be described using the non relativistic formalism.
While at IA level this problem can be easily circumvented replacing the
non relativistic kinetic energy with its relativistic counterpart,
including the effects of FSI in the response of Eq.(\ref{L0})
involves further approximations, needed to obtain the particle spectral function
at large $|{\bf q}|$.

\section{Particle spectral function at large momentum}

A systematic scheme to include corrections to Eq.(\ref{IA}) and take into
account FSI, originally proposed in Ref.\cite{gangofsix}, is discussed
in Ref.\cite{marianthi}. The main effects of FSI on the response are i) a
shift in energy, due to the mean field of the spectator nucleons and ii) a
redistributions of the strength, due to the coupling of the one particle-one hole
final state to $n$ particle-$n$ hole final states.
                                  
In the simplest implementation of the approach of Refs.\cite{gangofsix,marianthi},
the response is obtained from the IA result according to
\beq
S({\bf q},\omega) = \int d\omega^\prime\ S_{IA}({\bf q},\omega^\prime)
 f_{{\bf q}}(\omega-\omega^\prime) \ ,
\label{S:fold}
\eeq
the folding function $f_{{\bf q}}$ being related to the particle spectral function
through
\beq
P_p({\bf k}+{\bf q},\omega-E) = \theta(k_F-|{\bf k}+{\bf q}|) \
 f_{|{\bf k}+{\bf q}|}(\omega-E-e^0_{|{\bf k}+{\bf q}|})
\label{f:fold}
\eeq
with $\epsilon^0_{|{\bf k}+{\bf q}|} = \sqrt{|{\bf k}+{\bf q}|^2+m^2}$.
In the absence of FSI, $f_{{\bf q}}$ shrinks to
a $\delta$-function and the IA result of Eq.(\ref{IA}) is recovered.

Obviously, at large ${\bf q}$ the calculation of $P_p({\bf k}+{\bf q},\omega-E)$ cannot
be carried out using a nuclear potential model. However, it can be
obtained form the measured NN scattering amplitude within the eikonal
approximation. The resulting folding function is
the Fourier transform of the Green function describing the
propagation of the struck particle, travelling in the direction of the $z$-axis
with constant velocity $v$:
\beq
f_{|{\bf q}|}(\omega) = \int \frac{dt}{2 \pi} \ {\rm e}^{i\omega t}
  {\rm e}^{i\int_0^t dt^\prime {\widetilde V}_{|{\bf q}|}(vt^\prime)} \ .
\eeq
where ${\bf k}+{\bf q} \approx {\bf q}$ and
\beq
{\widetilde V}_{|{\bf q}|}(z) =  \langle 0 |  \frac{1}{A}
\sum_{j>i} \Gamma_{|{\bf q}|}({\bf r}_{ij} + {\bf z}) | 0  \rangle \ .
\label{aver:V}
\eeq
In the above equation, $\Gamma_{|{\bf q}|}$ is the Fourier transform
of the NN scattering amplitude at incident momentum $|{\bf q}|$ and momentum
transfer $|{\bf t}|$, $A_{|{\bf q}|}(k)$, parameterized according to
\beq
A_{|{\bf q}|}(p) =
\frac{|{\bf q}|}{4 \pi} \sigma (i + \alpha){\rm e}^{-\beta p^2} \ .
\label{NN:ampl}
\eeq
In principle, the total cross section $\sigma$, the slope $\beta$ and the
ratio between the real and the imaginary part, $\alpha$, can be extracted
from NN scattering data. However, the modifications of the scattering amplitude
due to the presence of the nuclear medium are known to be sizable, and must
be taken into account. The calculation of these corrections within the
framework of NMBT is discussed in Ref.\cite{papi}.

In Eq.(\ref{aver:V}), the expectation value is evaluated in the {\em correlated}
ground state. It turns out that NN correlation, whose
effect on $P_h$ is illustrated in Fig. \ref{fig1}, also
affect the particle spectral function and, as a consequence, the folding function
of Eq. (\ref{f:fold}). Neglecting all correlations
\beq
{\widetilde V}_{|{\bf q}|}(z) \rightarrow {\widetilde V}^0_{|{\bf q}|}
 = \frac{1}{2} v \rho \sigma(i + \alpha) \ ,
\eeq
and the quasi-particle approximation
\beq
P_p({\bf q},\omega-E) =
\frac{1}{\pi} \frac{ {\rm Im}\  {\widetilde V}^0_{|{\bf q}|} }
{ \left[ \omega-E-e^0_{|{\bf q}|}-{\rm Re}\ {\widetilde V}^0_{|{\bf q}|} \right]^2
 + \left[ {\rm Im}\ {\widetilde V}^0_{|{\bf q}|} \right]^2 }
\eeq
is recovered.

\begin{figure}[th]
\centerline{\psfig{file=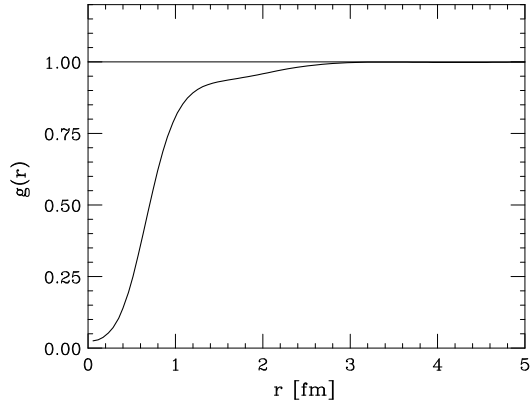,width=7cm}}
\vspace*{8pt}
\caption{Radial distribution function of nuclear matter at equilibrium
density, obtained from CBF perturbation theory using a realistic
hamiltonian \cite{gofr}. \label{gofr} }
\end{figure}

Correlations induce strong density fluctuations, preventing two nucleon from
coming close to one another. The joint probability of finding two particles
at positions ${\bf r_1}$ and ${\bf r_2}$ can be written
\beq
\rho({\bf r_1},{\bf r_2}) = \langle 0 | \sum_{j>i} \delta({\bf r}_i - {\bf r}_1)
\delta({\bf r}_j - {\bf r}_2) | 0 \rangle = \rho^2 g(|{\bf r_1}-{\bf r_2}|) \ .
\eeq
The above equation defines the {\em radial distribution function} $g(r)$, which describes 
correlation effects. Figure \ref{gofr} shows the typical shape of the radial distribution function
resulting from the CBF calculation of Ref. \cite{gofr}. 

The effect of correlation on FSI can be easily understood keeping in mind that 
the response is only sensitive to rescattering taking place within a distance
$\sim 1/|{\bf q}|$ of the primary interaction vertex \footnote{Note that this is no longer 
true in the case in which the hadronic final state is also observed.}
As the probability of finding a spectator within the range of the repulsive core of
the NN force ($\lsim 1$ fm) is small, the probability that the struck particle
rescatter against one of the spectators within a length $\sim 1/|{\bf q}|$
is also very small at large $|{\bf q}|$.
Hence, inclusion of correlations leads to a significant suppression of FSI.

\begin{figure}[th]
\centerline{\psfig{file=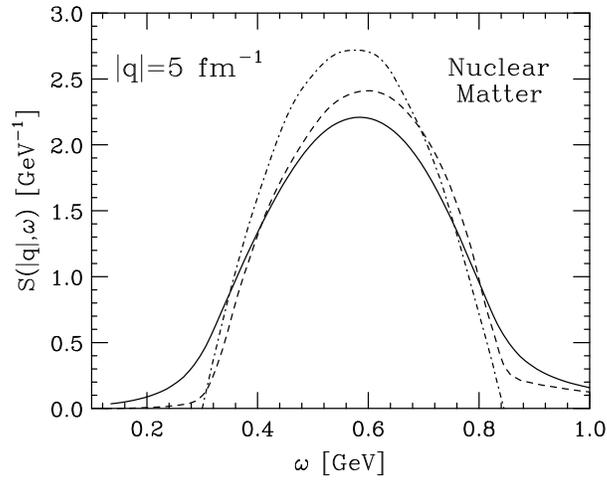,width=8cm}}
\vspace*{8pt}
\caption{Nuclear matter $S({\bf q},\omega)$, defined in Eq.(\ref{S:fold}), as a function
of $\omega$ at $|{\bf q}|=5$ fm$^{-1}$.  The solid and dashed lines have been obtained
from the spectral function of Ref.\protect\cite{BFF1}, with and without inclusion
of FSI, respectively. The dot-dash line corresponds to the FG model (shifted in such
a way as to account for nuclear matter binding energy), 
respectively (taken from Ref. \cite{IAPAP}). \label{fsicorr} }
\end{figure}

Fig. \ref{fsicorr} shows the $\omega$ dependence of the nuclear matter response of
Eq.(\ref{S:fold}) at $|{\bf q}|=5$ fm$^{-1}$. The solid and dashed lines have been
obtained using the spectral function of Ref.\cite{BFF1}, with and without inclusion
of FSI according to the formalism of Ref.\cite{gangofsix}, respectively.
For reference, the results of the FG model are also shown by the dot-dash line.
The two effects of FSI, energy shift and redistribution of the strength from the region of
the peak to the tails, clearly show up in the comparison between solid and dashed lines.

\chapter{Impulse Approximation regime}
\label{SF}
As pointed out in the previous Chapter, the nuclear response has been extensively 
investigated carrying out inclusive electron scattering experiments. 

The first attempts to provide a quantitative estimate the of the measured electron-nucleus 
cross section were based on oversimplified models of nuclear dynamics. At the end of the 
seventies, Moniz suggested that the target may be described as a degenerate 
gas of protons and neutrons at given constant density $\rho$ \cite{Moniz}, the 
effect of the interactions being crudely taken into account by an average binding 
energy $\overline{\epsilon}$. 
Despite its simplicity, the FG model of Ref. \cite{Moniz} was able to give a fairly accurate 
account of the electron-nucleus cross section in the region of the quasi-elastic peak, corresponding to 
$x_B = Q^2/2m\omega \ \sim$ 1, where $x_B$ is the Bjorken variable, $Q^2 = \mq^2-\omega^2$, and $\bf{q}$, 
$\omega$ and $m$ denote the momentum and energy transfer and the nucleon mass, respectively. 

In the past twenty years, with the availability of new data, extending in the region of high $\mq$ 
and low $\omega$, corresponding to $x_B \gg 1$, the limits of the FG model, and more generally of all 
independent particle models, became apparent. Away from the quasi elastic peak 
correlation effects, not included in the mean field picture, become more and more important and 
the FG model is not longer able to describe the measured cross section.

The experimental investigation of the neutrino-nucleus cross section involves additional 
difficulties due to the low counting rates and the lack of neutrino beams of fully specified
properties. However, a quantitative understanding of the weak nuclear response is needed in 
a variety of different fields, ranging from nuclear astrophysics to the analysis of 
neutrino oscillation experiments.

Electron scattering data provide a stringent test for validation of theoretical models of the nuclear 
response, in view of their application to the case of weakly interacting probes.
For example, the success of the FG model in explaining electron scattering in the quasi elastic region 
at $\mq \lsim $ 500\ MeV prompted its extension to neutrino scattering \cite{Moniz2}.
 
In this Chapter we will review the application of the formalism on NMBT to the calculation of 
the electromagnetic and charged current weak cross sections in the region of large momentum transfer, 
where the IA is expected to be safely applicable. 

\section{Electron-nucleus cross section}

The differential cross section of the process
\beq
e + A \rightarrow e^\prime + X \ , 
\label{process:e}
\eeq
in which an electron carrying initial four-momentum $k\equiv(E_e,{\bf k})$
scatters off a nuclear target to a state of four-momentum
$k^\prime\equiv(E_e^\prime,{\bf k}^\prime)$, the target final state 
being undetected, can be written in Born approximation as (see, e.g., 
Ref. \cite{IZ})
\beq
\frac{d^2\sigma}{d\Omega_{e^\prime} dE_e^\prime}=\frac{\alpha^2}{Q^4} 
\frac{E_e^\prime}{E_e}  \ \Lmunu\WmunuA \ ,
\label{e:cross:section}
\eeq
where $\alpha$ is the fine structure constant.
The leptonic tensor, that can be written, neglecting the lepton mass, as
\beq
\Lmunu = 2 \left[ k_\mu k_\nu^\prime + k_\nu k_\mu^\prime -g_{\mu\nu} (k k^\prime) \right] \ ,
\eeq
is completely determined by electron
kinematics, whereas the nuclear tensor $\WmunuA$ contains all the information 
on target structure. Its definition involves the initial
and final hadronic states $ |0\rangle$ and $|X \rangle$, carrying 
four-momenta $p_0$ and $p_X$, respectively, as well as the nuclear 
electromagnetic current operator
$J^\mu$:
\beq
\WmunuA = \sum_X \langle 0 | J^\mu | X\rangle 
 \langle X | J^\nu | 0\rangle \delta^{(4)}(p_0+q-p_X) \ ,
\label{e:hadrten}
\eeq
where the sum includes all hadronic final states. Comparison with Eq.(\ref{def:densresp})
shows that the above tensor is the generalization of the nuclear response to 
the case of vector interaction. 

Calculations of $\WmunuA$ at moderate momentum transfers $( {\bf |q|} < 0.5\, {\rm GeV})$
can be carried out within nuclear many-body theory (NMBT), using
non-relativistic wave functions to describe the initial and final
states and expanding the current operator in powers of ${\bf |q|}/m$, 
$m$ being the nucleon mass (see, e.g., Ref. \cite{Golak95,Efros94,Carlson98}). On the other hand, 
at higher values of 
${\bf |q|}$, corresponding to beam energies larger than $\sim 1$ \ GeV, 
the description of the final states $|X\rangle $ 
in terms of non-relativistic nucleons is no longer accurate. 
Calculations of $\WmunuA$ in this regime require a set of simplifying 
assumptions, allowing one to take into account the relativistic motion of 
final state particles carrying momenta $\sim {\bf q}$ as well as the occurrence 
of inelastic processes, leading to the 
appearance of hadrons other than protons and neutrons.

\section{The impulse approximation}

As stated in Chapter \ref{response},
the main assumptions underlying the impulse approximation (IA) scheme are that 
i) as the spatial resolution of a probe delivering momentum ${\bf q}$ 
is $\sim 1/|{\bf q}|$, at large enough $|{\bf q}|$ the target nucleus is seen 
by the probe as a collection of individual nucleons and ii) 
the particles produced at the interaction vertex and the recoiling (${\rm A}-1$)-nucleon 
system evolve independently of one another, which amounts to neglecting {\it both} 
statistical correlations due to Pauli blocking and dynamical Final State Interactions (FSI), i.e. 
rescattering processes driven by strong interactions.

\begin{figure}[hbt]
\centerline
{\psfig{figure=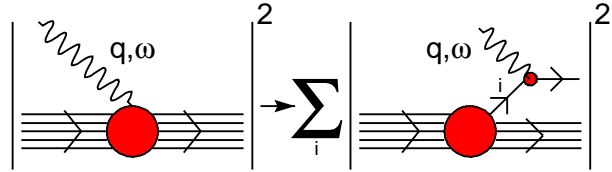,angle=00,width=9.00cm}}
\vspace*{-.2in}
\caption{\small 
Schematic representation of the IA scheme, in which
the nuclear cross section is replaced by the incoherent sum of
cross sections describing scattering off individual bound nucleons, the
recoiling $({\rm A}-1)$-nucleon system acting as a spectator.}
\label{fig:1}
\end{figure}

In the IA regime the scattering process off a nuclear target reduces to the
incoherent sum of elementary processes involving only one nucleon, as
schematically illustrated in Fig. \ref{fig:1}. 

Within this picture, the nuclear current can be written as a sum of one-body 
currents
\beq
J^\mu \rightarrow \sum_i j_i^\mu \ ,
\label{currIA}
\eeq
while the final state reduces to the direct product of the hadronic state 
produced at the electromagnetic vertex, carrying momentum 
${\bf p}_x$ and the $(A-1)$-nucleon residual 
system, carrying momentum ${\bf p}_{\cal R}= {\bf q}-{\bf p}_x$ (for simplicity, 
we omit spin indices)
\beq
|X\rangle \rightarrow |x,{\bf p}_x\rangle
\otimes |{\cal R},{\bf p_{\cal R}}\rangle \ .
\label{resIA}
\eeq
Using Eq. (\ref{resIA}) we can rewrite the sum in Eq. (\ref{e:hadrten}) replacing
\beq
\nonumber
\sum_X | X \rangle \langle X | \rightarrow \sum_{x} 
\int d^3p_x  | x,{\bf p}_x \rangle \langle {\bf p}_x,x | 
\  \sum_{{\cal R}} d^3p_{{\cal R}} 
| {\cal R}, {\bf p}_{{\cal R}} \rangle \langle {\bf p}_{{\cal R}}, {\cal R} | \ .
\label{sumn}
\eeq
Substitution of Eqs. (\ref{currIA})-(\ref{sumn}) into Eq. (\ref{e:hadrten}) and
insertion of a complete set of free nucleon states, satisfying 
\beq
\int d^3p\  | {\rm N},  {\bf p}\rangle\langle {\bf p},  {\rm N} |=I \ ,
\eeq
results in the factorization of the current matrix element 
\beq
\langle 0 | J^\mu | X\rangle =
\left( \frac{m}{\sqrt{{\bf p}_{{\cal R}}^2 + m^2}} \right)^{1/2}
 \langle 0 | {\cal R}, {\bf p}_{{\cal R}} ; {\rm N},-{\bf p}_{{\cal R}} \rangle 
\sum_i \langle -{\bf p}_{\cal R},N | j^\mu_i | x,{\bf p}_x \rangle \ ,
\label{matelnew}
\eeq
leading to
\bea
\nonumber
\WmunuA & = & \sum_{x,{\cal R}} \int d^3p_{{\cal R}}\ d^3p_x 
| \langle 0 | {\cal R},{\bf p}_{{\cal R}};{\rm N},-{\bf p}_{{\cal R}} \rangle |^2 
\  \frac{m}{ E_{{\bf p}_{\cal R}} } \\
\label{hadrtenIA}
& \times &  
\sum_i \ 
\langle -{\bf p}_{{\cal R}},{\rm N}| j^\mu_i | x,{\bf p}_x \rangle
\langle {\bf p}_x ,x | j^\nu_i | {\rm N},-{\bf p}_{{\cal R}} \rangle \\
\nonumber
& \times & 
\delta^{(3)}({\bf q}-{\bf p}_{{\cal R}}-{\bf p}_x)
\delta(\omega+E_0-E_{\cal R}-E_x),
\eea
where $E_{{\bf p}_{\cal R}} = 
\sqrt{|{\bf p}_{{\cal R}}|^2 + m^2}$. Finally, using the identity
\beq
\delta(\omega+E_0-E_{\cal R}-E_x) = \int dE \delta(E-m+E_0-E_{\cal R}) 
\ \delta (\omega-E+m-E_x) \ ,
\label{deltaIA}
\eeq
and the definition of the target spectral function given in the previous 
Chapter\footnote{As we will consider a target having ${\rm N}={\rm Z}={\rm A}/2$,
the spectral functions describing proton and neutron removal
will be assumed to be the same.},
\beq
P({\bf p}, E) = \sum_{\cal R}
|\langle 0|{\cal R},-{\bf p};{\rm N},{\bf p} \rangle |^2  
 \ \delta(E-m+E_0-E_{\cal R}) \ ,
\label{specfunIA}
\eeq
we can rewrite Eq.(\ref{e:hadrten}) in the form 
\beq
\WmunuA({\bf q},\nu) = \sum_i
\int d^3p\ dE\ w_i^{\mu\nu}({\widetilde q})  
\left(\frac{m}{E_{{\bf p}}}\right)P({\bf p}, E) \ ,
\label{hadrten2}
\eeq
with $E_{{\bf p}} = \sqrt{|{\bf p}^2|+m^2}$ and
\beq
\nonumber
w_i^{\mu\nu} = \sum_x \langle {\bf p},{\rm N}| j^\mu_i | x,{\bf p}+{\bf q} \rangle
\langle {\bf p}+{\bf q},x | j^\nu_i | {\rm N},{\bf p} \rangle  
\ \delta({\widetilde \omega} + \sqrt{{\bf p}^2 + m^2} - E_x )\ .
\label{nucl:tens}
\eeq
Note that the factor $(m/\sqrt{|{\bf p}_{{\cal R}}|^2 + m^2})^{1/2}$ in Eq.(\ref{matelnew}) 
takes into account the implicit covariant normalization of $\langle -{\bf p}_{\cal R},N |$ in 
the matrix element of $j_i^\mu$. 

The quantity defined in the above equation is the tensor describing electromagnetic 
interactions of the $i$-th nucleon {\it in free space}. Hence, Eq. (\ref{nucl:tens}) 
shows that in the IA scheme the effect 
of nuclear binding of the struck nucleon is accounted for by the replacement
\beq
q \equiv (\omega,{\bf q}) \rightarrow {\widetilde q} \equiv ({\widetilde \omega},{\bf q}) \ ,
\label{def:qt}
\eeq
with (see Eqs. (\ref{hadrtenIA}) and (\ref{specfunIA}))
\bea
\nonumber
{\widetilde \omega} & = & E_x - \sqrt{{\bf p}^2 + m^2} \\ 
\nonumber
       & = & \omega + E_0 - E_{\cal R} - \sqrt{{\bf p}^2 + m^2} \\
       & = & \omega - E  + m - \sqrt{{\bf p}^2 + m^2} \ ,
\label{def:nut}
\eea
in the argument of $w_i^{\mu\nu}$. This procedure essentially amounts to assuming 
that:
i) a fraction $\delta \omega$ of the energy transfer 
 goes into excitation energy of the spectator system and ii) the elementary 
scattering process can be described as if it took place in free space with 
energy transfer ${\widetilde \omega} = \omega - \delta \omega$. This interpretation 
emerges most naturally
in the $|{\bf p}| \ll m$ limit, in which Eq. (\ref{def:nut}) yields $\delta \omega = E$. 

Collecting together all the above results we can finally 
rewrite the doubly differential nuclear cross section in the form
\bea
\nonumber
& & \frac{d\sigma_{IA}}{d\Omega_{e^\prime} dE_{e^\prime}} = \int \ d^3p \ dE \ 
P({\bf p},E)\ \left[ Z \frac{d\sigma_{ep}}{d\Omega_{e^\prime} dE_{e^\prime}}
 \right. \\
 &  & \ \ \ \ \ \ \ \ \ \ \ \ \ \ \ \ \ \ \ \ + 
\left. (A-Z) \frac{d\sigma_{en}}{d\Omega_{e^\prime} dE_{e^\prime}} 
\right] \delta(\omega-E+m-E_x) ,
\label{csIA}
\eea
where $d\sigma_{eN}/d\Omega_{e^\prime} dE_{e^\prime}$ ($N \equiv n,p$ denotes 
a proton or a neutron) is the cross section 
describing the elementary scattering process 
\beq
e(k) \ + \ N(p) \rightarrow e^\prime(k^\prime) 
+ x(p + {\widetilde q}) \ ,
\eeq
given by
\beq
\frac{d\sigma_{eN}}{d\Omega_{e^\prime} dE_{e^\prime}} = \frac{\alpha^2}{Q^4}
\frac{E_e^\prime}{E_e}  \  \frac{m}{E_{\bf p}} \Lmunu w^{\mu\nu}_N \ ,
\eeq
stripped of both the flux factor and the energy conserving $\delta$-function.

\section{The nuclear spectral function}

Non-relativistic NMBT provides a fully consistent computational 
framework that has been employed to obtain the spectral functions of 
the few-nucleon systems, having A$=3$ \cite{dieperink,cps,sauer} 
and 4 \cite{ciofi4,morita,bp}, 
as well as of nuclear matter, i.e. in the limit A $\rightarrow \infty$ 
with Z=A/2 \cite{BFF1,pkebbg}. Calculations based on 
G-matrix perturbation theory have also been carried out 
for oxygen \cite{geurts16,polls16}.

The spectral functions of different nuclei, ranging from Carbon to Gold, 
have been modeled using the
Local Density Approximation (LDA) \cite{bffs}, in which the experimental
information obtained from nucleon knock-out measurements is combined 
with the results of theoretical calculations of the nuclear matter 
$P({\bf p},E)$ at different densities \cite{bffs}.

Nucleon removal from shell model states 
has been extensively studied by coincidence $(e,e^\prime p)$ experiments 
(see, e.g., Ref. \cite{book}). The corresponding measured spectral function 
is usually parameterized in the factorized form 
\beq
P_{MF}({\bf p},E) = \sum_n Z_n\ |\phi_n({\bf p})|^2 F_n(E-E_n) \ ,
\label{P:MF}
\eeq
where $\phi_{n}({\bf p})$ is the momentum-space wave function of the 
single particle shell mode state $n$ (e.g. Woods-Saxon wave functions), 
whose energy width is described by the function $F_n(E-E_n)$ (e.g. a lorentzian). 
The normalization of  the $n$-th state is 
given by the so called spectroscopic factor $Z_n < 1$, and the sum in 
Eq. (\ref{P:MF}) is extended to 
all occupied states. Typically, $P_{MF}({\bf p},E)$ vanishes at 
$E$ larger than $\sim 30$ MeV and $|{\bf p}|$ larger than $\sim 250$ MeV.
Note that in absence of NN correlations the full spectral function 
could be written as in Eq. (\ref{P:MF}), with 
$F_n(E-E_n) \equiv \delta(E-E_n)$ and $Z_n \equiv 1$.

Strong dynamical NN correlations give rise
to virtual scattering processes leading to the excitation of the participating
nucleons to states of energy larger than the Fermi energy, thus depleting
the shell model states within the Fermi sea. As a consequence, the spectral function
associated with nucleons belonging to correlated pairs extends to 
the region of $|{\bf p}| \gg p_F$ {\it and} $E \gg e_F$, where $e_F$ denotes
the Fermi energy, typically $\lsim 30$ MeV. 

The correlation contribution to $P({\bf p},E)$ of uniform nuclear matter 
has been calculated by Benhar {\it et al} for a wide range of density values \cite{bffs}.
Within the LDA scheme, the results of Ref. \cite{bffs} can be used to obtain the corresponding 
quantity for a finite nucleus of mass number $A$ from 
\beq
P_{corr}({\bf p},E) = \int d^3r\ \rho_A({\bf r})
P^{NM}_{corr}({\bf p},E;\rho = \rho_A({\bf r})) \ ,
\label{P:corr}
\eeq
where $\rho_A({\bf r})$ is the nuclear density distribution and
$P^{NM}_{corr}({\bf p},E;\rho)$ is the correlation component of
the spectral function of uniform nuclear matter at density $\rho$.

Finally, the full LDA nuclear spectral function can be written
\beq
P_{LDA}({\bf p},E) = P_{MF}({\bf p},E) + P_{corr}({\bf p},E) \ ,
\label{P:LDA}
\eeq
the spectroscopic factors $Z_n$ of Eq. (\ref{P:MF}) being constrained 
by the normalization requirement
\beq
\int d^3p\ dE\ P_{LDA}({\bf p},E) = 1\ .
\label{P:norm}
\eeq

\begin{figure}[hbt]
\centerline{\psfig{figure=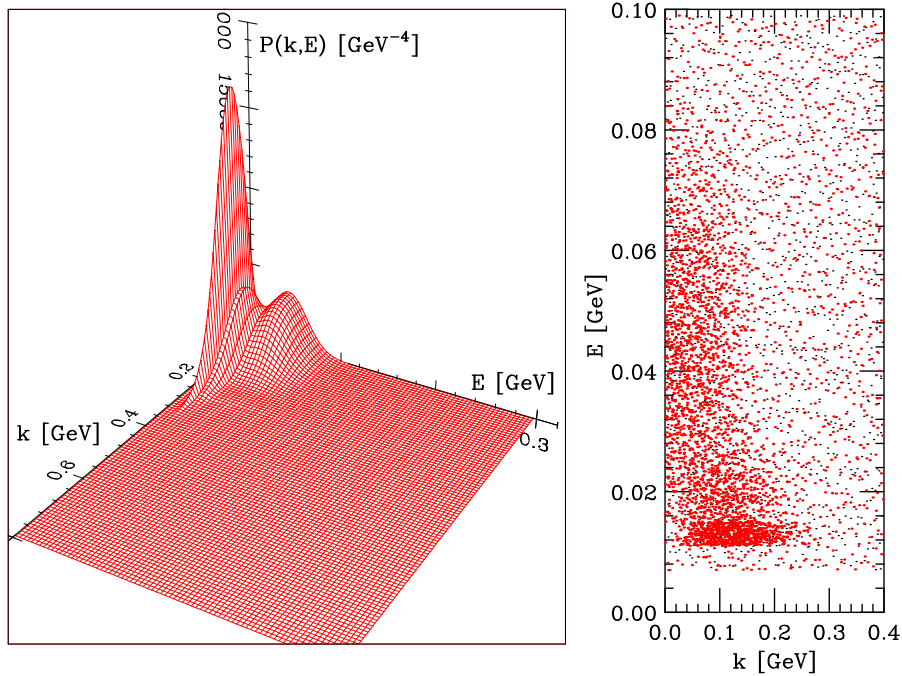,angle=00,width=12.00cm}}
\caption{\small  
Three-dimensional plot (left panel) and scatter plot 
(right panel) of the oxygen spectral function obtained using the LDA
approximation described in the text.}
\label{pke:O}
\end{figure}

The LDA spectral function of $^{16}O$ obtained combining the nuclear matter 
results of Ref. \cite{bffs} and the Saclay $(e,e^\prime p)$ data \cite{eep16O}
is shown in Fig. \ref{pke:O}. The shell model contribution $P_{MF}({\bf p},E)$ 
accounts for $\sim$ 80 \% of its normalization, whereas the remaining $\sim$ 20 \% 
of the strength, accounted for by $P_{corr}({\bf p},E)$, is located at high 
momentum ($|{\bf p}| \gg p_F$) {\it and} large removal energy ($E \gg e_F$). 
It has to be emphasized 
that large $E$ and large ${\bf p}$ are strongly correlated. For example, 
$\sim$ 50 \% of the strength at $|{\bf p}|$ = 320 MeV is located at 
$E >$ 80 MeV.

The LDA scheme rests on the premise that short range nuclear dynamics is
unaffected by surface and shell effects. The validity of this assumption
is confirmed by theoretical calculations of the nucleon momentum 
distribution, defined as
\beq
\label{def1:nk}
n({\bf p}) = \int dE\ P({\bf p},E) = \langle 0 | a^\dagger_{\bf p} a_{\bf p} | 0 \rangle \ ,
\label{def2:nk}
\eeq
where $a^\dagger_{\bf p}$ and $a_{\bf p}$ denote the creation and annihilation 
operators of a nucleon of momentum ${\bf p}$. The results clearly show that 
 for A$\ge 4$ the quantity $n({\bf p})/A$ becomes nearly independent of
$A$ in the region of large $|{\bf p}|$ ($\gsim 300$ MeV), where NN correlations 
dominate (see, e.g., Ref. \cite{rmp0}).

\begin{figure}[hbt]
\centerline{\psfig{figure=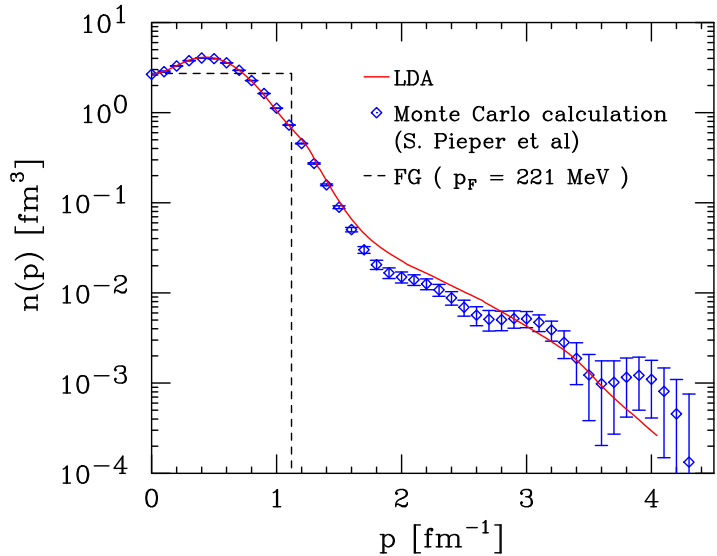
,angle=00,width=9.5cm}}
\caption{\small  
Momentum distribution of nucleons in the oxygen ground state.
Solid line: LDA approximation. Dashed line: FG model with Fermi momentum
$p_F = 221$ MeV. 
Diamonds: Monte Carlo calculation carried out by S.C. Pieper \protect\cite{steve} 
using the 
wave function of Ref. \protect\cite{16Owf}.}
\label{nk:O}
\end{figure}

In Fig. \ref{nk:O} the nucleon momentum distribution 
of $^{16}$O, obtained from Eq. (\ref{def1:nk}) using the LDA spectral function 
of Fig. \ref{pke:O}, is compared to the one resulting from a 
Monte Carlo calculation \cite{steve}, carried out using the definition of 
Eq. (\ref{def2:nk}) and a highly realistic many-body wave function \cite{16Owf}. 
For reference, the FG model 
momentum distribution corresponding to Fermi momentum $p_F$ = 221 MeV, currently used 
in the analysis of neutrino oscillation experiments (see, e.g. Ref.\cite{K2K}), is also 
shown by the dashed line. It clearly appears that the $n({\bf p})$ obtained
from the spectral function is close to that of Ref.\cite{steve}, while the FG
distribution exhibits a completely different behavior.
 
A direct measurement of the correlation component of the 
spectral function of $^{12}C$, obtained measuring the
$(e,e^\prime p)$ cross section at missing momentum and energy up to $\sim$~800~MeV
and $\sim~200$~MeV, respectively, has been recently carried out at Jefferson Lab by the
E97-006 Collaboration \cite{E97-006}. The data resulting from the preliminary
analysis appear to be consistent with the theoretical predictions based 
on LDA.  

\section {Comparison to electron scattering data}
\label{eA-results}

We have employed the formalism described in the previous Sections to compute
the inclusive electron scattering cross section off oxygen at 
$0.2 \lsim Q^2 \lsim 0.6 $ GeV$^2$ \cite{PRD,NPBPC}. 

The IA cross section has been obtained using the LDA spectral function shown 
in Fig. \ref{pke:O} and the nucleon tensor defined by Eq. (\ref{nucl:tens}),
that can be written as
\beq 
w_N^{\mu\nu} = w^N_1 \left( -g^{\mu\nu} + \frac{\qt^\mu \qt^\nu}{\qt^2} 
  \right) + \frac{w^N_2}{m^2} \left(p^\mu - \frac{(p \qt)}{\qt^2}q^\mu \right)
                   \left(p^\nu - \frac{(p \qt)}{\qt^2}q^\nu \right) \ ,
\eeq
where $p\equiv (E_{\bf p},{\bf k})$ and the off-shell four momentum transfer 
$\qt$ is defined by Eqs. (\ref{def:qt}) and (\ref{def:nut}). The two structure 
functions $w^N_1$ and $w^N_2$
are extracted from electron-proton and electron-deuteron scattering data. In the case of 
quasi-elastic scattering they are simply related to the electric 
and magnetic nucleon form factors, $G_{E_N}$ and $G_{M_N}$, through
\beq
w^N_1 = -\frac{\qt^2}{4m^2}\ \delta\left({\widetilde \omega} + \frac{\qt^2}{2m} \right) 
\ G_{M_N}^2 \ ,
\eeq
\beq
w^N_2 = \frac{1}{1 - \qt^2/4 m^2} \ 
\delta\left({\widetilde \omega} + \frac{\qt^2}{2m} \right) 
\left( G_{E_N}^2 - \frac{\qt^2}{4m^2} G_{M_N}^2 \right)\ .
\eeq
Numerical calculations have been carried out using 
the H\"ohler-Brash parameterization of the form factors \cite{Hohler76,Brash02}, 
resulting from a fit which includes the recent Jefferson Lab data \cite{Jones00}. 

In the kinematical region under discussion, inelastic processes, mainly 
quasi-free $\Delta$ resonance production, are also known to play a role. To include these 
contributions in the calculation of the inclusive cross section, we have adopted the 
Bodek and Ritchie parameterization of the proton and neutron structure functions \cite{br}, 
 covering both the resonance and deep inelastic region.

\begin{figure}[hbt]
\centerline
{\psfig{figure=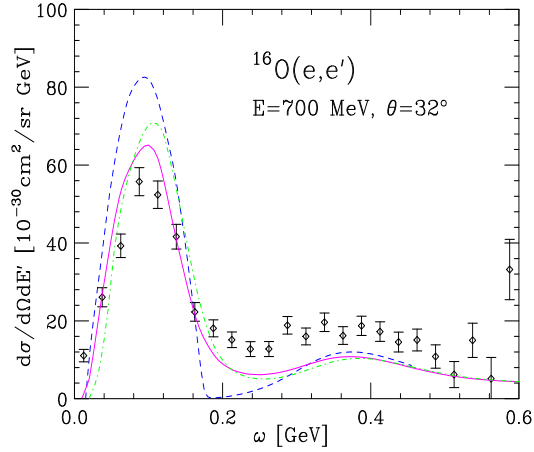,angle=00,width=7.0cm,height=6.0cm}}
\caption{\small Cross section of the process $^{16}O(e,e^\prime)$
at beam energy 700 MeV and electron scattering angle 32$^\circ$. 
Solid line: full calculation, with inclusion of final state interaction effects. 
Dot-dash line: IA calculation, carried out
neglecting FSI effects. Dashed line: FG model with $p_F = 225$ MeV and 
$\epsilon = 25$ MeV. The experimental data are from Ref.\protect\cite{LNF}.}
\label{fig:ee1}
\end{figure}

\begin{figure}[hbt]
\centerline
{\psfig{figure=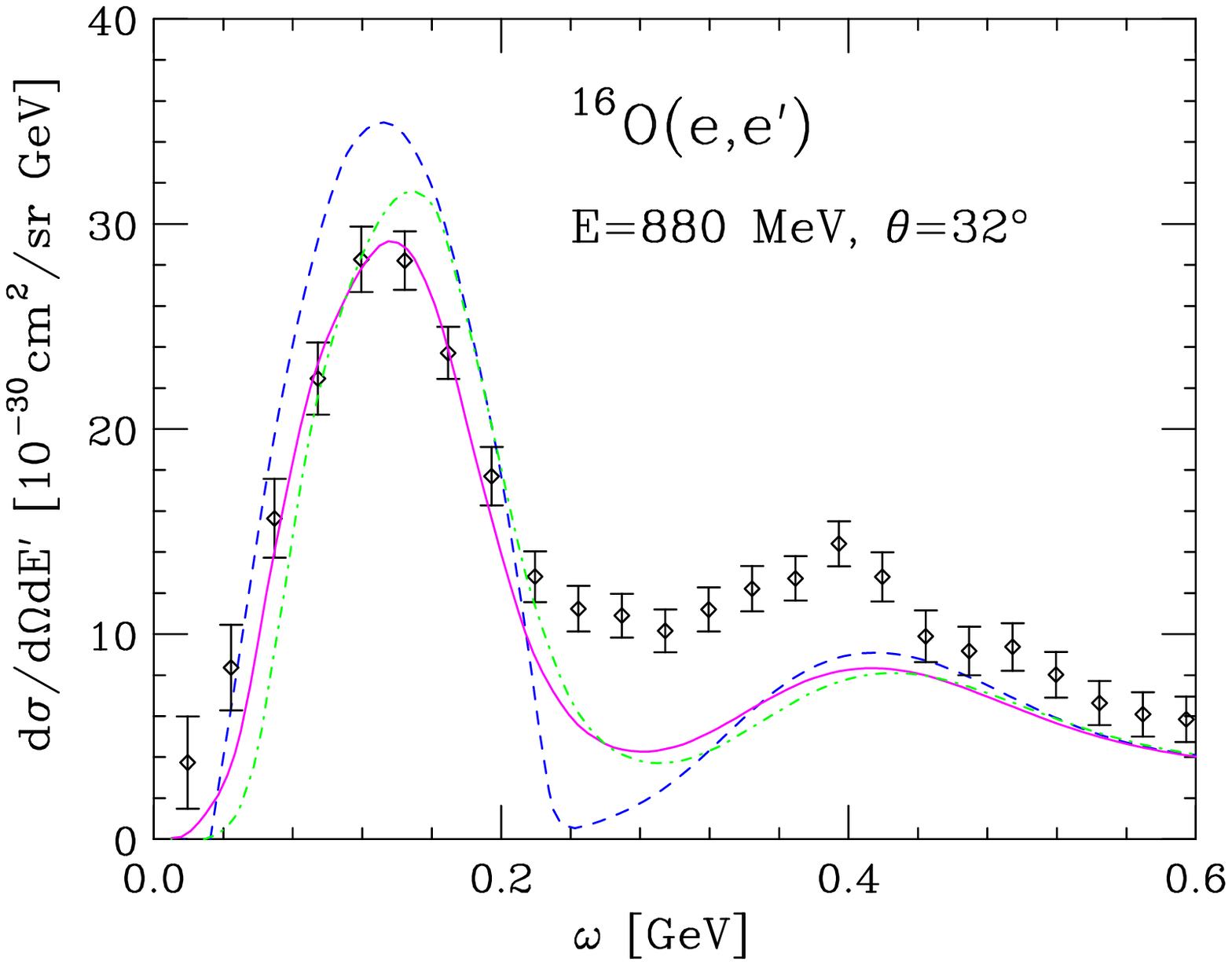,angle=00,width=7.0cm,height=6.0cm}}
\caption{\small 
Same as in Fig. \protect\ref{fig:ee1}, but for beam energy 880 MeV.}
\label{fig:ee2}
\end{figure}

\begin{figure}[hbt]
\centerline
{\psfig{figure=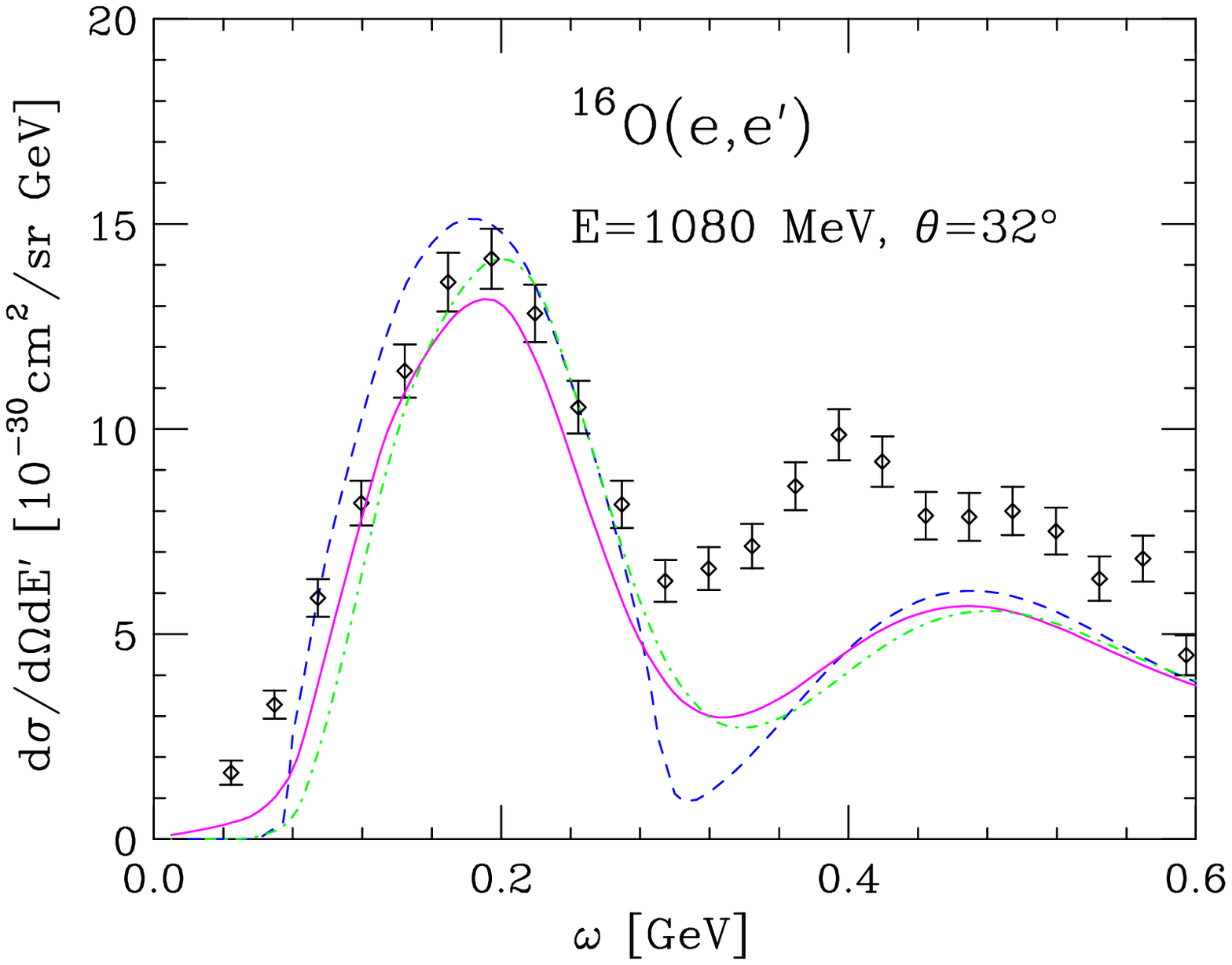,angle=00,width=7.0cm,height=6.0cm}}
\caption{\small
Same as in Fig. \protect\ref{fig:ee1}, but for beam energy 1080 MeV.}
\label{fig:ee3}
\end{figure}

\begin{figure}[hbt]
\centerline
{\psfig{figure=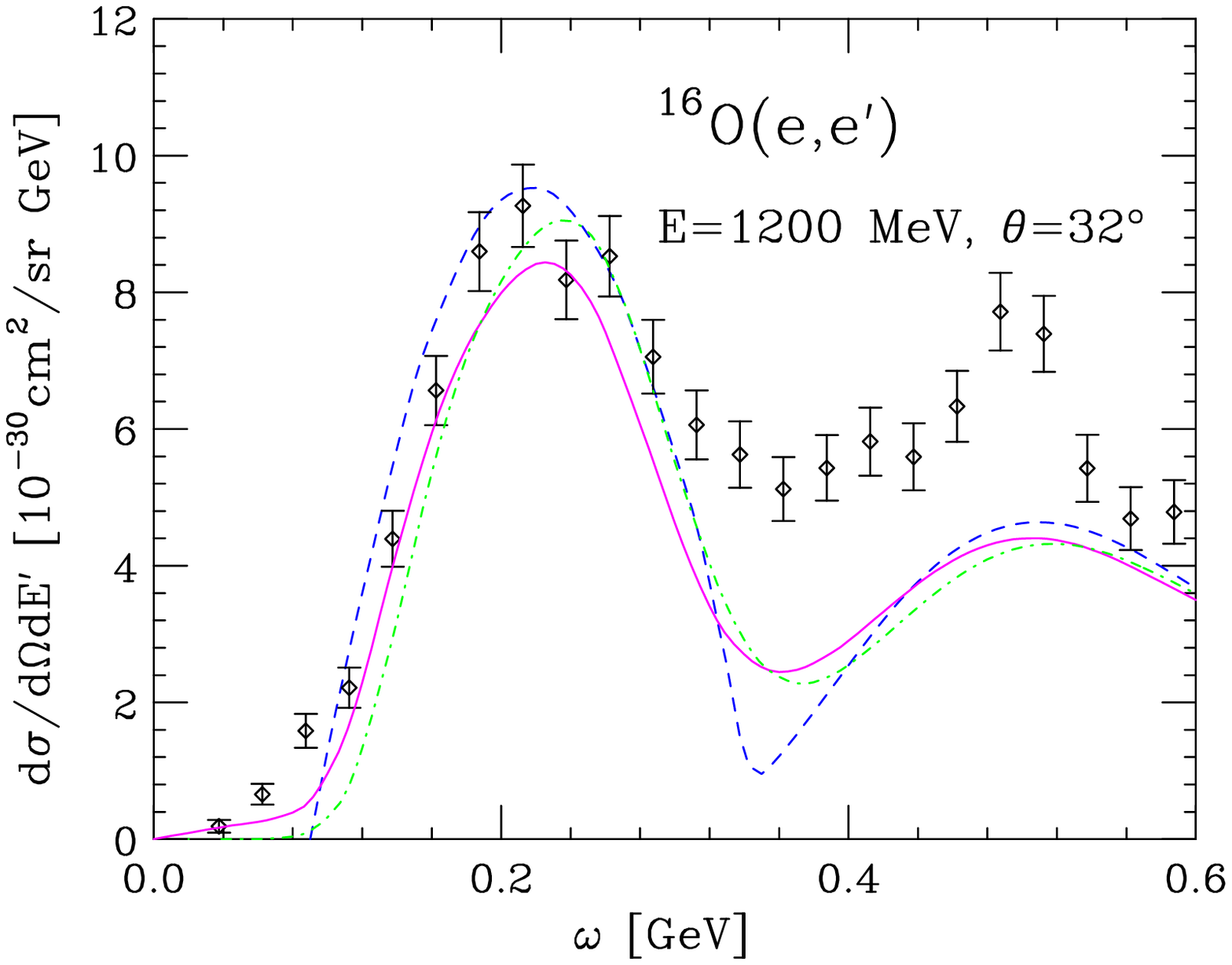,angle=00,width=7.0cm,height=6.0cm}}
\caption{\small
Same as in Fig. \protect\ref{fig:ee1}, but for beam energy 1200 MeV.}
\label{fig:ee4}
\end{figure}

In Figs. \ref{fig:ee1}-\ref{fig:ee4} the results of our calculations
are compared to the data of Ref. \cite{LNF}, corresponding to beam energies
700, 880, 1080 and 1200 MeV and electron scattering angle 32$^\circ$. For 
reference, the results of the FG model corresponding to Fermi 
momentum $p_F = 225$ MeV and average removal energy $\epsilon = 25$ MeV
are also shown. The results including FSI effects have been obtained from the 
approach described in Chapter \ref{response}, using the gaussian parameterization of
Eq.(\ref{NN:ampl}), with the parameter values resulting from the fit of Ref. \cite{oneillNN} 

Overall, the approach described in the previous Sections, {\it involving no
adjustable parameters}, provides a fairly accurate account of the measured
cross sections in the region of the quasi-free peak. On the other hand, the
FG model, while yielding a reasonable description 
at beam energies 1080 and 1200 MeV, largely overestimates the data at lower energies. 
The discrepancy at the top of the quasi-elastic peak turns out to be $\sim$ 25 \% 
and $\sim$ 50 \% at 880 and 700 MeV, respectively.

The results of NMBT and FG model also turn out to be sizably different
in the dip region, on the right hand side of the quasi-elastic peak, while
the discrepancies become less pronounced at the $\Delta$-production peak.
However, it clearly appears that, independent of the employed approach and beam 
energy, theoretical 
results significantly underestimate the data at energy transfer larger than the
pion production threshold.

\begin{figure}[hbt]
\centerline
{\psfig{figure=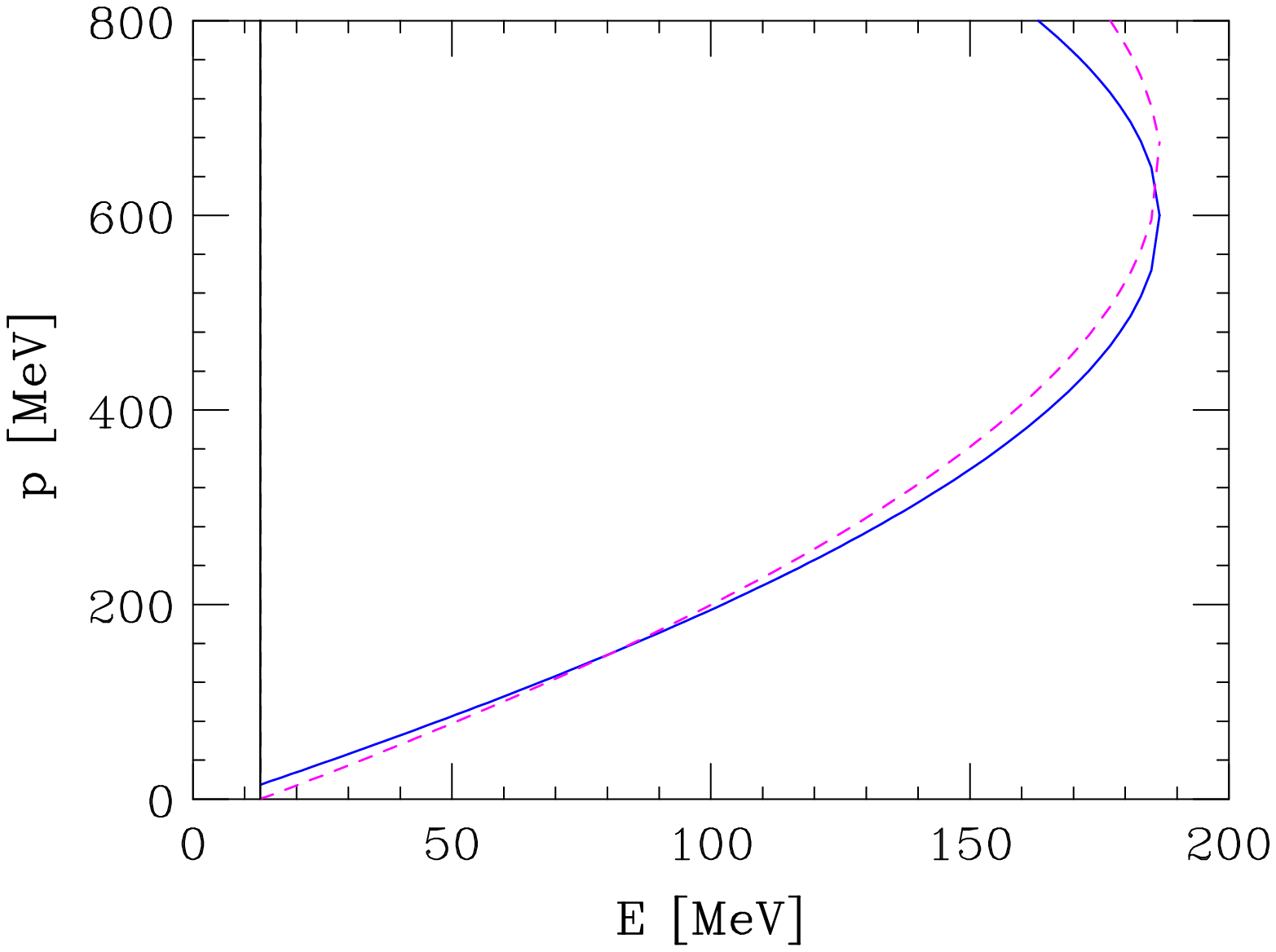 ,angle=00,width=7.00cm,height=6.00cm}}
\caption{\small 
The solid and dashed lines enclose the integration regions in the 
$({\bf p},E)$ plane relevant to the calculation of the IA cross section at 
the top of the quasi-elastic and $\Delta$ production peak, respectively, for beam 
energy 1200 MeV and scattering angle 32$^\circ$.}
\label{nu:0}
\end{figure}

In view of the fact that the quasi-elastic peak is correctly reproduced
(within an accuracy of $\sim$ 10 \%), the failure of NMBT to reproduce the data 
at larger $\omega$ may be ascribed to deficiencies in the description of the 
elementary electron-nucleon cross section. In fact, as illustrated 
in Fig. \ref{nu:0}, the calculation of the IA 
cross section at the quasi-elastic and $\Delta$ production peak involves 
integrations of $P({\bf p},E)$ extending over regions of the 
$({\bf p},E)$ plane almost exactly overlapping one another. 

\begin{figure}[hbt]
\centerline
{\psfig{figure=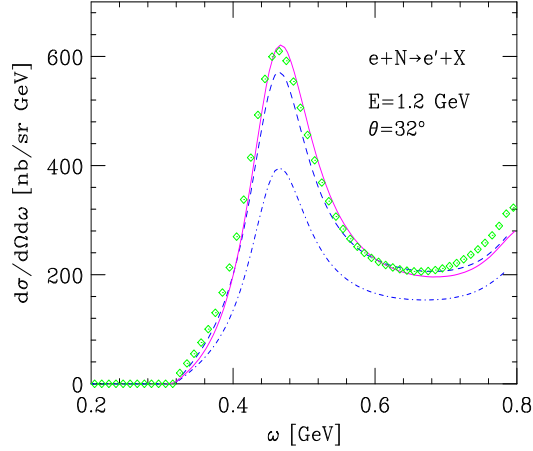,angle=00,width=7.00cm,height=6.00cm}}
\caption{\small 
Cross section of the process $e+N \rightarrow e^\prime + X$ above pion 
production threshold, at beam energy 1200 MeV and scattering angle 32$^\circ$. 
Solid line: H2 fit of Ref. \protect\cite{thia3} for $ep$ scattering; dashed line:
fit of Ref. \protect\cite{br} for $ep$ scattering; diamonds: fit of 
Ref. \protect\cite{christy} for $ep$ scattering; 
dot-dash line: fit of Ref. \protect\cite{br} for $en$ scattering.  }
\label{eN:fit}
\end{figure}

To gauge the uncertainty associated with the description of the nucleon structure 
functions $w_1^N$ and $w_2^N$, we have compared the electron-proton cross sections 
obtained from the model of Ref. \cite{br} to the ones obtained from the model
developed in Refs. \cite{thia1,thia2,thia3} and from a global fit \cite{christy}
 including recent Jefferson Lab data \cite{christy2}. The results of 
Fig. \ref{eN:fit} show that at $E_e=1200$ MeV and $\theta=32^\circ$ the 
discrepancy between the different models is not large, being $\sim$ 15 \% at 
the $\Delta$ production peak.  
It has to be noticed, however, that the models of Refs. \cite{br,thia1,thia2,thia3,
christy} have all been obtained fitting data taken at electron 
beam energies larger than 2 GeV, so that their use in the kinematical regime 
discussed in this work involves a degree of extrapolation.

On the other hand, the results obtained using the approach described in this 
paper and the nucleon structure functions of Ref. \cite{br}
are in excellent agreement with the measured $(e,e^\prime)$ cross sections
at beam energies of few GeV \cite{bffs}. 

\begin{figure}[hbt]
\centerline
{\psfig{figure=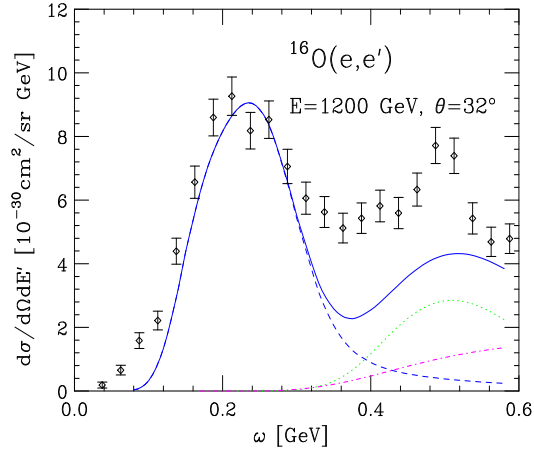 ,angle=00,width=7.00cm,height=6.00cm}}
\caption{\small 
IA cross section of the process $^{16}O(e,e^\prime)$ at beam energy 1200 MeV and
scattering angle 32$^\circ$.
Dashed line: quasi-elastic; dots: quasi-free $\Delta$ production;
dashes: nonresonant background; solid line: total. The experimental data are
from Ref. \protect\cite{LNF}.}
\label{back}
\end{figure}

Figure \ref{eN:fit} also shows the prediction of the Bodek and Ritchie fit for
the neutron cross section, which turns out to be much smaller than the proton one.
The results of Ref. \cite{BenMel1} suggest that extrapolating the Bodek and Ritchie fit
to the low $Q^2$ region relevant to tha analysis of the data of Ref. \cite{LNF} may 
lead to sizably underestimate the neutron contributions.
On the other hand, the fit of Ref. \cite{br} consistently includes 
both resonant and nonresonant contributions to the nuclear cross section. In this regard, 
it has to be pointed out that the nonresonant background is not negligible.
As illustrated in Fig. \ref{back}, for beam energy 1200 MeV and scattering
angle 32$^\circ$ it provides $\sim$ 25 \% of the cross section
at energy transfer corresponding to the $\Delta$ peak.

\section{Neutrino-nucleus cross section}
\label{nuA}

The Born approximation cross section of the weak charged current process
\beq
\nu_\ell + A \rightarrow \ell^- + X\ ,
\label{nu:process}
\eeq
can be written in the form (compare to Eq. (\ref{e:cross:section}))
\beq
\frac{d\sigma}{d\Omega_\ell dE_\ell} = \frac{G^2}{32 \pi^2}\
\frac{|{\bf k}^\prime|}{|{\bf k}|}\
 L_{\mu \nu} W^{\mu \nu}\ ,
\label{nu:cross:section}
\eeq
where $G=G_F \cos \theta_C$, $G_F$ and $\theta_C$ being Fermi's coupling constant and
Cabibbo's angle, $E_\ell$ is the energy of the final state
lepton and ${\bf k}$ and ${\bf k}^\prime$ are the neutrino and charged lepton
momenta, respectively. Compared to the corresponding quantities appearing in
Eq. (\ref{e:cross:section}), the tensors $L_{\mu \nu}$ and
$W^{\mu \nu}$ include additional terms resulting from the
presence of axial-vector components in the leptonic and hadronic
currents (see, e.g., Ref. \cite{walecka}).

Within the IA scheme, the cross section of Eq. (\ref{nu:cross:section}) can be cast
in a form similar to that obtained for the case of electron-nucleus scattering
(see Eq. (\ref{csIA})). Hence, its calculation requires the nuclear spectral function
and the tensor describing the weak charged current interaction of a free nucleon,
$w_N^{\mu\nu}$. In the case of quasi-elastic scattering, neglecting the contribution
associated with the pseudoscalar form factor $F_P$, the latter can be written
in terms of the nucleon Dirac and Pauli form factors $F_1$ and $F_2$, related to the
measured electric and magnetic form factors $G_{E}$ and $G_{M}$ through
\beq
F_{1} = \frac{1}{1 - q^2/4 m^2} \left( G_{E} - \frac{q^2}{4 m^2} G_{M} \right)
\eeq
\beq
F_{2} = \frac{1}{1 - q^2/4 m^2} \left( G_{M} - G_{E} \right) \ ,
\eeq
and the axial form factor $F_A$.

\begin{figure}[hbt]
\centerline{\psfig{figure=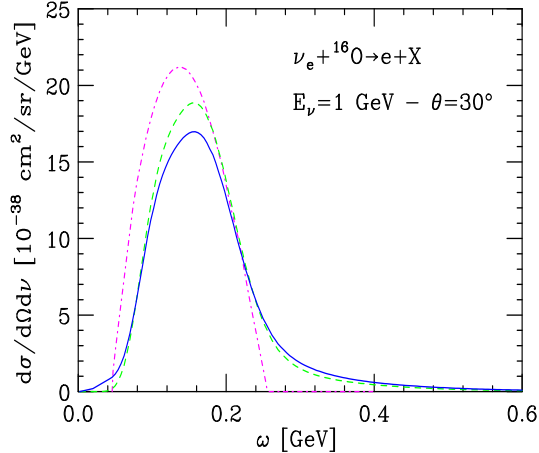
,angle=00,width=7.00cm,height=6.00cm}}
\caption{\small 
Differential cross section $d\sigma/d\Omega_e d\nu$
for neutrino energy $E_\nu = 1 $ GeV
and electron scattering angle $\theta_e = 30^\circ$.
The IA results are represented by the dashed line, while the solid line
corresponds to the full calculation, including the effects of FSI.
The dotted line shows the prediction of the FG model
with Fermi momentum $k_F = 225$ MeV and average separation energy
$\epsilon = 25$ MeV.}
\label{nu:1}
\end{figure}

Figure \ref{nu:1} shows the calculated cross section of the process $^{16}O(\nu_e,e)$,
corresponding to neutrino energy $E_\nu = 1 $ GeV and electron scattering
angle $\theta_e = 30^\circ$, plotted as a function of the energy transfer
$\nu = E_\nu - E_e$.
Numerical results have been obtained using the spectral function of Fig. \ref{pke:O}
and the dipole parameterization for the form factors, with an axial mass of 1.03 GeV.

Comparison between the solid and dashed lines shows that
the inclusion of FSI results in a sizable redistribution of the IA strength,
leading to a quenching of the quasi-elastic peak and to the enhancement of the tails.
For reference, we also show the cross section predicted by the FG
 model with Fermi momentum $p_F = 225$ MeV and average separation energy
$\epsilon = 25$ MeV. Nuclear dynamics, neglected in the oversimplified
picture in terms of noninteracting nucleons, clearly appears to play a relevant role.

\begin{figure}[hbt]
\centerline{\psfig{figure=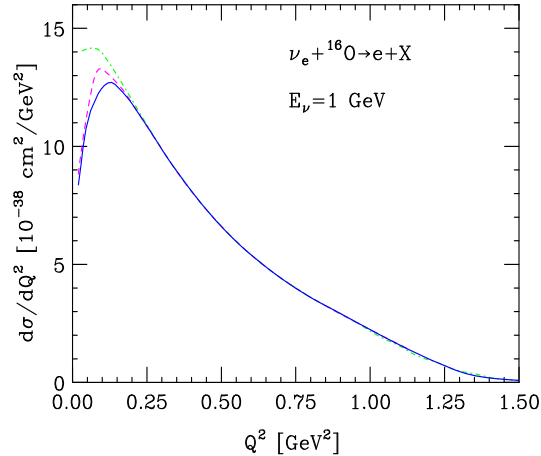
,angle=00,width=7.00cm,height=6.00cm}}
\caption{\small 
Differential cross section $d\sigma/dQ^2$
for neutrino energy $E= 1$ GeV. The dot-dash line shows the IA results,
while the solid and dashed lines have been obtained using the modified
spectral function of Eq. (\protect\ref{pauli1}), with and without inclusion
of FSI, respectively.
.}
\label{nu:2}
\end{figure}

It has to be pointed out that the approach described in
Chapter \ref{response}, while including dynamical
correlations in the final state, does not take into account statistical
correlations, leading to Pauli blocking of the phase space available to the
knocked-out nucleon.

A rather crude prescription to estimate the effect of Pauli blocking amounts to
modifying the spectral function through the replacement
\beq
P({\bf p},E) \rightarrow P({\bf p},E)
\theta(|{\bf p} + {\bf q}| - {\overline p}_F)
\label{pauli1}
\eeq
where ${\overline p}_F$ is the average nuclear Fermi momentum, defined as
\beq
{\overline p}_F = \int  d^3r\ \rho_A({\bf r}) p_F({\bf r}),
\label{local:kF}
\eeq
with $p_F({\bf r})=(3 \pi^2 \rho_A({\bf r})/2 )^{1/3}$, $\rho_A({\bf r})$ being the
nuclear density distribution. For oxygen, Eq. (\ref{local:kF}) yields
${\overline p}_F = 209$ MeV. Note that, unlike the spectral function, the
quantity defined in Eq. (\ref{pauli1})
does not describe intrinsic properties of the target only, as it depends
explicitly on the momentum transfer.

\begin{figure}[ht]
\begin{center}
\includegraphics[scale=0.65]{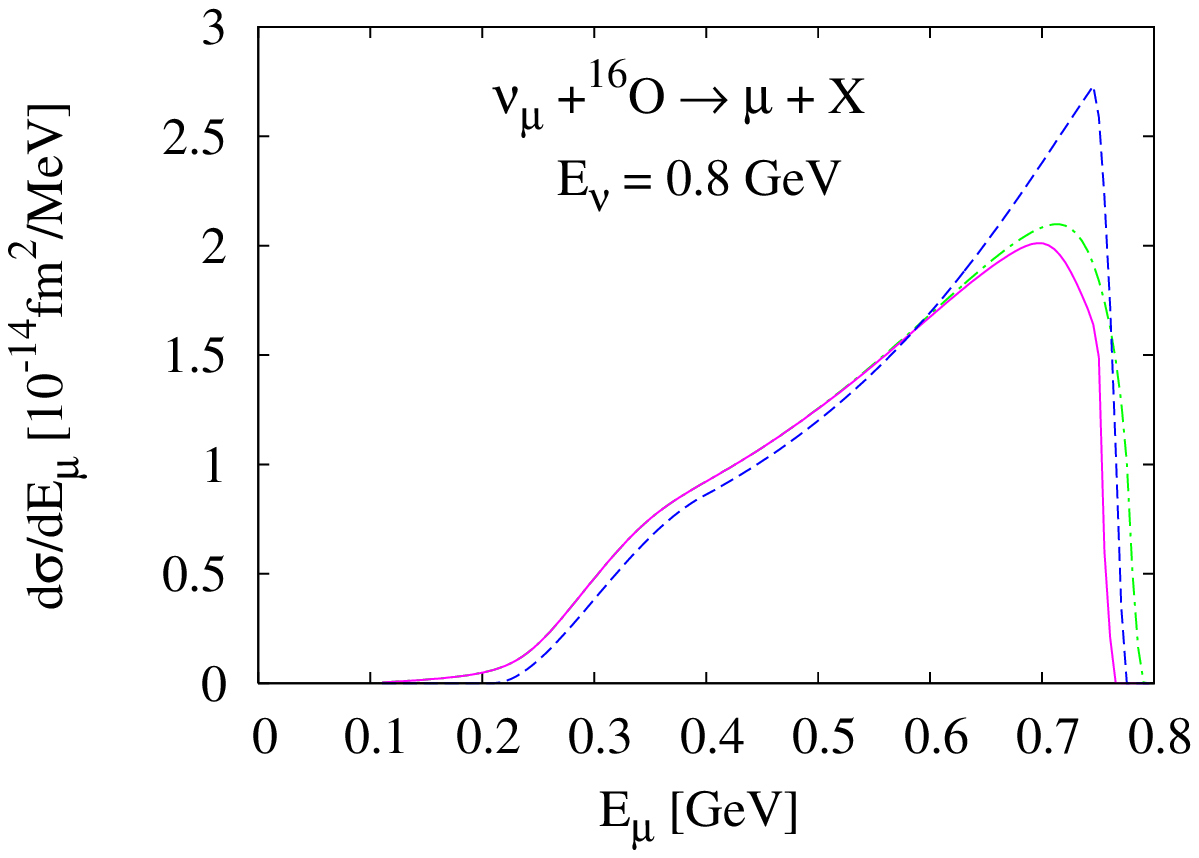}
\includegraphics[scale=0.65]{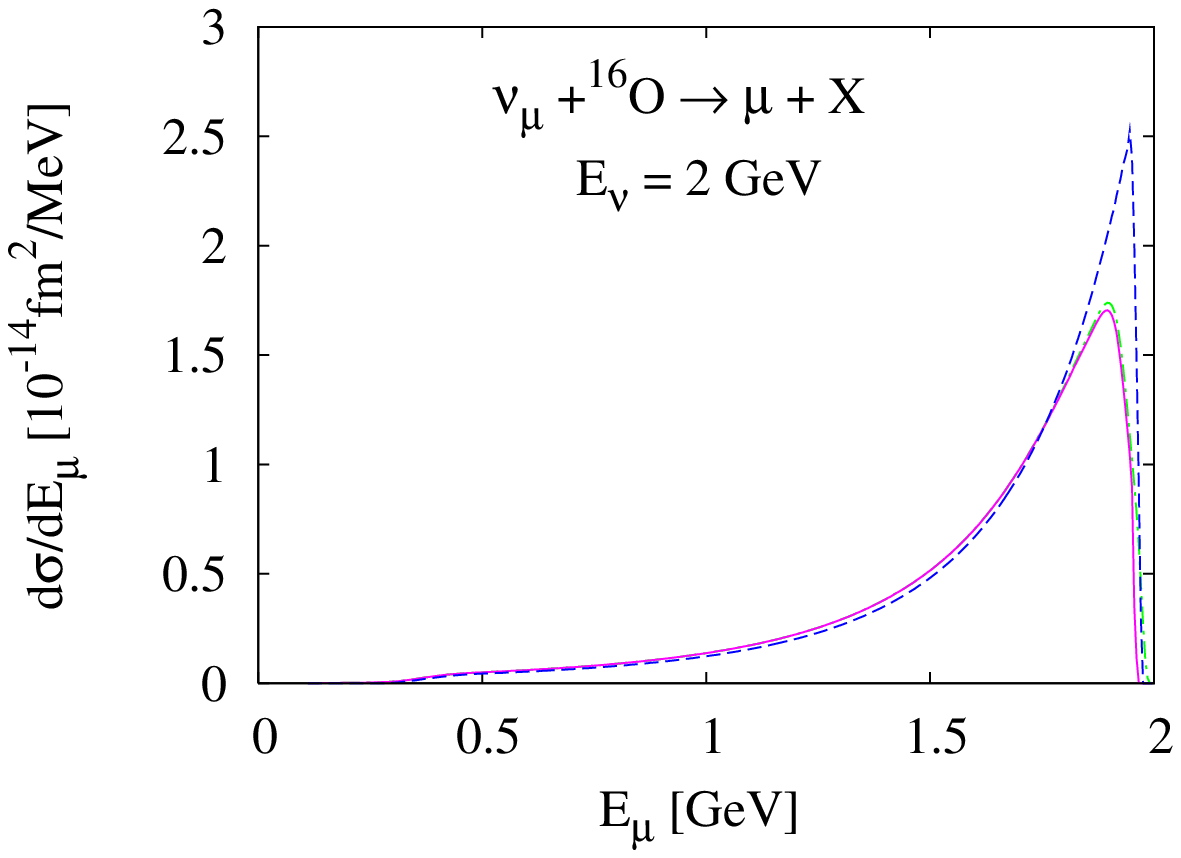}
\end{center}
\caption{
Quasi-elastic differential cross section $d\sigma/dE_\mu$
as a function of the scattered energy $E_\mu$ for the neutrino energy $E =$ 0.8
and 2.0 GeV. The solid line shows IA calculation with Pauli
blocking as in Eq. (40), the dot-dash line IA calculation without Pauli
blocking, and the dashed line FG model.}
\label{del}
\end{figure}

The effect of Pauli blocking is hardly visible in the differential cross section
shown in Fig. \ref{nu:1}, as the kinematical setup corresponds to
$Q^2 > 0.2$ GeV$^2$ at the quasi-elastic peak. The same is true for the 
electron scattering cross sections discussed in the previous Section. 
On the other hand, this effect becomes very large at lower $Q^2$.

Figure \ref{nu:2} shows the calculated differential cross section $d\sigma/dQ^2$
for neutrino energy $E_\nu= 1$ GeV. The dashed and dot-dash lines correspond to the
IA results with and without inclusion of Pauli blocking, respectively. It
clearly appears that the effect of Fermi statistic in suppressing scattering
shows up at $Q^2 < 0.2$ GeV$^2$ and becomes very large at lower $Q^2$. The results of
the full calculation, in which dynamical FSI are also included, are displayed as a full line.
The results of Fig. \ref{nu:2} suggest that Pauli blocking and FSI may explain
the deficit of the measured cross section at low $Q^2$ with respect to the
predictions of Monte Carlo simulations \cite{Ishida}.

Figure \ref{del} shows the $\nu_\mu$-nucleus cross sections
as a function of the scattered muon energy, by comparing
the cross sections calculated by FG, and by the use of the spectral
function with and without Pauli blocking.  Figure \ref{del} shows
that FG yields a larger high-energy peak contribution than the other
two. This is {\it not} due to the Pauli blocking, but due to the
nuclear correlation effects in the spectral function: the muons
tend to be scattered with a higher energy. This effect should show
up in the forward angle cross section and  {\it may have a direct
effect on neutrino oscillation measurements}.

\chapter{Low momentum transfer regime}
\label{RESP}
In this chapter we focus on the nuclear matter response to weak interactions in the regime of low 
momentum transfer ($\mq \sim$ 10 MeV), where the non relativistic approximation is expected to be 
applicable. 
Within this approach, the initial and final states can be obtained from NMBT, while the
weak current entering the definition of the tensor $W^{\mu\nu}$ (see Eq.(\ref{e:hadrten}) and 
(\ref{nu:cross:section}) )
 is expanded in powers of $\magq/m$. At leading order, the resulting response can 
be written in the simple form (compare to Eq.(\ref{def:densresp}))
\beq
S(\qm,\omega)=\frac{1}{N}\sum_n\langle 0|O_{\qm}^\dagger|n\rangle\langle
n|O_{\qm}|0\rangle\delta(\omega+E_0-E_n) \ .
\label{ris:def}
\eeq
where, in the case of charged current interactions, $O_{\qm}$ is the operator corresponding to 
Fermi or Gamow-Teller transitions.

The calculation of the nuclear matter response has been carried out using CBF states, obtained from the 
states of the noninteracting system trough the transformation (\ref{def:corr_states}), and the two-body 
cluster approximation for the weak transition matrix elements (see Chapter \ref{CBF}). 
The effect of long range correlations, which are known to play a critical role at low momentum transfer, 
has been also included in our scheme in a fully consistent fashion, using the effective interaction 
defined in Chapter \ref{CBF} and the Tamm-Dancoff (TD) approximation \cite{FetterWalecka,Boffi}. 
We will restrict our discussion to the case of Fermi transitions. The extension to Gamow-Teller 
transitions is trivial.

\section{Non relativistic reduction of the weak charged current}

The starting point of our calculation is the non relativistic reduction of the weak charged current 
operator. Basically, one needs to expand in powers of $\mq/m$ the matrix elements
\bea
\nonumber
\langle \ppm\primo|J_{\mu V}^+|\ppm\rangle&=&\ub
(\ppm\primo,s\primo\eta\primo) \Gamma_\mu u(\ppm,s)\  \eta_{t\primo}^\dagger \tau^+ \eta_t \\
\langle \ppm\primo|J_{\mu A}^+|\ppm\rangle&=&\ub
(\ppm\primo,s\primo)\Gamma_{\mu A}\tau^+u(\ppm,s)  \ \eta_{t\primo}^\dagger \tau^+ \eta_t \ ,
\label{cdc}
\eea
where 
\beq
\Gamma^\mu = \gamma^\mu F_1 + i \sigma^{\mu\nu}q_\nu \frac{F_2}{2m} + q^\mu F_S \ ,
\label{matGamv}
\eeq
with $\sigma^{\mu\nu} = i[\gamma_\mu,\gamma_\nu]/2$, $\gamma^\mu$ being the Dirac matrices, and
\beq
\Gamma^\mu_A = \gamma^\mu\gamma^5 F_A + q^\mu\gamma^5 F_P +
 i \gamma^5 \sigma^ {\mu\nu}q_\nu F_T \ .
\label{matGama}
\eeq
In the above equations, $\gamma^5 = i \gamma^0 \gamma^1 \gamma^2 \gamma^3$, $u(\ppm,s)$ is 
the spinor describing a Dirac fermion with momentum ${\bf p}$ and 
spin polarization $s$, $\eta_t$ is the Pauli spinor specifying the isospin state of the nucleon,
$\tau ^+ = (\tau^1 + i \tau^2)/2$ is the isospin raising operator and the 
form factors $F_1$, $F_2$, $F_S$, $F_A$, $F_P$ and $F_T$ are functions of the squared four momentum 
transfer $q^2$.

From the definition
\begin{equation} 
u(\ppm,s) = N_\ppm \left(\begin{array}{c}
\chi_s \\
\\
\frac{\bbox{\sigma}\cdot\textbf{p}}{E_\ppm + m}\chi_s
\end{array}
\right) \ ,
\label{spDnr}
\end{equation}
where $N_{\ppm}$ is a normalization constant and $\chi_s$ is a Pauli spinor, it follows that, 
to zero-th order in $\mq/m$, we can write
\beq
\langle \ppm\primo|J_{0 V}^+(0)|\ppm \rangle
\simeq F_1(0) \ \chi_{s\primo}^\dagger \chi_s \ \eta_{t\primo}^\dagger\tau^+\eta_t \ .
\label{J0}
\eeq
Due to the antidiagonal structure of the matrix $\gamma^5$, the corresponding matrix element 
of the axial vector current, 
$\langle \ppm \primo|J_{0 A}^+(0)| \ppm\rangle$, does not have any zero-th order contributions.

In conclusion, denoting $g_V=F_1(0)$ the non relativistic reduction 
amounts to making the replacement
\beq
\langle {\bf r}_i^\prime | J^+_0V(\qm) | {\bf r}_i \rangle 
\rightarrow  \ O^F_i(\qm)= g_V \delta( {\bf r}_i - {\bf r}_i^\prime ) \ 
 \ {\rm e}^{i\qm{\bf r}_i} \tau_i^+ \ ,
\label{ofdef}
\eeq 
which defines the Fermi transition operator 
$O^F_i(\qm)$ in coordinate space.

The axial part of the current contributes through the $\mu = i = 1,2,3$ components
\bea
\nonumber
\langle \ppm\primo|\J_A|\ppm\rangle&=&\ub(\ppm\primo,s\primo) 
[F_A\gam\gamma_5+F_P\gamma_0\qm\gamma_5]u(\ppm,s \ \eta_{t\primo} \tau^+ \eta_t \\
&=&u^\dagger(\ppm\primo,s\primo)
[F_A\gamma_0\gam\gamma_5+F_P\gamma_0\qm\gamma_5]\tau^+u(\ppm,s) \ \eta_{t\primo} \tau^+ \eta_t \ .
\label{corass}
\eea
In the above equation, the zero-th order term proportional to $F_P$ vanishes because 
$\gamma_0\gamma_5$ is antidiagonal. As for the term proportional to $F_A$ we find
instead
\beq
\gamma_0\gam\gamma_5=\left(\begin{array}{cc}
                           I&0\\
			   0&-I
			   \end{array}
			   \right)
			   \left(\begin{array}{cc}
                           0&\bbox{\sigma}\\
			   -\bbox{\sigma}&0
			   \end{array}
			   \right)
		           \left(\begin{array}{cc}
                           0&I\\
			   I&0
			   \end{array}
			   \right)=
                           \left(\begin{array}{cc}
                           \bbox{\sigma}&0\\
			   0&\bbox{\sigma}
			   \end{array}
			   \right) \ .
\label{prodmatr}
\eeq
Finally, the vector part of $\J^+$ does not contribute at zero-th order. 

Making use of Eq.(\ref{prodmatr}), we can then write 
\beq
\langle {\bf r}_i^\prime | \J^+_0(\qm) | {\bf r}_i \rangle \rightarrow 
{\bf O}^{GT}_i(\qm) = g_A \delta( {\bf r}_i - {\bf r}_i^\prime ) \ {\rm e}^{i\qm{\bf r}_i}
\bbox{\sigma}_i\tau_i^+ \ ,
\label{ogtdef}
\eeq
with $g_A=F_A(0)$, which defines the operator inducing Gamow-Teller transitions.

\section{Correlated matrix elements and effective operators}

Using correlated states implies severe difficulties in the explicit calculation of the weak 
matrix element. In the FG model, the nuclear response is non vanishing only when the final 
nuclear state differs from the initial state for the presence of a particle excited outside 
the Fermi sea and a hole in the Fermi sea. In the presence of correlations, which can 
induce virtual nucleon-nucleon scattering processes leading to excitation of nucleons to
states outside the Fermi sea, more complex scenarios must also be considered. 
For example, if the initial state has a two particle-two hole component, the final state can 
be a three particle-three hole state or, if the probe interacts with an 
excited nucleon, a two particle-two hole state.

In the following we will consider only the dominant transition, between the {\em correlated} 
ground state and a {\em correlated} one particle-one hole ($ph$) state. The corresponding 
Weak matrix element can be written 
\beq
M_{ph}=\frac{\langle ph|F^\dagger O F|0\rangle}
{\langle ph|F^\dagger F|ph\rangle^\met\langle 0|F^\dagger F|0\rangle^\met} \ ,
\label{Mph}
\eeq
where $F$ is the correlation operator defined in Eq.(\ref{def:F}). Here the kets $|0\rangle$ 
and $|ph\rangle$ correspond to the ground and one particle-one hole Fermi Gas states, 
respectively, 
and $O=\sum_i O_i$, $O_i$ being the Fermi or the Gamow-Teller transition operator 
(see Eqs.(\ref{ofdef})) and (\ref{ogtdef})).

In calculating the weak matrix element, we will use the two-body cluster approximation 
discussed in Chapter \ref{CBF}. Let us define
\beq
g_{ij}=f_{ij}-1 \ ,
\eeq  
with $f_{ij}$ defined as in Eq,(\ref{def:corrf}). 
Note the the $g_{ij}$ is short ranged, and therefore its matrix elements are small. 

At two-body level, the cluster expansion Eq.(\ref{Mph}) yields
\bea
\nonumber
\langle ph |F^\dagger O F |0\rangle & \simeq & \langle ph|(1+\sum_{j>i}\g)O 
(1+\sum_{j>i}\g)|0\rangle\\
\nonumber
&=&N\langle ph|O_1|0\rangle+\coppie\langle ph|\{O_1+O_2,g_{12}\}|0\rangle\\
&+&\coppie\langle ph|g_{12}(O_1+O_2)g_{12}|0\rangle \ ,
\label{Nph}
\eea
where $\{A,B\}=AB+BA$. 
The above equation, suggest the definition of an {\em effective operator} $ O^{eff}_{12}$, 
acting on Fermi Gas states, reminiscent of the effective interaction of Chapter \ref{CBF}. 
From 
\beq
\langle ph|F^\dagger O F|0\rangle=\langle ph|O^{eff}|0\rangle .
\label{oeff:def}
\eeq
it follows that, at the two-body cluster level (compare to Eq.(\ref{Nph}))
\beq
\frac{1}{N} O^{eff}_{12}=O_1 +\frac{N-1}{2}\{O_1+O_2.g_{12}\}+\frac{N-1}{2}[g_{12}(O_1+O_2)g_{12}] \ .
\label{oeff:tb}
\eeq
Note that the $O^{eff}$ is a two-body operator, as it includes screening effects arising from 
nucleon-nucleon correlations. 

Replacing Eq.(\ref{Nph}) into Eq.(\ref{Mph}) leads to
\beq
M_{ph}\simeq N_1+N_2+N_3
\label{mph2}
\eeq
where
\bea
N_1&=&N\langle ph|O_1|0\rangle\label{A} \ ,\\
N_2&=&\coppie\langle ph|\{O_1+O_2,g_{12}\}|0\rangle\label{B} \ ,\\
N_3&=&\coppie\langle ph|g_{12}(O_1+O_2)g_{12}|0\rangle\label{C} \ .
\eea

The wave function of the FG ground state $\Psi_0(R) = \langle R | 0 \rangle$ can be written as a 
determinant according to
\beq
\Psi_0(X)=\frac{1}{\sqrt{N!}}\left|\begin{array}{ccc}
                                    \phi_1(1)&\cdots&\phi_N(1)\\
				    \vdots&\ddots&\vdots\\
				    \phi_1(N)&\cdots&\phi_N(N)\\
				    \end{array}
			     \right| \ ,
\label{fosf}
\eeq
where $\phi_m(n)$ is the wave function describing the n-th particle in the state $m$, 
with momentum ${\bf k}_m$ and spin and isospin projections $s_m$ and $t_m$, respectively
\beq
\phi_m(n)=\frac{1}{\sqrt{V}}{\rm e}^{i {\bf k}_m\rmo_n}\chi_{s_m}\eta_{t_m} \ .
\label{fopl}
\eeq
One particle-one hole states, and the corresponding wave functions 
$\Psi_{ph}(R) = \langle 0 | ph \rangle$, can be built from the ground state through
\beq
|ph\rangle=a^\dagger_p a_h|0\rangle 
\label{ph:deffromgs}
\eeq
using Eq.(\ref{fosf}).

As $g_{ij}$ is a two-body operator, it turns out to be convenient rewriting the ground 
and one particle-one hole wave functions, $\Psi_0(R)$ and $\Psi_{ph}(R)$, in the form
(see Eq.(\ref{fosf}))
\bea
\Psi_0(R)&=&\frac{1}{\sqrt{N(N-1)}}\sum_{\alpha\beta}(-1)^{n_\alpha +n_\beta}
\phi_\alpha(1)\phi_\beta(2)\Phi_{\gamma\neq\alpha,\beta}(3,\cdots ,N)
\label{fosf2c},\\
\nonumber
\Psi_{ph}(R)&=&\frac{1}{\sqrt{N(N-1)}}\left\{\sum_{\alpha\neq
h}\sum_{\beta\neq\alpha,h}(-1)^{n_\alpha+n_\beta}\right.
\phi_\alpha(1)\phi_\beta(2)\Phi_{p,\gamma\neq\alpha,\beta,h}(3,\cdots, N)\\
\nonumber
&+&
\sum_{\alpha\neq h}(-1)^{\nanb}\phi_\alpha(1)\phi_p(2)\Phi_{\gamma\neq\alpha,h}
(3\cdots N) \\ 
\nonumber
& + & \left. \sum_{\alpha\neq h}(-1)^{\nanb+1}\phi_\alpha(2)\phi_p(1)
\Phi_{\gamma\neq\alpha,h}(3,\cdots N)\right\} \ ,
\label{foph2c}
\eea
where $\Phi_{p,\gamma\neq\alpha,\beta}(3,\cdots, N)$ denotes the wave function of a $N-2$-particle 
system, with a particle in the state $p$ and holes in the state $\gamma \neq\alpha, \ \beta$. 

The (N-2)-particle wave functions satisfy the following orthonormalization relations
\beq
\int\ d^3r_3\cdots d^3r_n \ \Phi^\dagger_{\gamma\neq\alpha,\beta}(3,\cdots ,N) 
\Phi_{p,\gamma\neq\alpha\primo,\beta\primo,h}(3,\cdots,N)  = 0 \ ,
\label{orto1}
\eeq
\beq
\int\ d^3r_3\cdots d^3r_n\Phi^\dagger_{\gamma\neq\alpha\beta}(3,\cdots ,N)
\Phi_{\gamma\neq\alpha\primo,h}(3,\cdots,N) = \delta_{\alpha \alpha\primo} 
  \  \delta_{\beta h} \ .
\label{orto2}
\eeq
With the help of the above equations, the weak matrix element can be reduced to an integral 
over the coordinates of two particles. 

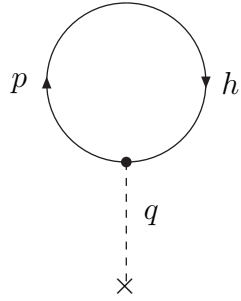
\begin{figure}
\begin{center}
\begin{picture}(300,100)(0,0)
\ArrowArcn(150,50)(30,90,270)
\ArrowArcn(150,50)(30,270,90)
\Text(110,50)[0]{$p$}
\Text(190,50)[0]{$h$}
\Text(160,0)[0]{$q$}
\Vertex(150,20){2}
\DashLine(150,20)(150,-25){3}
\Text(150,-27)[1]{$\times$}
\end{picture}
\end{center}
\caption{Diagram associated with the contribution $N_1$ (see Eq.(\ref{A})).
\label{diag1}}
\end{figure}

The next section will be devoted to the details of the calculation of the 
Fermi transition matrix element. We will start writing it as a sum of 
contributions corresponding to the physical processes schematically represented by 
the diagrams of Figs.~\ref{diag1}-\ref{diag5}. 
The different contributions will be labeled by three indices
\begin{itemize}
\item
the first index, $B$ or $C$, denotes the order (first and second, respectively) in 
the correlation $g$;
\item
the value of the second index, 1 or 2, indicates that the Fermi operator 
acts on particle 1 (any nucleon inside the Fermi-sea in the initial state) or on particle 2 
(the active particle, carrying momentum $\hm$ and spin projection $\sigma_h$ in the ground state);
\item
the third index specifies the direct ($d$) and  exchange ($e$) contributions to the 
matrix elements. Note that exchange terms carry an additional minus sign.
\end{itemize}

As the Fermi operator involves the isospin raising operator, the only 
non vanishing matrix elements are those in which the particle excited outside the Fermi-sea 
is a proton and the active particle in the initial state is a neutron. 

\subsection{Fermi Transition}

We will now discuss the explicit calculation of the weak matrix element. 
 Since the corresponding operator does not induce spin transitions, 
all contributions to the matrix element involve a Kronecker delta expressing the 
condition $\sigma_p=\sigma_h$, where $\sigma_p$ and $\sigma_h$ are the spin 
projections of the particle and hole states, respectively. In order to simplify 
the notation, this factor will be omitted.

From Eq.(\ref{A}) we obtain
\bea
\nonumber
N_1 & = & N\langle ph|O_1|0\rangle = \rho\int d^3r_1 e^{-i\ppm\rmo_1}e^{i\qm\rmo_1}
e^{i\hm\rmo_1}\chi^\dagger_p\chi_h\eta^\dagger_p\tau^+\eta_h \\
& = &\rho \ (2\pi)^3\delta^{(3)}(\ppm-\hm-\qm) \ ,
\label{A2}
\eea
where $\chi_{p(h)} = \chi_{\sigma_{p(h)}}$ and $\eta_{p(h)} = \eta_{\tau_{p(h)}}$.

The above equation shows that $N_1$, the leading term of the expansion, is just 
the Fermi transition matrix element in the absence of correlations. 

Io order to write down the complete expression of $N_2$, let us consider the 
matrix element
\bea
B1&=&\coppie\int dR\Psi^\dagger_{ph}\{O_1,g_{12}\}\Psi_0(R)\\
\nonumber
&=&\met\sum_\alpha\int
d^3r_1d^3r_2[\phi_\alpha(1)\phi_p(2)-\phi_p(1)\phi_\alpha(2)]\{O_1,g_{12}\}
\phi_\alpha(1)\phi_h(2) \ ,
\label{B1}
\eea
in which we have used Eqs.(\ref{fosf2c})-(\ref{foph2c}) and the sum is extended over the Fermi sea. The right hand side 
of Eq.(\ref{B1}) has both direct and exchange contributions; for the direct term 
we can write
\bea
\nonumber
B1d&=&\met\frac{1}{V^2} \sum_\alpha \sum_n \int d^3r_1d^3r_2 e^{-i({\bf k}_\alpha \rmo_1
+\ppm\rmo_2)}e^{i\qm\rmo}e^{i({\bf k}_\alpha\rmo_1 +\hm\rmo_2)}g_n(r)\\
&\times&\langle\alpha p|\{\tau_1^+,O^{12}_n\}|\alpha h\rangle \ ,
\label{B1d0}
\eea 
where the $g_n(r)$ are the radial functions entering the definition of $g_{12}$ 
and $r=|\rmo_1-\rmo_2|$. 
From now on, the ket $|\alpha\beta\rangle$ indicates a state in which particles 1 and 2 
have spin-isospin $\alpha$ and $\beta$, respectively.
Carrying out the change of variables
\beq
\left\{\begin{array}{l}
        \rmo=\rmo_1-\rmo_2\\ \\
	{\bf R}=(\rmo_1+\rmo_2)/2 \ ,
	\end{array}
\right. 
\label{change}
\eeq
and integrating over ${\bf R}$ we obtain
\beq
B1d=\frac{\rho}{4V}\del \sum_{\sigma_\alpha,\tau_\alpha} \sum_n \int d^3r
e^{i\qm\rmo}g_n(r)\langle\alpha p|\{\tau_1^+,O^{12}_n\}|\alpha h\rangle,
\eeq
where we have used 
\beq
\sum_\alpha\rightarrow\sum_{k_\alpha,\sigma_\alpha,\tau_\alpha}\rightarrow
\frac{N}{4}\sum_{\sigma_\alpha,\tau_\alpha} \ .
\eeq

Let us now focus on the spin-isospin matrix element. From the definition of the six 
operators $O^{12}_n$ (\ref{pot2}) and 
\bea
\nonumber
\langle\alpha|\tau^+|\alpha\rangle & = & 0  \ , \\
\nonumber
\sum_{\sigma_\alpha}\langle\alpha p|\tre|\alpha h\rangle  & = & 0 \ ,\\
\nonumber
\sum_{\sigma_\alpha}\langle\alpha p|S_{12}|\alpha h\rangle & = & 0 \ , \\	
\eea
it follows that the only non-vanishing contribution is given by
\beq
\sum_{\sigma_\alpha,\tau_\alpha}\langle\alpha p|\{\tau_1^+,\due\}|\alpha
h\rangle=8 \ .
\eeq
Note that the previous result was obtained exploiting the anticommutator
\beq
\{\tau_1^+,\due\}=2\tau_2^+.
\eeq
The final expression of $B1d$ turns out to be
\beq
B1d=\frac{\rho}{4V}\del \ 8\int d^3re^{i\qm\rmo} \ g_2(r).
\eeq 
\begin{figure}
\begin{center}
\begin{picture}(300,100)(0,0)
\ArrowArcn(200,50)(20,0,180)
\ArrowArcn(200,50)(20,180,0)
\ArrowArcn(100,50)(20,0,180)
\ArrowArcn(100,50)(20,180,0)
\Text(100,78)[0]{$p$}
\Text(100,20)[0]{$h$}
\Text(200,78)[0]{$k_\alpha$}
\Vertex(120,50){2}
\Text(125,40)[0]{2}
\Vertex(180,50){2}
\Text(175,40)[0]{1}
\DashLine(180,50)(180,5){3}
\Text(190,22)[0]{$q$}
\Text(181,5)[1]{$\times$}
\Photon(120,50)(180,50){4}{5}
\end{picture}
\end{center}
\vspace*{-.3 in}
\caption{Diagram associated with the contribution $B1d$ (see Eq.(\ref{B1d0})).
\label{diag2}}
\end{figure}

Let us now consider the exchange term. Performing again the change of variables of 
Eq.(\ref{change}), we find the expression
\beq
B1e=\met\frac{1}{V^2}\del\sum_{\alpha,n} \int d^3re^{-i(\hm+{\bf
k}_\alpha)\rmo} g_n(r)\langle p\alpha|\{\tau_1^+,O^{12}_n\}|\alpha h\rangle \ ,
\label{B1e1}
\eeq
which, after summing over $k_\alpha$, becomes
\beq
B1e=\frac{\rho}{4V}\del\sum_{\sigma_\alpha,\tau_\alpha,n}\int
d^3re^{-i\hm\rmo}g_n(r)\ell(k_Fr) 
\langle p\alpha|\{\tau_1^+,O^{12}_n\}|\alpha h\rangle \ ,
\label{B1e2}
\eeq
where
\beq
\ell(k_Fr)=\frac{4}{\rho}\int_0^{k_F}\frac{d^3k}{(2\pi)^3}e^{i{\bf k}\rmo}\\
=3 \ \frac{\sin(k_Fr)-(k_F r)\cos(k_Fr)}{(k_Fr)^3} \ ,
\eeq
is the Slater function.

For the spin-isospin matrix element, using the completeness relation
\beq
\sum_{\sigma_\alpha,\tau_\alpha}|\alpha\rangle\langle\alpha|=1 \ .
\label{iden}
\eeq
we can write:
\bea
\nonumber
\suma\langle p\alpha|\{1,\tau_1^+\}|\alpha h\rangle=2 \ ,\\
\nonumber
\suma\langle p\alpha|\{\due,\tau_1^+\}|\alpha h\rangle=2 \ ,\\
\nonumber
\suma\langle p\alpha|\{\tre,\tau_1^+\}|\alpha h\rangle=6 \ ,\\
\nonumber
\suma\langle p\alpha|\{\quattro,\tau_1^+\}|\alpha h\rangle=6 \ ,\\
\nonumber
\suma\langle p\alpha|\{S_{12},\tau_1^+\}|\alpha h\rangle=0 \ ,\\
\nonumber
\suma\langle p\alpha|\{S_{12}\due,\tau_1^+\}|\alpha h\rangle=0 \ .
\eea
Collecting all the above results we finally obtain
\beq
B1e=\frac{\rho}{4V}\del\int
d^3re^{-i\hm\rmo}\ell(k_Fr)[2g_1+2g_2+6g_3+6g_4].
\label{B1e}
\eeq 
\begin{figure}
\vspace*{-.4 in}
\begin{center}
\begin{picture}(300,100)(0,0)
\ArrowArc(150,10)(50,52,125)
\ArrowArc(150,90)(50,232,305)
\Photon(120,50)(180,50){3}{5}
\Vertex(120,50){2}
\Text(115,50)[0]{2}
\Vertex(180,50){2}
\Text(185,50)[0]{1}
\DashLine(180,50)(180,5){3}
\Text(190,22)[0]{$q$}
\Text(181,5)[1]{$\times$}
\end{picture}
\end{center}
\vspace*{-.4 in}
\caption{Diagram associated with the contribution $B1e$ (see Eq.(\ref{B1e})).
\label{diag3}}
\end{figure}
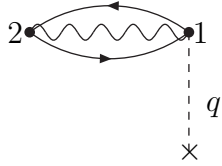
Note that, in order to simplify the notation, in the above equation and in the 
rest of this Section the functional dependence of the $g_n$'s on $r$ is omitted.

The contributions $B2d$ and $B2e$ can be obtained from $B1d$ and $B1e$ through the 
substitution
\beq
e^{i\qm\rmo_1}\tau_1^+\rightarrow e^{i\qm\rmo_2}\tau_2^+ \ .
\label{sost12}
\eeq
For the direct term, $B2d$, we find
\beq
B2d=\frac{\rho}{4V}\del \ 8\int d^3r \ g_1(r) \ ,
\label{b2d}
\eeq
\begin{figure}
\vspace*{-.3 in}
\begin{center}
\begin{picture}(300,100)(0,0)
\ArrowArcn(200,50)(20,0,180)
\ArrowArcn(200,50)(20,180,0)
\ArrowArcn(100,50)(20,0,180)
\ArrowArcn(100,50)(20,180,0)
\Text(100,78)[0]{$p$}
\Text(100,20)[0]{$h$}
\Text(200,78)[0]{$k_\alpha$}
\Vertex(120,50){2}
\Text(125,40)[0]{2}
\Vertex(180,50){2}
\Text(175,40)[0]{1}
\DashLine(120,50)(120,5){3}
\Text(130,22)[0]{$q$}
\Text(121,5)[1]{$\times$}
\Photon(120,50)(180,50){4}{5}
\end{picture}
\end{center}
\vspace*{-.3 in}
\caption{Diagram associated with the contribution $B2d$ (see Eq.(\ref{b2d})).
\label{diag4}}
\end{figure}
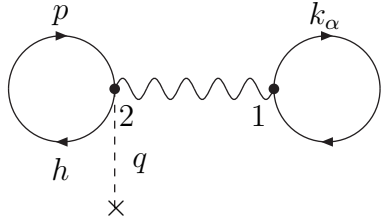
while the exchange term, $B2e$, reads
\beq
B2e=\frac{\rho}{4V}\del\int
d^3re^{-i\ppm\rmo}\ell(k_Fr)[2g_1+2g_2+6g_3+6g_4] \ .
\label{B2Ne}
\eeq

The contributions of Eq.(\ref{C}), $N_3$, can again be obtained from the 
corresponding contributions of Eq.(\ref{A}), $N_1$, through the substitution
\beq
\sum_n g_n \langle \alpha\primo\beta\primo|\{O_n^{12},\tau^+_i\}|\alpha\beta\rangle
\rightarrow\sum_{n,n\primo} g_ng_{n\primo} \langle\alpha\primo\beta\primo|
O_n^{12}\tau_i^+O_{n\primo}^{12}|\alpha\beta\rangle \ .
\label{sostxc}
\eeq
As the six operators $O_n^{12}$ form an algebra (see Appendix \ref{pauli}), we can write
\beq
O_n^{12}O_m^{12}=\sum_r K_{nmr}O_r^{12} \ .
\label{algop}
\eeq
Exploiting this property, the product of two or more operators can be rewritten as 
a linear combination of the six operators. We will also make use of the following relations
\bea
(\bbox{\sigma}{\bf A})(\bbox{\sigma}{\bf B})&=&({\bf AB})+
i\bbox{\sigma}({\bf A}\times {\bf B}),\label{fonsigma}\\
(\tre)^2&=&3-2(\tre),\label{quadsig}\\
S_{12}(\tre)&=&(\tre) S_{12}=S_{12},\label{S123}\\
S_{12}^2&=&6+2(\tre)-2S_{12}.\label{S12q}
\eea
The first equation of the above group follows from the properties of the Pauli matrices. 
The second can be obtained directly from the first, while the third and fourth can be 
easily derived writing the tensor $S_{12}$ in the form:
\beq
S_{12}=\sum_{ij}(3\hat{r}_i\hat{r}_j-\delta_{ij})\sigma_1^i\sigma_2^j,
\label{S12v}
\eeq
where $\hat{r}$ is the unit vector in the direction of $\rmo$.
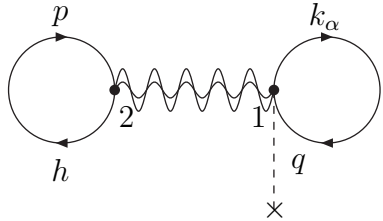
\begin{figure}
\vspace*{-.3 in}
\begin{center}
\begin{picture}(300,100)(0,0)
\ArrowArcn(200,50)(20,0,180)
\ArrowArcn(200,50)(20,180,0)
\ArrowArcn(100,50)(20,0,180)
\ArrowArcn(100,50)(20,180,0)
\Text(100,78)[0]{$p$}
\Text(100,20)[0]{$h$}
\Text(200,78)[0]{$k_\alpha$}
\Vertex(120,50){2}
\Text(125,40)[0]{2}
\Vertex(180,50){2}
\Text(175,40)[0]{1}
\DashLine(180,50)(180,5){3}
\Text(190,22)[0]{$q$}
\Text(181,5)[1]{$\times$}
\Photon(120,50)(180,50){8}{5}
\Photon(120,50)(180,50){3}{5}
\end{picture}
\end{center}
\vspace*{-.3 in}
\caption{Diagram associated with the contribution $C1d$ (see Eq.(\ref{C1d})).
\label{diag5}}
\end{figure}
Using Eqs.(\ref{fonsigma})-(\ref{S12q}) we find
\bea
\nonumber
C1d & = & \frac{\rho}{4V}\del \\
& & \ \ \times \int
d^3re^{i\qm\rmo}[8g_1g_2+8g_2^2+24g_4^2+48g_6^2+24g_3g_4+48g_5g_6] \ ,
\label{C1d}
\eea
and 
\bea
\nonumber
C1e & = & \frac{\rho}{4V}\del\int d^3r e^{-i\hm\rmo}\ell(k_Fr)[g_1^2+2g_1g_2+6g_1g_3
 + 6g_1g_4 \\
 & + & g_2^2-3g_3^2 - 3g_4^2+12g_5^2+12g_6^2+6g_2g_3+6g_2g_4-6g_3g_4+24g_5g_6] \ .
\label{C1e}
\eea

Following the same procedure and using the substitution (\ref{sost12}) we also 
obtain
\beq
C2d=\frac{\rho}{4V}\del\int\ d^3r [8g_1^2-4g_2^2+12g_3^2-12g_4^2+24g_5^2-24g_6^2] \, 
\label{C2d} 
\eeq
\bea
\nonumber
C2e&=&\frac{\rho}{4V}\del\int\ d^3r e^{-i\ppm\rmo}
[2g_1^2+2g_1g_2+6g_1g_3+6g_1g_4 \\
&& + g_2^2-3g_3^2 + 12g_5^2+12g_6^2+6g_2g_3+6g_2g_4-6g_3g_4+24g_5g_6].
\label{C2e}
\eea

\section{Calculation of the response}

In the previous Sections, we have defined all the ingredients needed to obtain 
the weak response of nuclear matter in the one particle-one hole channel. For 
reasons that will become apparent in the next Section, devoted
to the discussion of the TD approximation, numerical calculations have been 
carried out in a cubic box of finite volume $L^3$, using a discrete set of one particle 
one-hole states. Within this scheme, the allowed nucleon momenta ${\bf k}$ are 
given by 
\beq
{\bf k}=\frac{2\pi}{L} (n^x {\hat x} + n^y {\hat y} + n^z {\hat z}) \ ,
\label{momlattice}
\eeq
where $n^x, \ n^y, \ n^z  = 0, \ \pm 1 , \ \pm 2,  \ \ldots$ and 
${\hat x},\ {\hat y}$ and ${\hat z}$ are 
unit vectors along the directions of the cartesian axes. 

As we are interested in the low momentum transfer regime ($\mq \ll k_F$), we determine the size of 
the normalization box by requiring that ${\bf q}$ be on the lattice, i.e. that 
\beq
\qm=\frac{2\pi}{L} (n_q^x {\hat x} + n_q^y {\hat y} + n_q^z {\hat z}) \ .
\label{qlattice}
\eeq
For fixed momentum transfer, i.e. for fixed $n_q^x,\ n_q^y,\  n_q^z$ and $\mq$, the above 
equation yields
\beq
L=2\pi\frac{n_q}{\mq} \ ,
\eeq
with
\beq
n_q=\sqrt{(n_q^x)^2+(n_q^y)^2+(n_q^z)^2} \ .
\label{nqdef}
\eeq
Obviously, in the $L\rightarrow\infty$ limit our procedure must reproduce the results obtained 
working with a continuum set of one particle-one hole states.

Let us rewrite the definition of the response using the effective operator, defined as in 
Eq.(\ref{oeff:def}), associated with Fermi transitions
\beq
S_F(\qm,\omega)=\frac{2}{N} \sum_{ |{\bf h}|\leq k_F,|{\bf p}|\geq k_F}
|\langle ph| O_{eff}^F(\qm)|0\rangle|^2 \delta(\omega + E_0 - E_{ph})
\delta(\ppm-\hm-\qm) \ .
\label{risdefgen}
\eeq
Note that the factor $2$ in the right hand side of the above equation comes from the 
sum over the particle and hole spin projections.

For the sake of illustration, we will first consider the noninteracting FG model.
In this case we find
\beq
|M_{ph}(\qm)|^2 = |\langle ph| O_{eff}^F(\qm)|0\rangle|^2 
\rightarrow |\langle ph| O^F(\qm)|0\rangle|^2 \ = 1 \ ,
\eeq
and
\beq
E_{ph} - E_0 = \omega_p - \omega_h  = \omega_{ph} \ ,
\eeq
with
\beq
\omega_k=\frac{|{\bf k}|^2}{2m} \ .
\label{FGenergy}
\eeq

The sum appearing in Eq.(\ref{risdefgen}) is extended to all lattice momenta 
\beq
{\bf h}_i=\frac{2\pi}{L} (n^x_i {\hat x} + n^y_i {\hat y} + n^z_i {\hat z}) 
\label{h}
\eeq
such that $|{\bf h}_i| \leq k_F$ and $|{\bf h}_i + {\bf q}| \geq k_F$, 
with ${\bf q}$ given by Eq.(\ref{qlattice}).

Obviously, the response obtained from the discrete set of final states consists of a collection of
delta function peaks located at 
$\omega = \omega_i = \omega_{|{\bf h}_i+{\bf q}|} - \omega_{h_i}$. 
A smooth function of $\omega$ has been obtained using a finite width gaussian 
representation of the energy conserving $\delta$ function, i.e. replacing
\beq
\delta(\omega - \omega_{i}) \rightarrow \frac{1}{\sigma \sqrt{\pi}} \ 
{\rm exp} \{ - [(\omega - \omega_{i})/\sigma ]^2 \} \ .
\eeq
For sufficiently small values of $\sigma$, the results obtained using this procedure
become independent of $\sigma$.

\begin{figure}[ht]
\centerline{\includegraphics[scale=0.45]{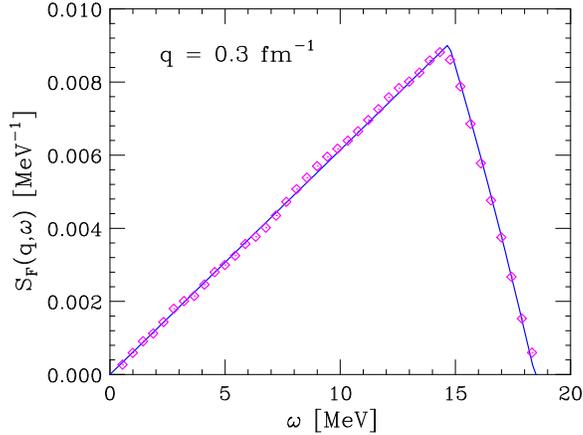}}
\caption{\small Comparison between the FG response at 
$|{\bf q}| = 0.3 \ {\rm fm}^{-1}$ evaluated in a cubic box of side $L= 78 \ {\rm fm}$
using a basis of $\sim$ 3000 states (diamonds) and that resulting from an 
analytical calculation using a continuum set of one particle-one hole states (solid
line).  \label{res1}}
\end{figure}

In Fig. \ref{res1} the FG response at $|{\bf q}| = 0.3 \ {\rm fm}^{-1}$, obtained 
from an analytical calculation using a continuum set of one particle-one hole
states, is compared to that resulting from the above procedure
with $(n_q^x, \ n_q^y, \ n_q^z ) \equiv (1,2,3)$. This choice corresponds to 
a normalization box of linear dimension $L= 78 \ {\rm fm}$, containing 
$\sim$ 78400 nucleons. The corresponding number of one particle-one hole
states in the basis is $\sim$ 3000. It clearly appears that the basis is large enough 
to reproduce the results obtained in the continuum limit.

\section{Correlated Hatree-Fock approximation}

\begin{figure}[ht]
\centerline{\includegraphics[scale=0.45]{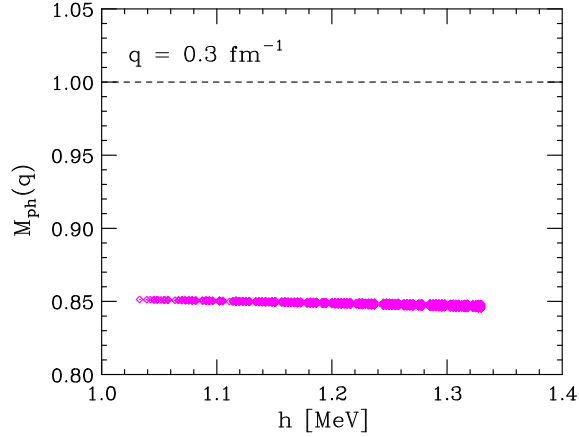}}
\caption{\small Fermi transition matrix element at $|{\bf q}|=0.3 \ {\rm fm}^{-1}$ 
as a function of the magnitude of hole momentum $|{\bf h}|$. The dashed horizontal 
line corresponds to the result of the FG model.  \label{quenching}}
\end{figure}

\begin{figure}[ht]
\centerline{ \includegraphics[scale=0.45]{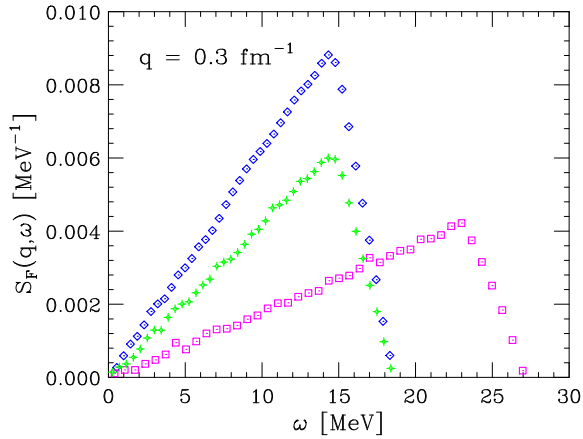} }
\caption{\small Nuclear matter response at $|{\bf q}|=0.3 \ {\rm fm}^{-1}$.
Diamonds: FG model. Crosses: results obtained using the effective operator
in the calculation of $M_{ph}$ and the FG spectrum of Eq.(\ref{FGenergy}).
Squares: correlated HF approximation described in the text. \label{res2}}
\end{figure}

The inclusion of interaction leads to sizable modifications of the FG response.
Correlation effects in the transition matrix elements, taken into account through
the use of the effective operator, produce a quenching of the Fermi transition 
matrix elements of $\sim$ 15 \%, largely independent of the hole momentum, as 
illustrated in Fig. \ref{quenching}. Note that the effect on the response is
larger, as its definition involves the transition probabilities, i.e. the 
squared matrix elements. This feature is apparent in Fig. \ref{res2},  
where the FG response, represented by diamonds, is compared to that obtained 
using the effective operator
in the calculation of $M_{ph}$ and the FG spectrum of Eq.(\ref{FGenergy}), 
represented by crosses.

An even larger modification is produced by interaction effects on the single particle
energies. Corrections to the kinetic energy spectrum (\ref{FGenergy}) have been 
calculated within the Hartree-Fock (HF) approximation, using the CBF effective 
interaction discussed in Chapter \ref{CBF} (see Eq.(\ref{HF})). Replacing the 
single particle energies of Eq.(\ref{FGenergy}) with the HF energies, shown by the 
solid line in Fig. \ref{sp:spectrum}, leads to a sizable broadening of $\omega$ region 
corresponding to non-vanishing response. 

The squares of Fig. \ref{res2} show the response evaluated within
the correlated HF approximation , i.e. using the CBF effective operator
and the HF spectrum obtained from the CBF effective interaction, at 
$\mq=0.3 \ {\rm fm}^{-1}$. Comparison with the diamonds and the crosses shows that 
using the HF spectrum leads to an increase of the upper limit of the energy transfer $\omega$ 
from $\sim$ 18 MeV to $\sim$ 27 MeV.

\begin{figure}[ht]
\centerline{ \includegraphics[scale=0.45]{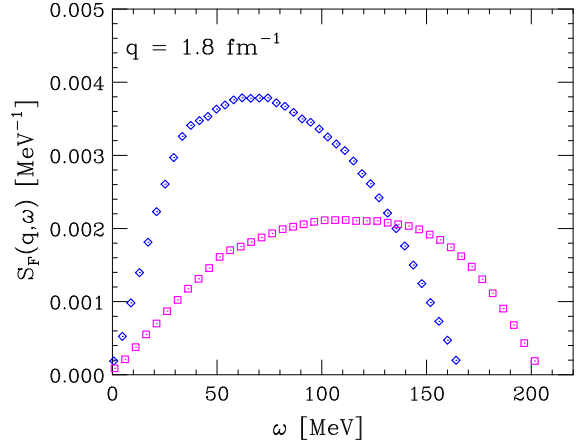}}
\caption{\small  Nuclear matter response at $|{\bf q}|=1.8 \ {\rm fm}^{-1}$. 
Diamonds and squares correspond to the FG model and the correlated HF approximation 
described in the text. \label{res3}}
\end{figure}

\begin{figure}[ht]
\centerline{ \includegraphics[scale=0.45]{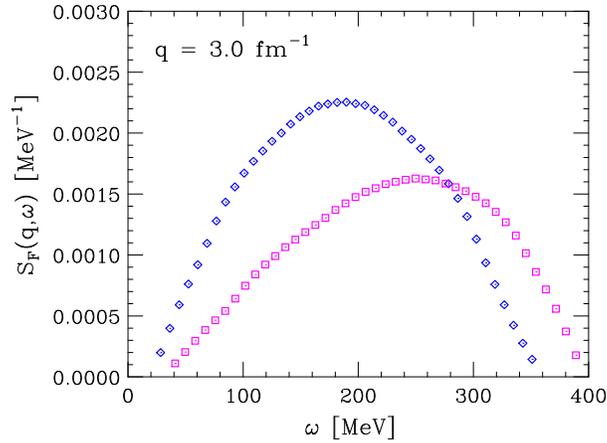}}
\caption{\small  Same as in Fig. \ref{res3}, for momentum transfer 
$|{\bf q}|=3.0 \ {\rm fm}^{-1}$. \label{res4}}
\end{figure}

In Figs. \ref{res3}-\ref{res4} the comparison between FG model and correlated HF
approximation is extended to larger values of the momentum 
 transfer $\mq=1.8, \ {\rm and}  \ 3.0 \ {\rm fm}^{-1}$, corresponding to the regions
$\mq \ge k_F$ and $\mq \ge 2 k_F$, respectively.
In the case $\mq=1.8, \ {\rm fm}^{-1}$ we have used a box of side $L = \sim 47 \ {\rm fm}$, 
corresponding to a basis of $\sim 3900$ states, while the calculation at 
$\mq=3.0, \ {\rm fm}^{-1}$ has been carried out with $L = \sim 44 \ {\rm fm}$ and 
$\sim 3300$ basis states.
From Figs. \ref{res2}-\ref{res4} it clearly appears that 
the quenching of the peak and the shift of the strength towards larger values of
$\omega$, resulting from the inclusion of interaction effects, are sizable
in all instances. 


\section{Tamm-Dancoff Approximation}
In the previous section we have discussed the nuclear response in the correlated 
HF approximation, in which the bare Fermi transition operator is replaced 
by the effective operator of Eq.(\ref{oeff:def}) and the final state is assumed to
be a one particle-one hole state. 

It is important to realize that the FG one particle-one hole states, while being 
eigenstates of the HF hamiltonian, defined as 
\beq
H_{HF} = \sum_i e_i \ ,
\label{def:HHF}
\eeq
with $e_i$ given by Eq.(\ref{HF}), are not eigenstates of the full nuclear 
hamiltonian. As a consequence, there is a residual interaction $V_{res}$ that can 
induce transitions between different one particle-one hole states, as long as 
their total momentum {\bf q}, spin and isospin are conserved. 

In order to include the effects of these transitions, we use the TD
approximation, which amounts to expanding the final state in the basis of
one particle-one hole states according to \cite{Boffi}
\beq
| f \rangle = | {\bf q}, \ T S M \rangle = \sum_{i} c^{TSM}_{i}
| {\bf h}_i ,\ {\bf p}_i = {\bf h}_i + {\bf q}, \ T S M \rangle \ ,
\label{tda:phexp}
\eeq
where $S$ and $T$ denote the total spin and isospin of the particle hole pair 
and $M$ is the spin projection. 

At fixed ${\bf q}$, the excitation energy of the state (\ref{tda:phexp}), 
$\omega^{f}$, is obtained solving the eigenvalue equation 
\beq
H | f \rangle = (E_0 + \omega^{f}) | f \rangle \ ,
\label{eigenprob}
\eeq
where $E_0$ is the ground state energy. Substituting Eq.(\ref{tda:phexp}) into 
Eq.(\ref{eigenprob}) and multiplying by $\langle {\bf h}_j, \ {\bf p}_j, \ T S M |$ 
from the left leads to
\bea
\nonumber
\sum_{i}  
\langle {\bf h}_j, \ {\bf p}_j, \ T S M | H | {\bf h}_i, \ {\bf p}_i, 
\ T S M \rangle \ c^{TSM}_i & = & \sum_i H^{TSM}_{ji} \ c^{TSM}_i \\ 
& = & (E_0 + \omega^{TSM}) c^{TSM}_j \ ,
\eea
with 
\beq
H^{TSM}_{ji} = (E_0 + e_{p_i} - e_{h_i}) \delta_{ji} 
 + \langle {\bf h}_j, \ {\bf p}_j, \ T S M | V_{res} 
| {\bf h}_i, \ {\bf p}_i, \ T S M_S \rangle \ ,
\label{Hmat1}
\eeq
and \cite{FetterWalecka}
\beq
\langle {\bf h}_j, \ {\bf p}_j, \ T S M | V_{res}
| {\bf h}_i, \ {\bf p}_i, \ T S M \rangle = 
\langle v_{eff} \rangle_D 
 - \langle v_{eff} \rangle_E \ .
\label{Hmat2}
\eeq
In the above equation, $v_{eff}$ is the CFB effective interaction discussed
in Chapter \ref{CBF}, and the direct and exchange contributions to the matrix
elements are given by
\beq
\langle v_{eff} \rangle_D = \frac{1}{V^2}  
\int d^3 r_1 d^3 r_2 \ {\rm e}^{-i({\bf p}_j {\bf r}_1 + {\bf h}_i {\bf r}_2)}
 \ \langle \alpha_{p_j} \alpha_{h_i} | v_{eff} | \alpha_{h_j} \alpha_{p_i} \rangle
 \ {\rm e}^{i({\bf h}_j {\bf r}_1 + {\bf p}_i {\bf r}_2)} \ ,
\label{vresd}
\eeq
\beq
\langle v_{eff} \rangle_E = \frac{1}{V^2}  
\int d^3 r_1 d^3 r_2 \ {\rm e}^{-i({\bf h}_i {\bf r}_1 + {\bf p}_j {\bf r}_2)}
 \ \langle \alpha_{h_i} \alpha_{p_j} | v_{eff} | \alpha_{h_j} \alpha_{p_i} \rangle
 \ {\rm e}^{i({\bf h}_j {\bf r}_1 + {\bf p}_i {\bf r}_2)} \ ,
\label{vrese}
\eeq
where the two-nucleon state $| \alpha_{h} \alpha_{p} \rangle$ describes a hole in the 
spin-isospin state $\alpha_{h}$ and a particle in the spin-isospin state $\alpha_p$, 
coupled in such a way as to obtain the assigned values of $T$ and $S$ and $M$.

We can now exploit the fact that, in the case of Fermi transitions, the requirement that
the matrix elements of $O^F_{eff}$ be non-vanishing implies that the particle hole pair
be in a $S=1$ state with $M= \pm 1$. As a consequence, $H^{TSM}_{ij}$,
 $c^{TSM}_i$ and $\omega^{TSM}_i$ become independent of $M$.

For any fixed $T$ and $S$, the diagonalization of the hamiltonian matrix defined in 
Eqs.(\ref{Hmat1})-(\ref{vrese})
determines the eigenvalues $\omega^{TS}_n$ and the the corresponding eigenvectors $c^{TS}_n$
with $n = 1, \ 2, \ \ldots, \ N_B$, $N_B$ being the number of states in the 
one particle-one hole basis.

Collecting all the above results, we can finally write the response in the TD
approximation as
\beq
S(\qm,\omega)= \frac{1}{2} \sum_{T=0,1} \sum_{M= \pm 1} \ \sum_{n} 
\left| \sum_{i}  (c^{T1}_n)_i 
\langle {\bf h}_i, \ {\bf p}_i, \ T 1 M | O_{eff}(\qm)|0\rangle \right|^2  
\delta(\omega-\omega^{T1}_n) \ ,
\label{tda:resp}
\eeq
where $(c^{T1}_n)_i$ denotes the $i$-th component of the eigenvector belonging to the 
eigenvalue $\omega^{T1}_n$.

\begin{figure}[ht]
\centerline{ \includegraphics[scale=0.60]{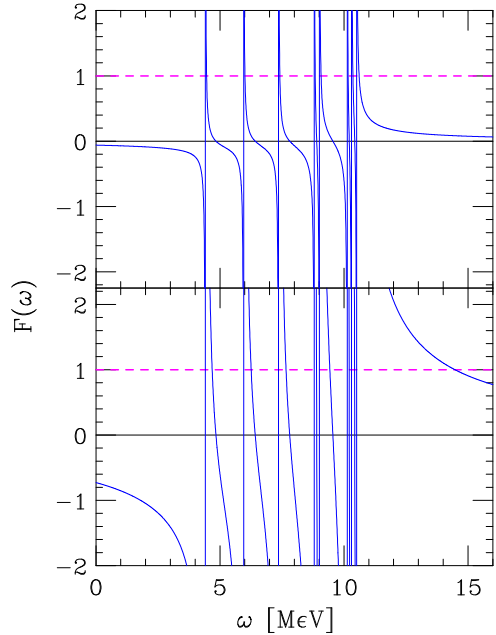}}
\caption{\small Graphical solution of Eq.(\ref{tda:omsp}) for $|{\bf q}|=0.3 \ {\rm fm}^{-1}$ 
and a basis consisting of 8 states. The upper and lower panel correspond to 
$\hat{v}_{eff}(q) = -0.06$ and $-0.7$ MeV, respectively.
 \label{plasmon}}
\end{figure}

The main features of the response in the TD approximation can be understood considering a 
simple model in which $v_{eff}$ is assumed to be central and spin-isospin independent,
and the exchange contribution to Eq.(\ref{Hmat2}) is neglected. As a result we can make the
replacement
\beq
\langle {\bf h}_j, \ {\bf p}_j, \ T S M | V_{res} | {\bf h}_i, \ {\bf p}_i, \ T S M \rangle 
 \rightarrow \frac{2}{L^3} \int d^3r \ v_{eff}(r) \ {\rm e}^{i {\bf q}{\bf r}}\ = \hat{v}_{eff}(q) \ ,
\eeq
leading to the eigenvalue equation
\beq
1= \hat{v}_{eff}(q) \sum_{i} \frac{1}{ \omega - e_{p_i}+e_{h_i} }  = \mathcal{F}(\omega)\ .
\label{tda:omsp}
\eeq
The above equation can be solved graphically plotting the right hand side as a 
function of $\omega$ and finding the intersections with the line $\mathcal{F}(\omega)=1$, as 
shown in Fig. \ref{plasmon} for $|{\bf q}|=0.3 \ {\rm fm}^{-1}$ and a basis consisting of 8 
states. The upper and lower panel correspond to $\hat{v}_{eff}(q) = -.06$ and 0.7 MeV, 
respectively. It appears that in the latter case the spectrum exhibits an eigenvalue 
lying well outside the particle hole continuum, corresponding to a collective excitation, 
reminiscent to the plasmon mode of the electron gas.

\begin{figure}[ht]
\includegraphics[scale=0.62]{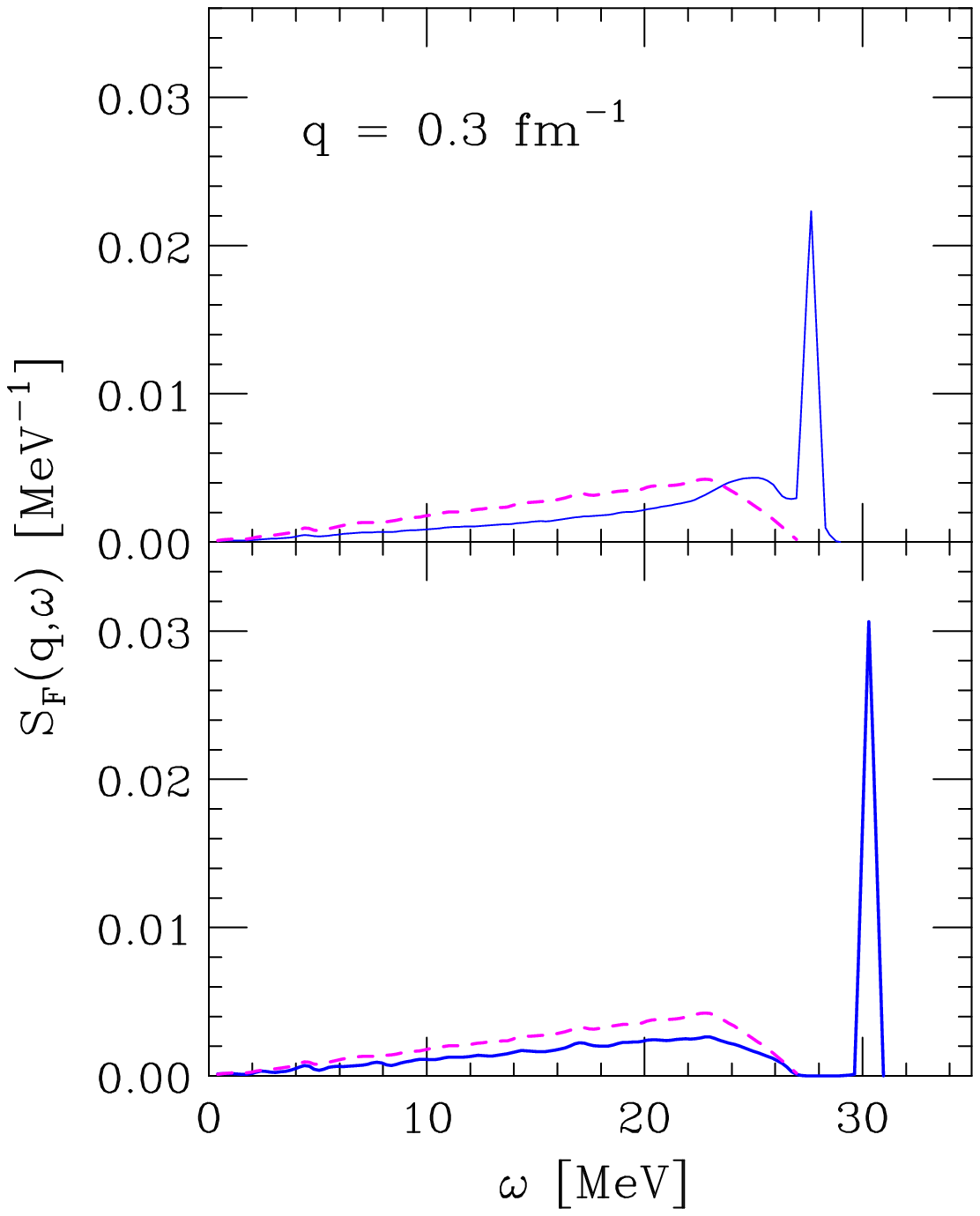}
 \includegraphics[scale=0.62]{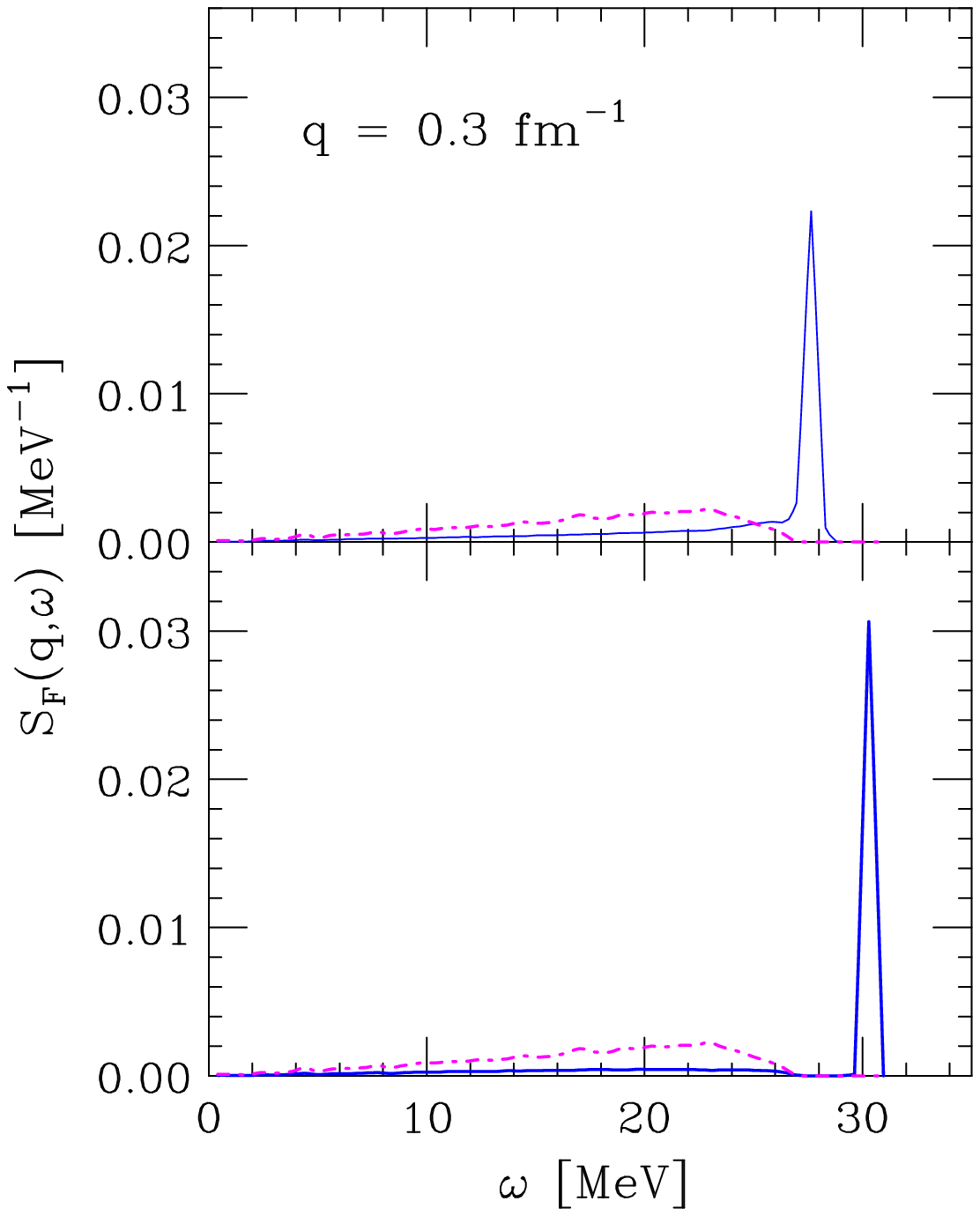}
\caption{\small Nuclear matter response in the TD approximation at 
$|{\bf q}|=0.3 \ {\rm fm}^{-1}$. The solid line in the upper left panel 
corresponds to the full calculation, while that in the lower left panel
has been obtained neglecting the exchange term in Eq.(\ref{Hmat2}). The dashed
lines show the results of the correlated HF approximation.
The right panels illustrate the contributions of the $T=0$ (solid lines) and 
$T=1$ (dash-dot lines) to the TD response, evaluated including direct and exchange
terms (lower right) or direct term only (upper right).  \label{res5}}
\end{figure}

The nuclear matter response at $|{\bf q}|=0.3 \ {\rm fm}^{-1}$ obtained using the 
TD approximation is shown in Fig. \ref{res5}.
The solid line in the lower left panel corresponds to the full calculation, including both 
the direct and exchange contributions to the matrix elements of the effective interactions
(see Eq.(\ref{Hmat2})), while the solid line in the upper left panel has been obtained 
neglecting the exchange part. In both left panels, the dashed line refers to the correlated
HF approximation. All calculation have been carried out using $\sim$ 3000 basis states.

The full TD response exhibits a sharp isolated peak corresponding to the 
collective mode, lying $\sim$ 4 MeV above the upper limit of the particle hole continuum.   
If only direct contributions are included, the peak is still clearly visible, but 
located at lower $\omega$, and not well separated from the continuum. 

The right panels of Fig. \ref{res5} illustrate the contributions to the TD response 
associated with different final states, corresponding to total isospin of the particle hole 
pair $T=0$ (solid lines) and $T=1$ (dashed lines). In the case of Fermi 
transitions, the total spin is $S=1$ and the two allowed spin projections, $M=\pm1$ give
equal contributions. The right lower and upper panels show the results of the full calculations
and those obtained neglecting the exchange terms.

It clearly appears that the excitation of the collective mode is due to the antisymmetric 
$T=0$ state, while the strength arising from the symmetric $T=1$ state lies within the 
particle hole continuum. 
                                                                                                    

Our correlated Hartree Fock results are in good agreement with those obtained by Cowell 
and Pandharipande using correlated states and a cluster expansion truncated at the 
two body level \cite{shannon2}. On the other hand, the results of Ref. \cite{shannon2}
suggest that the inclusion of the exchange term of Eq.(\ref{Hmat2}) is essential for
the excitation of the coherent state, while our results show that the corresponding peak 
survives if exchange contributions are neglected. This discrepancy is likely to be 
ascribed to differences in the effective interactions.
\chapter*{Summary \& Outlook}
\addcontentsline{toc}{chapter}{Summary \& Outlook}
\markboth{Summary \& Outlook}{}
\markright{Summary \& Outlook}{}
\label{concl}
We have carried out calculations of the charged current weak response of nuclei and 
nuclear matter in both the low momentum transfer and the impulse approximation regimes.
The quantitative understanding of this quantity is required in many areas of physics,
 ranging from simulations of supernov$\ae$ explosions and neutron star
cooling to the analysis of neutrino oscillation experiments.

The calculation has been performed using a many-body approach based on 
a realistic nuclear hamiltonian, including two- and three-nucleon interactions, 
yielding a good description of the properties of both the two-nucleon systems and
uniform nuclear matter.

At beam energies around 1 GeV, the formalism based on the target spectral 
function has been tested comparing the calculated cross 
sections to electron scattering data in the kinematical region corresponding to 
quasielastic scattering and $\Delta$-production, relevant to many long baseline
oscillation experiments. The results suggest that, while the quasi elastic cross
section is understood at the level of $\sim$ 10 \%, the accurate description of
the resonance region will require the extension of the available fits of the nucleon 
structure functions down to $Q^2 \lsim$ 0.4 GeV$^2$, as proposed in Ref. \cite{BenMel1}. 

In the high energy domain the differences between our results and the prediction
of the FG model, widely used in the analysis of experimental data, mostly arise
from the effects of short range nucleon-nucleon correlation. 
On the other hand, the response at low $\magq$ (of the order of tens of MeV) is 
known to be sizably affected by long range correlations, giving rise to the excitation 
of collective modes.

The response associated with Fermi transitions at low momentum transfer has been
calculated from an effective interaction, derived using the formalism of 
correlated basis functions and the cluster expansion technique. Our work improves 
upon existing effective interaction models 
in that it includes the effects of many-nucleon forces, which become sizable, 
at high density. 

The energy per nucleon of both symmetric nuclear matter and pure neutron matter, obtained
from our effective interaction model, turns out to be in fairly good agreement with 
the results of highly refined many body calculations, based on similar dynamical 
models. A comparable agreement with the results available in the literature has also 
been found for single particle properties, e.g. the effective mass, and the spin 
susceptibility.
The emerging picture suggests that our approach can be regarded as an {\em effective 
theory} that captures the relevant physics, 
allowing one to obtain reasonable estimates of a number of different quantities 
using low order perturbation theory in the Fermi gas basis.   

The responses calculated within the correlated HF approximation show that the 
inclusion of short range correlations leads 
to a significant quenching of the transition matrix elements and shifts the strength
towards larger values of the energy transfer, $\omega$, for all values of $\magq$.
At low momentum transfer, long range correlations have been taken into account 
within the TD approximation. The excitation of the coherent state can be clearly 
seen in our results at $\magq = 0.3 \ {\rm fm}^{-1}$.

The most straightforward extension of our approach is the calculation of the 
Gamow Teller response. The inclusion of finite temperature effects, needed
to extract the observables relevant to supernov$\ae$ and proto-neutron star physics, 
also appears to be doable, in the low temperature region $T \ll m_\pi$. 

As a final remark, it has to be pointed out that, while our approach can and should be 
further developed, the possible improvements only pertain the structure of the effective 
interaction and the inclusion of perturbative corrections, and {\em do not} involve going 
to higher order in the cluster expansion. 

Although the contribution of clusters involving more than two nucleons is known to be, 
in general, non negligible, effective theories are in fact {\em designed} 
to provide {\em lowest order} results reasonably accounting for the available data.

In this context, however, the rationale underlying the choice of the two-nucleon cluster approximation, employed in our work, deserves a comment.

While being likely to be reasonable for use in nuclear matter limit, our procedure does not preserve the normalization of the many-body wave function. We plan to investigate quantitavely this issue using alternative truncation schemes, designed to guarantee the correct normalization at finite order in the cluster expansion \cite{obcluster,cocluster}.

The most obvious improvement is the inclusion in $v_{\rm eff}$ of the 
non static components of NN potential, which are known to be needed to reproduce 
scattering data. 
 
On the other hand, inclusion of higher order terms in the perturbative expansion,
which is expected to be rapidly convergent, is necessary to take into account 
more complex mechanisms, that play a role in determining 
several properties of many-body systems as, for example, the effective mass at the 
Fermi surface \cite{FFP}.
\newpage
\newpage
\begin{appendix}
\addcontentsline{toc}{chapter}{Appendices}
\chapter{Properties of the operators $O^n_{ij}$}
\label{pauli}
In this Appendix, we discuss the properties of the six operators defined in
Eq.(\ref{pot2}), as well as some useful properties of the Pauli matrices.

\section{Pauli matrices}
In the standard representation, in which $\sigma^3$ is chosen to be diagonal,
the threee Pauli matrices are given by (we specialize here to the spin
matrices $\sigma^i$: analog properties obviously hold for the isospin matrices
$\tau^i$)
\beq
\sigma^1 = \left( \begin{array}{rr} 0 & \phantom{-}1 \\ 1 & 0 \end{array}
\right) \ \ , \ \
 \sigma^2 = \left( \begin{array}{rr} 0 & -i \\ i & 0 \end{array} \right) \ \ , \
 \ 
\sigma^3 = \left( \begin{array}{rr} 1 & 0 \\ 0 & -1 \end{array} \right) \ .
\label{Pauli:mat}
\eeq

The Pauli matrices satisfy
\bea
\sigma^i\sigma^j               & = & \delta_{ij}+i\epsilon_{ijk}\sigma^k \ ,
\label{Pauli:prod1} \\
\epsilon_{ijk}\sigma^j\sigma^k & = & 2i\sigma^i \ , \label{Pauli:prod2}
\eea
that can be put in the form
\bea
[\sigma^i,\,\sigma^j]    & = & 2i\epsilon_{ijk}\sigma^k \ , \label{Pauli:comm}  \\
\{\sigma^i,\,\sigma^j\}  & = & 2\delta_{ij}             \ , \label{Pauli:acomm} 
\eea
where $\epsilon_{ijk}$ is the totally antisymmetric tensor and
$i,j,k=1,2,3$. The first properties shows that the Pauli matrices are the
generators of an $SU(2)$ algebra.

\section{Projection operators}
Let now ${\bm \sigma}_1$ and ${\bm \sigma}_2$ be the vectors of Pauli matrices
for particle 1 and 2, respectively 
(i.e. ${\bm \sigma}_1\equiv \left\{ \sigma_1^1,\,\sigma_1^2,\,\sigma_1^3\right\}$). From the properties
(\ref{Pauli:prod1})-(\ref{Pauli:prod2}), it follows that
\beq
\sigsig^2 = 3 - 2\sigsig \ .
\label{sigsigsq}  
\eeq
As $\sigsig$ is a scalar quantity, we can interpret the above equation as an
algebraic one, with solutions $\sigsig = -3$ and $\sigsig = 1$. They correspond to the states
of total spin $S=0$ (spin singlet channel) and $S=1$ (spin triplet channel),
respectively. It is thus useful introducing the operators $P_{2S+1}$ (and the
analog $\Pi_{2T+1}$ for the isospin states), defined as
\bea
P_{(S=0)} \equiv P_1 & = & \frac{1-\sigsig}{4} \ , \\
P_{(S=1)} \equiv P_3 & = & \frac{3+\sigsig}{4} \ ,
\eea
which project onto states of definte total spin 0 or 1, respectively:
\beq
P_{2S+1} \vert S^\prime \rangle = \delta_{SS^\prime} \vert S^\prime \rangle \ ,
\label{proj:prop}
\eeq
The projection operators satisfy to
\bea
P^2_{2S+1}   & = & P^{\phantom{2}}_{2S+1} \ , \\
P_1 + P_3    & = & \openone \ , \\
P_1 P_3 = P_3 P_1   & = & 0          \ ,
\eea
where $\openone$ is the two-dimensional identity matrix.

\section{Spin and isospin exchange operators}

Consider the two-nucleon spin states (or the analog isospin states)
\bea
\vert 0\, 0 \rangle & = &\frac{1}{\sqrt{2}} \left( \vert \uparrow \downarrow
  \rangle - \vert  \downarrow \uparrow \right) \ , \nonumber \\
\vert 1 -1 \rangle & = & \vert \downarrow \downarrow \rangle \ , \nonumber \\
\vert 1\, 0 \rangle & = & \frac{1}{\sqrt{2}} \left( \vert \uparrow \downarrow
  \rangle + \vert  \downarrow \uparrow \right) \ , \nonumber \\
\vert 1\, 1 \rangle & = & \vert \uparrow \uparrow \rangle \ , \nonumber
\eea
where $\vert 0\, 0 \rangle \equiv \vert S=0 \, M_S=0 \rangle$ etc., and the inverse
relations
\bea
\vert \uparrow \uparrow \rangle   & = & \vert 1\, 1 \rangle \ , \nonumber \\
\vert \uparrow \downarrow \rangle & = & \frac{1}{\sqrt{2}} \left( \vert 1\, 0
  \rangle + \vert 0\, 0 \rangle \right) \ , \nonumber \\
\vert \downarrow \uparrow \rangle & = & \frac{1}{\sqrt{2}} \left( \vert 1\, 0
  \rangle - \vert 0\, 0 \rangle \right) \ , \nonumber \\
\vert \downarrow \downarrow \rangle & = & \vert 1\, -1 \rangle \ . \nonumber
\eea

From properties (\ref{proj:prop}), and from
\bea
\left( P_3 - P_1 \right) \vert \uparrow \uparrow \rangle =  \vert \uparrow
\uparrow \rangle
& , & \left( P_3 - P_1 \right) \vert \downarrow \downarrow \rangle = \vert
\downarrow \downarrow \rangle \ , \nonumber \\
\left( P_3 - P_1 \right) \vert \uparrow \downarrow \rangle = \vert \downarrow
\uparrow \rangle & , & \left( P_3 - P_1 \right) \vert \downarrow
\uparrow \rangle = \vert \uparrow \downarrow \rangle \ , \nonumber 
\eea
it follows that $P_{\sigma}\equiv P_3 - P_1$ is the spin-exchange operator,
satisfying
\beq
P_{\sigma} \vert S\,M_S \rangle = (-)^{S+1} \vert S\,M_S \rangle \ .
\eeq
A similar exchange operator can be defined for isospin, $P_{\tau}\equiv
\Pi_3-\Pi_1$, with
\beq
P_{\tau} \vert T\,M_T \rangle = (-)^{T+1} \vert T\,M_T \rangle \ .
\eeq
Combining the above results we find
\beq
P_{\sigma\tau} \equiv P_{\sigma}P_{\tau} = \frac{1}{4}\,\Big( 1 + \sigsig
\Big)\Big( 1 + \tautau \Big) \ ,
\eeq
with
\beq
P_{\sigma\tau} \vert S\,M_S,\,T\, M_T \rangle = (-)^{S+T}\vert S\,M_S,\,T\,
M_T \rangle \ .
\eeq

\section{The tensor operator $S_{12}$}

The tensor operator $S_{12}$ is defined as
\beq
S_{12} \equiv \frac{3}{r^2}\; \left( {\bm \sigma}_1 \cdot {\bf r} \right)
\,\left( {\bm \sigma}_2 \cdot {\bf r} \right) - \sigsig \ ,
\label{tens:op}
\eeq
where ${\bf r}$ is the relative coordinate of particels 1 and 2 while $r=\vert
{\bf r}\vert$.

Making use of Eq.(\ref{Pauli:prod1}), it can be shown that
\beq
S_{12}\sigsig = \sigsig S_{12} = S_{12} \ .
\label{S12sigsig}
\eeq

As we saw, $\sigsig = $ 1 on triplet states, while $\sigsig =$ $-3$ on singlet
states. The above equation thus implies that the tensor operator only acts on
triplet states and
\beq
\left[ S_{12},\;P_3 \right] = 0 \ .
\eeq

Moreover,
\beq
S^2_{12} = 6 - 2S^{\phantom{2}}_{12} + 2 \sigsig \ .
\label{S12sq}
\eeq 

The tensor operator is a function of ${\bf r}$ satisfying 
\bea
{\bm \nabla}S_{12} & = &  \frac{3}{r^2} \Bigg[ {\bm \sigma}_1 \left( {\bm \sigma}_2 \cdot {\bf r}
 \right) + {\bm \sigma}_2 \left( {\bm \sigma}_1 \cdot {\bf r}
 \right) - 2 \frac{{\bf r}}{r^2}\left( {\bm \sigma}_1 \cdot {\bf r} \right)
\,\left( {\bm \sigma}_2 \cdot {\bf r} \right)  \Bigg] \ , \label{nabla:S12} \\
\nabla^2 S_{12} & = & -\frac{6}{r^2} S_{12} \ . \label{nablasq:S12}
\eea

For any function $u(r)$, Eq.(\ref{nabla:S12}) implies
\beq
\left( {\bm \nabla}u \right)\cdot\left( {\bm \nabla}S_{12} \right) =
\frac{du}{dr}\;\frac{{\bf r}}{r}\cdot\left( {\bm \nabla}S_{12} \right) = 0 \ .
\eeq
Moreover
\bea
\left( {\bm \nabla} S_{12}  \right)^2 & = & \frac{6}{r^2} \left( 8 - S_{12}
\right) \ , \\
\left[ S_{12},\; \left( {\bm \nabla} S_{12}  \right) \right] & = & 
\frac{36}{r^2}\;i\,\left( {\bf S}\times{\bf r} \right) \ , \\
\left[ S_{12},\; \left( {\bm \nabla} S_{12}  \right) \right]{\bm \nabla} & = &
\frac{36}{r^2}\;\left( {\bf L}\cdot{\bf S}\right) \ ,
\eea
where ${\bf S}=\left({\bm \sigma}_1+{\bm \sigma}_2\right)/2$ and ${\bf L} =
{\bf r}\times {\bf p}= -i \left({\bf r}\times {\bm \nabla}\right)$ is the
orbital angular momentum operator of the relative motion.

From Equation (\ref{nablasq:S12}),  we can calculate 
\beq
\left[ S_{12},\;\nabla^2 S_{12} \right] = 0 \ , 
\eeq
and
\beq
\left( {\bm \nabla} S_{12} \right)\;\left[ S_{12},\;{\bm \nabla} \right] =
 - \left( {\bm \nabla} S_{12} \right)^2 \ . 
\eeq

\section{Operator algebra}
Equations (\ref{sigsigsq}), (\ref{S12sigsig}) and (\ref{S12sq}) show that the six operators
\beq
O^{1,\ldots,6} =  1,\;\tautau,\;\sigsig,\;\sigsig\tautau,\;S_{12},\;S_{12}\tautau \  \ ,
\label{six:op}
\eeq
close an algebra, i.e. they satisfy
\beq
O^{i}O^{j} = \sum_k K^{k}_{ij} O^{k} \ .
\label{six:alg}
\eeq

The coefficients $ K^{k}_{ij}$ are easily obtained by calculating
\[
O^{1}O^{i} = O^{i}O^{1} = O^{i} \Longrightarrow K^{k}_{1i} = K^{k}_{i1} =  
\delta^{k}_i \ 
\]
\[
O^{2}O^{2} = 3O^{2}-2O^{2} \Longrightarrow K^{k}_{22} =
                          3\delta^{k}_1 - 2\delta^{k}_2\ ,    
\]
\[
O^{2}O^{3} = O^{3} O^{2} = O^{4} \Longrightarrow K^{k}_{23} = 
          K^{k}_{32} = \delta^{k}_4\ ,  
\]
\[
O^{2}O^{4} = 3O^{3}-2O^{4}  \Longrightarrow  K^{k}_{24} = 
          K^{k}_{42} = \delta^{k}_3 - 1 \delta^{k}_4   \ , 
\]
\[
O^{2}O^{5} = O^{5}O^{2} = O^{6} \Longrightarrow K^{k}_{25} = 
          K^{k}_{52} = \delta^{k}_6\ , 
\]
\[
O^{2}O^{6} = O^{6}O^{2} = 3O^{5}-2O^{6}  \Longrightarrow 
          K^{k}_{26} = K^{k}_{62} = 3\delta^{k}_5 -2\delta^{k}_6 \ , 
\]
\[
O^{3}O^{3} = 3O^{1}- 2O^{3} \Longrightarrow 
          K^{k}_{33} = 3\delta^{k}_1 - 2\delta^{k}_3\ , 
\]
\[
O^{3}O^{4} = O^{4}O^{3} = 3O^{2}- 2O^{4}  \Longrightarrow 
          K^{k}_{34} =  K^{k}_{43} = 3\delta^{k}_2 -  2\delta^{k}_4 \ ,
\]
\[ 
O^{3}O^{5} = O^{5}O^{3} = O^{5} \Longrightarrow 
          K^{k}_{35} = K^{k}_{53} = \delta^{k}_5 \ , 
\]
\[
O^{3}O^{6} = O^{6}O^{3} = O^{6} \Longrightarrow 
          K^{k}_{36} = K^{k}_{63} = \delta^{k}_6 \ , 
\]
\[
O^{4}O^{4} = 9O^{1} -6O^{2}-6O^{3} +4O^{4} \Longrightarrow 
          K^{k}_{44} = 9\delta^{k}_1 -6\delta^{k}_2 -6\delta^{k}_3
          +4\delta^{k}_4 \ , 
\]
\[
O^{4}O^{5} = O^{5}O^{4} = O^{6}     \Longrightarrow 
          K^{k}_{45} = K^{k}_{54} = \delta^{k}_6 \ , 
\]
\[
O^{4}O^{6} = O^{6}O^{4} = 3O^{5}- 2O^{6} \Longrightarrow 
          K^{k}_{46} = K^{k}_{64} = 3\delta^{k}_5 - 2\delta^{k}_6\ , 
\]
\[
O^{5}O^{5} = 6O^{1} +2O^{3} -2O^{5}  \Longrightarrow 
          K^{k}_{55} = 6\delta^{k}_1 +2\delta^{k}_3  -2\delta^{k}_5 \ , 
\]
\[
O^{5}O^{6} = O^{6}O^{5} = 6O^{2} +2O^{4} -2O^{6} 
 \Longrightarrow K^{k}_{56} = K^{k}_{65} = 6\delta^{k}_2 +2\delta^{k}_4
-2\delta^{k}_6 \ , 
\]
\bea
\nonumber
O^{6}O^{6}  & = & 18O^{1} -12O^{2} + 6O^{3} -4O^{4} -6O^{6} +4O^{6} \\
\nonumber
& & \ \ \ \ \ \ \ \ \ \ \ \ \ \ \ \ 
 \Longrightarrow K^{k}_{66} = 18\delta^{k}_1 -12\delta^{k}_2
 +6\delta^{k}_3 -4\delta^{k}_4 -6\delta^{k}_4 +4\delta^{k}_6 \ . 
\eea

\section{Matrix elements}

Finally, we report a number of expectation values of operators
involving Pauli matrices, in two-nucleon states of definite total spin
and isospion, $\vert S \; M_S ,\;T \; M_T \rangle$.
\bea
\langle P_{2S^\prime+1}\Pi_{2T^\prime+1} \rangle & = &
\delta_{SS^\prime}\delta_{TT^\prime} \ , \\
\langle P_{2S^\prime+1}\Pi_{2T^\prime+1} P_{\sigma\tau} \rangle & = &
(-)^{S+T}\delta_{SS^\prime}\delta_{TT^\prime} \ , \\
\sum_{S M_S} \delta_{S^\prime 1} \langle S_{12}
P_{2S^\prime+1}\Pi_{2T^\prime+1} \rangle & = & \delta_{S^\prime 1} \delta_{TT^\prime} \sum_{M_S} \langle 1\;M_S \vert S_{12} \vert
1\;M_S \rangle = 0 \ , \\
\sum_{S M_S} \delta_{S^\prime 1} \langle S_{12}
P_{2S^\prime+1}\Pi_{2T^\prime+1} P_{\sigma\tau}\rangle & = & 0 \ . 
\eea

\section{More matrix elements}
The explicit expressions for the matrices entering Eq.(\ref{energy:nm}), defined by
\beq
A^{i}_{\lambda\mu} = \langle \lambda\mu \vert O^{i}_{12} \vert \lambda\mu
\rangle \ \ , \ \
B^{i}_{\lambda\mu} = \langle \lambda\mu \vert O^{i}_{12} \vert \mu\lambda
\rangle \ ,
\eeq
where $\vert \lambda \mu \rangle$ denotes the two-nucleon spin-isospin state,
can be easily obtained from the above properties of the six operators $O^{n\leq6}$.

We find
\bea
A^{1} & = & \left(
\begin{array}{llll}
\phantom{-}1 & \phantom{-}1 & \phantom{-}1 & \phantom{-}1 \\
\phantom{-}1 & \phantom{-}1 & \phantom{-}1 & \phantom{-}1 \\
\phantom{-}1 & \phantom{-}1 & \phantom{-}1 & \phantom{-}1 \\
\phantom{-}1 & \phantom{-}1 & \phantom{-}1 & \phantom{-}1
\end{array}
\right) \ , \\
A^{2} & = & \left(
\begin{array}{rrrr}
1 & 1 & -1 & -1 \\
1 & 1 & -1 & -1 \\
-1 & -1 & 1 & 1 \\
-1 & -1 & 1 & 1
\end{array}
\right) \ , \\
A^{3} & = & \left(
\begin{array}{rrrr}
1 & -1 & 1 & -1 \\
-1 & 1 & -1 & 1 \\
1 & -1 & 1 & -1 \\
-1 & 1 & -1 & 1
\end{array}
\right) \ , \\
A^{4} & = & \left(
\begin{array}{rrrr}
1 & -1 & -1 & 1 \\
-1 & 1 &  1 & -1\\
-1 & 1 &  1 & -1\\
1 & -1 & -1 & 1
\end{array}
\right) \ , \\
A^{5} & = & \left(
\begin{array}{rrrr}
1 & 1 & -1 & -1 \\
1 & 1 & -1 & -1 \\
-1 & -1 & 1 & 1 \\
-1 & -1 & 1 & 1
\end{array}
\right) \left( 3\cos^2{\theta}-1 \right) = A^{2} \left( 3\cos^2{\theta}-1
\right) \ , \\
A^{6} & = & \left(
\begin{array}{rrrr}
1 & -1 & -1 & 1 \\
-1 & 1 &  1 & -1\\
-1 & 1 &  1 & -1\\
1 & -1 & -1 & 1
\end{array}
\right) \left( 3\cos^2{\theta}-1 \right) = A^{4} \left( 3\cos^2{\theta}-1
\right) \ , \\
B^{1} & = & \left(
\begin{array}{rrrr}
\phantom{-}1 & \phantom{-}0 & \phantom{-}0 & \phantom{-}0 \\
\phantom{-}0 & \phantom{-}1 & \phantom{-}0 & \phantom{-}0 \\
\phantom{-}0 & \phantom{-}0 & \phantom{-}1 & \phantom{-}0 \\
\phantom{-}0 & \phantom{-}0 & \phantom{-}0 & \phantom{-}1
\end{array}
\right) \ , \\
B^{2} & = & \left(
\begin{array}{rrrr}
\phantom{-}1 & \phantom{-}0 & \phantom{-}2 & \phantom{-}0 \\
\phantom{-}0 & \phantom{-}1 & \phantom{-}0 & \phantom{-}2 \\
\phantom{-}2 & \phantom{-}0 & \phantom{-}1 & \phantom{-}0 \\
\phantom{-}0 & \phantom{-}2 & \phantom{-}0 & \phantom{-}1
\end{array}
\right) \ , \\
B^{3} & = & \left(
\begin{array}{rrrr}
\phantom{-}1 & \phantom{-}2 & \phantom{-}0 & \phantom{-}0 \\
\phantom{-}2 & \phantom{-}1 & \phantom{-}0 & \phantom{-}0 \\
\phantom{-}0 & \phantom{-}0 & \phantom{-}1 & \phantom{-}2 \\
\phantom{-}0 & \phantom{-}0 & \phantom{-}2 & \phantom{-}1
\end{array}
\right) \ , \\
B^{4} & = & \left(
\begin{array}{rrrr}
\phantom{-}1 & \phantom{-}2 & \phantom{-}2 & \phantom{-}4 \\
\phantom{-}2 & \phantom{-}1 & \phantom{-}4 & \phantom{-}2 \\
\phantom{-}2 & \phantom{-}4 & \phantom{-}1 & \phantom{-}2 \\
\phantom{-}4 & \phantom{-}2 & \phantom{-}2 & \phantom{-}1
\end{array}
\right) \ , \\
B^{5} & = & \left(
\begin{array}{rrrr}
1 & -1 & 0 & 0 \\
-1 & 1 & 0 & 0 \\
0 & 0 & 1 & -1 \\
0 & 0 & -1 & 1
\end{array}
\right) \, \left( 3\cos^2{\theta}-1 \right) \ , \\
B^{6} & = & \left(
\begin{array}{rrrr}
1 & -1 & 2 & -2 \\
-1 & 1 & -2 & 2 \\
2 & -2 & 1 & -1 \\
-2 & 2 & -1 & 1
\end{array}
\right) \, \left( 3\cos^2{\theta}-1 \right) \ ,
\eea
where $\theta$ is the angle between ${\bf r}$ and the $z$ axis.

\section{Change of representation}

In this Section we discuss the different representation for the operators of
the ``$v_6$'' algebra. A generic operator $x$ can be written as
\beq
x = \sum_{p=1}^6 x^p_{ij}O^p = x_c + x_\tau \tautau + x_\sigma \sigsig +
x_{\sigma\tau} \sigsig \tautau + x_t S_{12} + x_{t\tau} S_{12}\tautau \ , 
\label{op:rep}
\eeq 
in the basis of operators (\ref{six:op}), or as
\beq
x = \sum_{TS} \left[ x_{T0} + \delta_{S1} x_{tT} S_{12} \right]
P_{2S+1}\Pi_{2T+1} \ ,
\label{TS:rep}
\eeq
in the ``TS-representation''.

The transformation matrix is given by
\beq
\left(
\begin{array}{rrrr}
1 & -3 & -3 &  9 \\
1 &  1 & -3 & -3 \\
1 & -3 &  1 & -3 \\
1 &  1 &  1 &  1
\end{array}
\right)
\left(
\begin{array}{c}
x_c \\ x_\tau \\ x_\sigma \\ x_{\sigma\tau}
\end{array}
\right)
 = 
\left(
\begin{array}{c}
x_{00} \\ x_{10} \\ x_{01} \\ x_{11}
\end{array}
\right) \ ,
\eeq
\beq
\left(
\begin{array}{rr}
1 & -3 \\
1 &  1
\end{array}
\right)
\left(
\begin{array}{c}
x_t \\ x_{t\tau}
\end{array}
\right)
 = 
\left(
\begin{array}{c}
x_{t0} \\ x_{t1}
\end{array}
\right) \ ,
\eeq
or
\beq
\left\{
\begin{array}{l}
x_{TS} = x_c + (4T-3) x_\tau + (4S-3) x_\sigma + (4S-3)(4T-3)
  x_{\sigma\tau} \ , \\
\ \ \ \ \\
x_{tT} = x_t + (4T-3) x_{tT} \ .
\end{array}
\right.
\eeq
The inverse transformation is given by
\beq
\frac{1}{16}
\left(
\begin{array}{rrrr}
 1 &  3 &  3 &  9 \\
-1 &  1 & -3 &  3 \\
-1 & -3 &  1 &  3 \\
 1 & -1 & -1 &  1
\end{array}
\right)
\left(
\begin{array}{c}
x_{00} \\ x_{10} \\ x_{01} \\ x_{11}
\end{array}
\right)
 = 
\left(
\begin{array}{c}
x_c \\ x_\tau \\ x_\sigma \\ x_{\sigma\tau}
\end{array}
\right) \ ,
\eeq
\beq
\frac{1}{4}
\left(
\begin{array}{rr}
 1 & 3 \\
-1 & 1
\end{array}
\right)
\left(
\begin{array}{c}
x_{t0} \\ x_{t1}
\end{array}
\right)
 = 
\left(
\begin{array}{c}
x_{t} \\ x_{t\tau}
\end{array}
\right) \ ,
\eeq
\chapter{Energy at two-body cluster level}
\label{2body}
The energy per particle at two-body cluster level can be written (see
Eqs.(\ref{def:deltaE2}) and (\ref{def:w}))
\beq
(\Delta E)_2 = \sum_{i<j}\:\langle
ij|\:\frac{1}{2}\bigg[f_{12},\left[\;t_1+t_2,\:f_{12}\right]\bigg] +
f_{12}v_{12}f_{12}\:|ij-ji\rangle \ ,
\eeq
with
\beq
t_i = -\frac{1}{2m}\nabla^2_i  \ \ , \ \ 
t_1 + t_2= -\frac{1}{m}\nabla^2_{\phantom{R}} -
\frac{1}{4m}\nabla^2_R \ , 
\eeq
where ${\bm \nabla}$ acts on the relative coordinate ${\bf r}$, while ${\bm
  \nabla}_R$ acts on the center of mass coordinate ${\bf R}$, defined as
\beq
{\bf r}={\bf r}_1-{\bf r}_2 \ \ , \ \ {\bf R}= \frac{1}{2}({\bf r}_1+{\bf r}_2)
\eeq
 respectively.

Including only the static part of the interaction, i.e. neglecting the spin-orbit components, 
both the correlation function $f_{12}$ 
and the two-nucleon potential $v_{12}$ are written as
\beq
f_{12}=\sum^6_{p=1}f^p(r_{12})O^p_{12} \ \ , \ \
v_{12}=\sum^6_{p=1}v^p(r_{12})O^p_{12} \ ,
\eeq
with the six operator $O^{n}_{12}$ listed in Eq.(\ref{pot2}), whose properties are discussed in 
Appendix \ref{pauli}..

The FG two-nucleon state is given by
\bea
\vert ij \rangle & = & \frac{1}{V}\;{\rm e}^{i({\bf k}_i\cdot{\bf r}_1+{\bf
    k}_j\cdot{\bf r}_2)}\;\vert S\,M_S,\,T\,M_T \rangle \nonumber \\
 &  = &  \frac{1}{V}\;{\rm e}^{i({\bf k}\cdot{\bf r}+{\bf K}\cdot{\bf
     R})}\;\vert S\,M_S,\,T\,M_T \rangle \ ,
\label{tns}
\eea
where
\bea
\vert {\bf k}_i\vert,\;\vert{\bf k}_j\vert & \leq & p_F \nonumber \\
{\bf k} = \frac{1}{2}({\bf k}_i-{\bf k}_j)  \  & , &  \ {\bf K} = {\bf k}_i+{\bf
  k}_j \ ,
\eea
and $\vert S\,M_S,\,T\,M_T \rangle$ denotes a state of total spin (isospin) $S \ (T)$ and
spin (isospin) projection $M_S \ (M_T)$.

We will discuss the potential and kinetic energy term separately.

\section{Potential energy}

Consider the operator
\beq
w_{12}=f_{12}v_{12}f_{12} \ ,
\eeq
and the decomposition of $f_{12}$ in the $TS$-representation (see
Eq.(\ref{TS:rep}))
\beq
f_{12}=\sum_{ST}\bigg[f_{ST}+\delta_{S1}f_{tT}S_{12}\bigg]
P^{\phantom{2}}_{2S+1}\Pi^{\phantom{2}}_{2T+1} \ .
\eeq
In the above equation, $P^{\phantom{2}}_{2S+1}$ and
$\Pi^{\phantom{2}}_{2T+1}$ are spin and isospin projection operators,
whose properties are given in Appendix \ref{pauli}. By writing the
corresponding decomposition for $w_{12}$ and $v_{12}$ and calculating 
\bea
w_{12} & = & \sum_{TS}
\Bigg\{\delta^{\phantom{2}}_{S0}f^2_{T0}v^{\phantom{2}}_{T0}+ \delta_{S1}
\Big\{\;v^{\phantom{2}}_{T1}\bigg[\;f^2_{T1}+8f^2_{tT}+
2\left(f^{\phantom{2}}_{T1}f^{\phantom{2}}_{tT}-f^2_{tT}\right)S_{12}\bigg] +  \nonumber\\
 & + &
 \!\!v^{\phantom{2}}_{tT}\bigg[16\left(f^{\phantom{2}}_{T1}f^{\phantom{2}}_{tT}-f^2_{tT}\right)
+\left(f^2_{T1}-4f^{\phantom{2}}_{T1}f^{\phantom{2}}_{tT}+12f^2_{t1}\right)S_{12}\bigg]\Big\}\Bigg\}
P^{\phantom{2}}_{2S+1}\Pi^{\phantom{2}}_{2T+1} \ , \nonumber
\eea
we can identify
\bea
w^{\phantom{2}}_{T0} & = & \phantom{2}v^{\phantom{2}}_{T0}\phantom{\Big(}f^2_{T0}\phantom{\Big)} \nonumber \\
w^{\phantom{2}}_{T1} & = & \phantom{2}
v^{\phantom{2}}_{T1}\Big(f^2_{T1}+8f^2_{tT}\Big)+16v^{\phantom{2}}_{tT}
\Big(f^{\phantom{2}}_{T1}f^{\phantom{2}}_{tT}-f^2_{tT}\Big)  \\
w^{\phantom{2}}_{tT} & = & 2v^{\phantom{2}}_{T1}\Big(f^{\phantom{2}}_{T1}
f^{\phantom{2}}_{tT}-f^2_{tT}\Big)+v^{\phantom{2}}_{tT}\Big(f^2_{T1}
-4f^{\phantom{2}}_{T1}f^{\phantom{2}}_{tT}+12f^2_{t1}\Big)  \ . \nonumber
\eea

After replacing 
\beq
\sum_{i<j} \longrightarrow \frac{1}{2}\,\sum_{i j} \ ,
\eeq
the potential energy contribution to $(\Delta E)_2$ reads
\bea
\langle w \rangle & = &
\frac{1}{2}\frac{1}{V^2}\sum_{SM_S}\sum_{TM_T}\sum_{{\bf k}_i{\bf
    k}_j}\sum_{S^\prime T^\prime}\Bigg\{\int \!d^3r_1d^3r_2 
 \bigg[w_{S^\prime T^\prime}(r)\langle P_{2S^\prime+1}\Pi_{2T^\prime+1}\rangle+ \nonumber \\
 & & \delta_{S^\prime 1}w_{tT^\prime}(r)\langle
 S_{12} P_{2S^\prime+1}\Pi_{2T^\prime+1}\rangle\bigg]- \int
 \!d^3r_1d^3r_2\,{\rm e}^{i({\bf k}_i\cdot{\bf r}-{\bf k}_j\cdot{\bf r})} \\
 & & \bigg[w_{S^\prime
   T^\prime}(r)\langle P_{2S^\prime+1}\Pi_{2T^\prime+1} P_{\sigma\tau}\rangle+\delta_{S^\prime1}w_{tT^\prime}(r)\langle S_{12}P_{2S^\prime+1}\Pi_{2T^\prime+1}P_{\sigma\tau}\rangle\bigg]\Bigg\} \ , \nonumber
\eea
where $P_{\sigma\tau}$ is the spin-isospin exchange operator defined in
Appendix \ref{pauli} and the expectation values $\< O \>$ are taken over
two-nucleon states of definite total spin and isospin $\vert S\, M_S,\, T\, M_T
\rangle$. Using
\beq
\int d^3r_1d^3r_2  =  \int d^3r\,d^3R = V\int d^3r \ ,
\eeq
the definition of the Slater function (\ref{slater:l}),
\beq
\sum_{\vert{\bf k}\vert \leq p_F}{\rm e}^{i{\bf k}\cdot{\bf
    r}}=\frac{V}{(2\pi)^3}\int_{\vert {\bf k}\vert \leq p_F}d^3k\,e^{i{\bf
    k}\cdot{\bf r}}=\frac{N}{\nu}\ell(p_Fr) \ ,
\eeq
and the results of Appendix \ref{pauli}, we finally obtain
\beq
\label{pot:en}
\langle w \rangle=
\frac{1}{2}\frac{1}{V^2}\frac{N^2}{\nu^2}V\sum_{ST}\left(2S+1\right)\left(2T+1\right)\int
d^3r\,w_{ST}(r)\left[1-(-1)^{S+T}\ell^2(p_Fr)\right] \ ,
\eeq
where $\nu$ denotes the degeneracy of the momentum eigenstates. 
In the case of symmetric nuclear matter ($\nu=4$)
\bea
\frac{1}{N}\langle w \rangle & = & \frac{\rho}{32}
\int d^3r \Big\{\big[w_{00}(r)+9w_{11}(r)\big]a_{-}(p_Fr)+ \nonumber \\
& & \phantom{\frac{\rho}{32}\int d^3r }\!\!+
\big[3w_{01}(r)+3w_{10}(r)\big]a_{+}(p_Fr)\big]\Big\} \ ,
\label{tbe:p}
\eea
where $\rho=N/V$ is the density and
\beq
a_{\pm}(x)=1\pm\ell^2(x) \ .
\eeq

\section{Kinetic energy}

Let us now discuss the kinetic contribution to the energy, given by
\beq
\frac{1}{2}\bigg[f_{12},\left[\;t_1+t_2,\:f_{12}\right]\bigg]=-\frac{1}{2m}\bigg[f_{12},\left[\;\nabla^2,\:f_{12}\right]\bigg]
\ . \label{en:kin}
\eeq

We consider spin-zero and spin-one channels separately.

\paragraph{Spin-zero channels}
In these channels, the relevant part of the correlation function is given by
\beq
f_{12} = \sum_T f_{T0}(r)\;P_1\Pi_{2T+1} \ .
\eeq

Making use of the results of Appendix \ref{pauli}, as well as of
\beq
\Big[f_{T0},\,\nabla^2 f_{T0}\Big]=0 \ \ , \ \
 \Big[f_{T0},\,\big({\bm \nabla}f_{T0}\big){\bm \nabla}\Big]=-\big({\bm
   \nabla}f_{T0}\big)^2 \ ,
\eeq
we find
\bea
\Big[f_{12},\left[\;\nabla^2,\:f_{12}\right]\Big] & = & \sum_{TT'}\Big[f_{T0}
\;P^{\phantom{2}}_1\Pi^{\phantom{2}}_{2T+1},\left[\nabla^2,\;f_{T0}\right]\;
P^{\phantom{2}}_1\Pi^{\phantom{2}}_{2T'+1}\Big] \nonumber \\
 & = & \sum_{TT'}\Big[f_{T0},
 \left[\nabla^2,\;f_{T0}\right]\Big]\;P^2_1\Pi^{\phantom{2}}_{2T+1}
\Pi^{\phantom{2}}_{2T'+1} \nonumber \\
 & = & \sum_T\Big[f_{T0},\, \big(\nabla^2 f_{T0}\big)+2\big({\bm
   \nabla}f_{T0}\big){\bm \nabla}\Big]\;P^{\phantom{2}}_1\Pi^{\phantom{2}}_{2T+1} \nonumber \\
 & = & 2\sum_T\Big[f_{T0},\,\big({\bm \nabla}f_{T0}\big){\bm \nabla}\Big]
\;P^{\phantom{2}}_1\Pi^{\phantom{2}}_{2T+1} \nonumber \\
 & = & -2\sum_T\big({\bm
   \nabla}f_{T0}\big)^2\;P^{\phantom{2}}_1\Pi^{\phantom{2}}_{2T+1} \ .
\eea

Finally,
\beq
-\frac{1}{2m}\bigg[f_{12},\left[\;\nabla^2,\:f_{12}\right]\bigg] = \frac{1}{m}\sum_T\big({\bm
   \nabla}f_{T0}\big)^2\;P^{\phantom{2}}_1\Pi^{\phantom{2}}_{2T+1} \ .
\label{tbe:ks}
\eeq

\paragraph{Spin-one channels}

In these channels, the correlation function is given by
\beq
f_{12}=\sum_T\bigg[f_{T1}(r)+f_{tT}(r)S_{12}\bigg]P^{\phantom{2}}_3\Pi^{\phantom{2}}_{2T+1}
\ .
\eeq
Relying once more on Appendix \ref{pauli}, we calculate
\[
\sum_{T'}\Big[\nabla^2,\,
\big(f_{T'1}+f_{tT'}S_{12}\big)P^{\phantom{2}}_3\Pi^{\phantom{2}}_{2T'+1}\Big]
  =  \sum_{T'}\Big\{\big[\nabla^2,\, f_{T'1}\big]+\big[\nabla^2,\,
f_{tT'}S_{12}\big]\Big\}
P^{\phantom{2}}_3\Pi^{\phantom{2}}_{2T'+1}
\]
\[
 =\sum_{T'}\Big\{\big(\nabla^2 f_{T'1}\big)+2\big({\bm \nabla} f_{tT'}\big){\bm
  \nabla}+\big(\nabla^2 f_{tT'}S_{12}\big)+2\big({\bm
  \nabla}f_{tT'}S_{12}\big){\bm
  \nabla}\Big\}P^{\phantom{2}}_3\Pi^{\phantom{2}}_{2T'+1}
\]
\[
  =  \sum_{T'} \Big\{ \big(\nabla^2 f_{T'1}\big) + 2\big({\bm \nabla}
f_{tT'}\big){\bm \nabla} + \big(\nabla^2 f_{tT'}\big)S_{12} +
\big(\nabla^2S_{12} \big)f_{tT'} 
\]
\beq
+ 2\big({\bm \nabla}f_{tT'}\big)\big({\bm \nabla}S_{12}\big) +2S_{12}\big({\bm
   \nabla}f_{tT'}\big){\bm \nabla} +2f_{tT'}\big({\bm \nabla}S_{12}\big){\bm
   \nabla}\Big\} 
P^{\phantom{2}}_3\Pi^{\phantom{2}}_{2T'+1} \ .
\label{graffe}
\eeq

Hence, the commutator in Eq.(\ref{en:kin}) can be rewritten as
\bea
\Big[f_{12},\left[\;\nabla^2,\:f_{12}\right]\Big] & = &
\sum_{TT'}\Big[\big(f_{T1}+f_{tT}S_{12}\big)
P^{\phantom{2}}_3\Pi^{\phantom{2}}_{2T+1},\left\{\ldots\right\}
P^{\phantom{2}}_3\Pi^{\phantom{2}}_{2T'+1} \Big] \nonumber \\
 & = & \sum_T\Big[f_{T1}+f_{tT}S_{12}, \left\{\ldots\right\}\Big]
P^{\phantom{2}}_3\Pi^{\phantom{2}}_{2T+1} \nonumber \\
 & = & \sum_T \Big( F^{(1)}_T + F^{(2)}_T \Big)
 P^{\phantom{2}}_3\Pi^{\phantom{2}}_{2T+1} \ ,
\eea
with
\beq
F^{(1)}_T = \Big[f_{T1}, \left\{\ldots\right\}\Big] \ \ , \ \ 
F^{(2)}_T = \Big[f_{tT}S_{12}, \left\{\ldots\right\}\Big] \ ,
\eeq
and
\bea
\Big\{ \ldots \Big\} & = & \Big\{ \big(\nabla^2 f_{T'1}\big) + 2\big({\bm \nabla}
f_{tT'}\big){\bm \nabla} + \big(\nabla^2 f_{tT'}\big)S_{12} +
\big(\nabla^2S_{12} \big)f_{tT'} \nonumber \\
 & + & 2\big({\bm \nabla}f_{tT'}\big)\big({\bm \nabla}S_{12}\big) +2S_{12}\big({\bm
   \nabla}f_{tT'}\big){\bm \nabla} +2f_{tT'}\big({\bm \nabla}S_{12}\big){\bm
   \nabla}\Big\} \ .
\eea
We find
\beq
F^{(1)}_T = -2\big({\bm \nabla}f_{T1}\big)^2 -2\big({\bm \nabla}f_{T1}\big)
\big({\bm \nabla}f_{tT}\big)S_{12} \ ,
\eeq
and
\bea
F^{(2)}_T & = & \Big[f_{tT}S_{12},\, 2\big({\bm \nabla}f_{T1}\big){\bm
  \nabla}\Big]  + \Big[ f_{tT}S_{12},\, 2S_{12}\big({\bm \nabla}f_{T1}\big){\bm \nabla}\Big] + \nonumber \\
&  & +\Big[ f_{tT}S_{12},\, 2f_{T1}\big({\bm \nabla}S_{12}\big){\bm \nabla}\Big] = \nonumber \\
 & = & -2\big({\bm \nabla}f_{T1}\big)\big({\bm
   \nabla}f_{tT}\big)S_{12}-2\big({\bm \nabla}f_{tT}\big)^2S^2_{12} + \nonumber \\
 & & +2f^2_{tT} \Big[S_{12}, \,\big({\bm \nabla}S_{12}\big){\bm \nabla}\Big]
 \nonumber \\
& = & -2\phantom{\Bigg[}\big({\bm \nabla}f_{T1}\big)\big({\bm
  \nabla}f_{tT}\big)S_{12}-2\big({\bm \nabla}f_{tT}\big)^2\big(8-2S_{12}\big)\phantom{\Bigg]}+ \nonumber \\
 & - &
 2f^2_{tT}\Bigg[\frac{36}{r^2}\;\big({\bf L}\cdot{\bf S}\big)+\frac{6}{r^2}\;\big(8-S_{12}\big)\Bigg] \ .
\eea

Collecting all pieces togheter, we find for the spin-one channels
\bea
-\frac{1}{2m}\bigg[f_{12},\left[\;\nabla^2,\:f_{12}\right]\bigg] & = & \frac{1}{m}\sum_T \Bigg\{
\big({\bm \nabla}f_{T1}\big)^2 +  \big({\bm \nabla}f_{T1}\big)\big({\bm
  \nabla}f_{tT} S_{12}\big)
 + \nonumber \\
 & & + \phantom{\Bigg[}\big({\bm \nabla}f_{tT}\big)^2\big(8-2S_{12}\big)\phantom{\Bigg]} +  f^2_{tT}\Bigg[\frac{36}{r^2}\;\big({\bf L}\cdot{\bf
   S}\big)+\frac{6}{r^2}\;\big(8-S_{12}\big)\Bigg]
\Bigg\}P^{\phantom{2}}_3\Pi^{\phantom{2}}_{2T+1} \nonumber \\
& = & \frac{1}{m}\sum_{TS}\Bigg\{\big( {\bm
  \nabla}f_{TS}\big)^2+\delta_{S1}\Big[2\big({\bm \nabla}f_{TS}\big)
\big({\bm \nabla}f_{tT}\big)S_{12}
+ \nonumber \\
 & & + \big({\bm \nabla}f^{\phantom{2}}_{tT}\big)^2
 S^2_{12}+f^{\phantom{2}}_{tT}\frac{36}{r^2}
\;\big({\bf L}\cdot{\bf S}\big)+\frac{6}{r^2}\;\big(8-S_{12}\big)\Big]\Bigg\}
P^{\phantom{2}}_{2S+1}\Pi^{\phantom{2}}_{2T+1} \nonumber \\
 & = & \sum_{TS} \Big\{t_{TS}(r) + \delta_{S1}\bigg[t_{tT}(r)S_{12} +
 t_{bT}(r)\big({\bf L}\cdot{\bf S}\big)\bigg]\Big\}
P^{\phantom{2}}_{2S+1}\Pi^{\phantom{2}}_{2T+1} \ ,
\eea
with
\bea
t_{T0} & = & \frac{1}{m} \big({\bm \nabla}f_{T0}\big)^2 \nonumber \\
t_{T1} & = & \frac{1}{m} \Big[\big({\bm \nabla}f_{T1}\big)^2 + 8\big({\bm
  \nabla}f_{tT}\big)^2 + \frac{48}{r^2}\;f^2_{tT} \Big] \nonumber \\
t_{tT} & = & \frac{1}{m} \Big[2\big({\bm \nabla}f_{T1}\big)\big({\bm \nabla}f_{tT}\big) 
- 2\big({\bm \nabla}f_{tT}\big)^2 -\frac{6}{r^2}\;f^2_{tT} \Big] \nonumber \\
t_{bT} & = & \frac{1}{m} \frac{36}{r^2}\;f^2_{tT} \ . \nonumber
\eea

\section{Final expression}
We can rewrite
\beq
\left(\Delta E\right)_2 = \sum_{i<j} \langle ij \vert W_{12} \vert ij-ji
\rangle \ ,
\eeq
with
\[
W_{12}  =  -\frac{1}{m}\bigg[f_{12},\left[\;\nabla^2,\:f_{12}\right]\bigg] +
f_{12}v_{12}f_{12}
\]
\[
  =  \sum_{TS} \Big\{W_{TS}(r) + \delta_{S1}\bigg[W_{tT}(r)S_{12} +
 W_{bT}(r)\big({\bf L}\cdot{\bf S}\big) \bigg]\Big\}
P^{\phantom{2}}_{2S+1}\Pi^{\phantom{2}}_{2T+1} \ ,
\]
where
\bea
W_{T0} & = & \frac{1}{m}\big({\bm \nabla}f^{\phantom{2}}_{T0}\big)^2 + v_{T0}f^2_{T0} \nonumber \\
W_{T1} & = & \frac{1}{m}\Big[\big({\bm \nabla}f_{T1}\big)^2 + 8\big({\bm
  \nabla}f_{tT}\big)^2 
+ \frac{48}{r^2}\;f^2_{tT} \Big]+ \nonumber \\
& & + v^{\phantom{2}}_{T1}\Big(f^2_{T1}+8f^2_{tT}\Big)+16v^{\phantom{2}}_{tT}
\Big(f^{\phantom{2}}_{T1}f^{\phantom{2}}_{tT}-f^2_{tT}\Big) \nonumber \\
W_{tT} & = & \frac{1}{m}\Big[2\big({\bm \nabla}f_{T1}\big)\big({\bm
  \nabla}f_{tT}\big)
 -2\big({\bm \nabla}f_{tT}\big)^2 -\frac{6}{r^2}\;f^2_{tT} \Big]+ \nonumber \\
 & &
 +2v^{\phantom{2}}_{T1}\Big(f^{\phantom{2}}_{T1}f^{\phantom{2}}_{tT}-f^2_{tT}\Big)
+v^{\phantom{2}}_{tT}\Big(f^2_{T1}-4f^{\phantom{2}}_{T1}f^{\phantom{2}}_{tT}+12f^2_{t1}\Big) \nonumber \\
W_{bT} & = & \frac{1}{m}\;\frac{36}{r^2}\;f^2_{tT} \ . \nonumber 
\eea

Making use of the expression for the expectation values given in Appendix
\ref{pauli}, we finally obtain (compare to Eq.(\ref{tbe:p}))
\bea
\frac{(\Delta E)_2}{N} & = & \frac{\rho}{32} \int\;d^3r
\Bigg\{\Big[W_{00}(r) + 9W_{11}(r)\Big]a_-(p_Fr)+ \nonumber \\
 & & +\Big[3W_{01}(r) + 3W_{10}(r)\Big]a_+(p_Fr)\Bigg\} \ .
\label{tbe:k}
\eea
\chapter{Euler-Lagrange equations for the correlation functions}
\label{corrfcn}
\section{Spin singlet channels: uncoupled equations}
In the spin-zero channels, the energy per partcicle of SNM, evaluated at two-body
cluster level, reads (compare to Eqs.(\ref{tbe:p}) and (\ref{tbe:ks}))
\bea
\frac{(\Delta E)_2}{N} & = &
\frac{\rho}{32}\;\left(2T+1\right)\int\;d^3r\left[\htm\big({\bm \nabla}f^{\phantom{2}}_{T0}\big)^2 + v_{T0}f^2_{T0}\right]a_{T0}(p_Fr) \nonumber \\
 & = & \frac{\rho}{32}\;\left(2T+1\right)\;4\pi\int\;r^2dr\left[\htm\big(f^\prime_{T0}\big)^2 + v_{T0}f^2_{T0}\right]a_{T0}(p_Fr) \nonumber \\
 & = &
 \textrm{const}\;\int^{\infty}_{0}dr\;F\left[f^{\phantom{^\prime}}_{T0},\;f^\prime_{T0}\right] \ , 
\eea
where $a_{TS}(x)=1-(-)^{T+S} \ell^2(x)$ and
\beq
F\left[f^{\phantom{^\prime}}_{T0},\;f^\prime_{T0}\right]  =  \Bigg[\big(f^\prime_{T0}\big)^2 + 
\mth \;v_{T0}f^2_{T0}\Bigg] \phi^2_{T0} \ ,
\eeq
with
\beq
\phi_{T0}  =  r\sqrt{a_{T0}} \ .
\label{eq:phi}
\eeq
The corresponding Euler-Lagrange (EL) equations for the unknown functions $f_{T0}$
are given by
\beq
\frac{d}{dr}\frac{\partial F}{\partial f^\prime_{T0}}-\frac{\partial F}{\partial
  f^{\phantom{^\prime}}_{T0}}=0 \ .
\eeq

From
\bea
\frac{\partial F}{\partial f^{\phantom{^\prime}}_{T0}} & = &
2\;\mth\;v_{T0}\;f^2_{T0}\;\phi^2_{T0} \ , \nonumber \\
\frac{\partial F}{\partial f^\prime_{T0}} & = & 2\;f^\prime_{T0}\;\phi^2_{T0} \ , \nonumber \\
\frac{d}{dr}\frac{\partial F}{\partial f^\prime_{T0}} & = &
2\;f^{\prime\prime}_{T0}\;\phi^2_{T0} + 4\;f^\prime_{T0}\;\phi^\prime_{T0}\;\phi_{T0} \ ,
\eea
we obtain
\beq
f^{\prime\prime}_{T0}\;\phi^2_{T0} + 2\;f^\prime_{T0}\;\phi^\prime_{T0}
-\mth\;v_{T0}\;f^2_{T0}\;\phi^2_{T0}=0 \ .
\label{EL0:f}
\eeq

Introducing
\beq
g_{T0} \equiv f_{T0}\;\phi_{T0} \ ,
\eeq
we can put Eq.(\ref{EL0:f}) in the form
\beq
g^{\prime\prime}_{T0} - \Bigg(\frac{\phi^{\prime\prime}_{T0}}{\phi_{T0}} + 
\mth\;v_{T0}\Bigg)\;g_{T0}=0 \ .
\eeq

Now we introduce a Lagrange multiplier, in order to fulfill the requirement
(see Eqs.(\ref{bound0:f})-(\ref{bound2:f}))
\beq
\left. g^\prime_{T0} \right|_{r=d}   = \left.  \phi^\prime_{T0} \right|_{r=d} \ .
\eeq
The resulting equation is Eq.(4) of  Ref.\cite{BCFR}
\beq
g^{\prime\prime}_{T0} - \Bigg(\frac{\phi^{\prime\prime}_{T0}}{\phi_{T0}} +
\mth\;\left(v_{T0}+\lambda\right)\Bigg)\;g_{T0}=0 \ ,
\eeq
to be integrated with the boundary conditions
\bea
\left. g_{T0} \right|_{r=0} & = & 0 \ \ , \\ 
\left. g_{T0} \right|_{r=d} & = & \left. \phi_{T0} \right|_{r=d} \ . 
\eea

\section{Spin triplet channels: coupled equations}

In the spin-one channels, the contribution to the energy is given by (see
Eqs.(\ref{tbe:p}) and (\ref{tbe:k}))
\bea
\frac{(\Delta E)_2}{N} & = & \frac{\rho}{32}\;\left(2T+1\right)
\int\;d^3r\Bigg\{\htm\left[\big({\bm \nabla} f^{\phantom{2}}_{T1}\big)^2 
+8\big({\bm \nabla} f^{\phantom{2}}_{tT}\big)^2 + \frac{48}{r^2}\;f^2_{tT}\right] + \nonumber \\
 & & + v_{T1}\bigg(f^2_{T1}+8f^2_{tT}\bigg)
 +16v_{tT}\bigg(f^{\phantom{2}}_{T1}f^{\phantom{2}}_{tT}
-f^2_{tT}\bigg)\Bigg\}\;a_{T1}(p_Fr) \nonumber \\
 & = & \textrm{const}\;\int^{\infty}_{0}dr
\;F\left[f^{\phantom{^\prime}}_{T1},\;f^{\phantom{^\prime}}_{tT};\;f^\prime_{T1},\;f^\prime_{tT}\right] \ ,
\eea
where
\bea
F\left[f^{\phantom{^\prime}}_{T1},\;f^{\phantom{^\prime}}_{tT};\;f^\prime_{T1},\;f^\prime_{tT}\right] 
& = &
\left(f^\prime_{T1}\right)^2\phi^2_{T1}+8\left(f^\prime_{tT}\right)^2\phi^2_{T1}+\frac{48}{r^2}
\;f^2_{tT}\phi^2_{T1}+ \ \ \ \ \ \ \ \ \ \ \ \ \ \ \ \ \ \nonumber \\
 & + & \mth\Big[v_{T1}\left(f^2_{T1}+8f^2_{tT}\right) 
+16v_{tT}\left(f^{\phantom{2}}_{T1}f^{\phantom{2}}_{tT}-f^2_{tT}\right)\Big] \ .
\eea

In this case we have two coupled EL equations
\beq
\left\{
\begin{array}{l}
\frac{d}{dr}\frac{\partial F}{\partial f^\prime_{T1}}-\frac{\partial F}{\partial
  f^{\phantom{^\prime}}_{T1}} 
 =  0 \nonumber \\
\ \ \ \ \ \\
\frac{d}{dr}\frac{\partial F}{\partial f^\prime_{tT}}-\frac{\partial F}{\partial
  f^{\phantom{^\prime}}_{tT}}  =  0 \ .
\end{array}
\right.
\eeq

Carrying out the derivativees as in the spin-zero channels and defining
\beq
g_{T1} \equiv f_{T1}\phi_{T1} \ \ , \ \  g_{tT} \equiv \sqrt{8}f_{tT}\phi_{T1}
\ ,
\eeq
we find
\beq
\left\{
\begin{array}{l}
g^{\prime\prime}_{T1} - \left(\frac{\phi^{\prime\prime}_{T1}}{\phi_{T1}} +
\mth\;v_{T1}\right)\;g_{T1}-\mth\sqrt{8}v_{tT}g_{tT}
 =  0  \\
\ \ \ \ \ \\
g^{\prime\prime}_{tT} - \left[\frac{\phi^{\prime\prime}_{T1}}{\phi_{T1}} + \mth\;\left(v_{T1}-
2v_{tT}\right)+\frac{6}{r^2}\right]\;g_{tT} - \mth\sqrt{8}v_{tT}g_{T1} = 0
\ .
\end{array}
\right.
\eeq

Finally, inclusion of the Lagrange multipliers needed to guarantee
\bea
\left. g^\prime_{T1}\right|_{r=d_1} & = & \left.\phi^\prime_{T1}\right|_{r=d_1} \ , \\
\left. g^\prime_{tT}\right|_{r=d_2} & = & \left.\phi^\prime_{T1}\right|_{r=d_2} \ ,
\eea
with, in general, $d_1\neq d_2$, leads to (compare to Eq.(5) of Ref.\cite{BCFR})
\beq
\left\{
\begin{array}{l}
g^{\prime\prime}_{T1} - \left[ \frac{\phi^{\prime\prime}_{T1}}{\phi_{T1}} +
\mth\;\left(v_{T1}+\lambda_1\right) \right] 
\;g_{T1}-\mth\left(\sqrt{8}v_{tT}+\lambda_2\right)g_{tT}  =  0 \\ 
 \ \ \ \\ 
g^{\prime\prime}_{tT} - \left[\frac{\phi^{\prime\prime}_{T1}}{\phi_{T1}} + \mth\;\left(v_{T1}-
2v_{tT}+\lambda_1\right)+\frac{6}{r^2}\right]\;g_{tT}
- \mth\left(\sqrt{8}v_{tT}+\lambda_2\right)g_{T1} = 0 \ , 
\end{array}
\right.
\eeq
with the boundary conditions
\bea
\left. g_{T1}\right|_{r=0}   & = &  0  \ , \\
\left. g_{T1}\right|_{r=d_1} & = &  \left. \phi_{T1}\right|_{r=d_1}  \ ,
\eea
and
\bea
\left. g_{tT}\right|_{r=0}    & = & 0 \ , \\
\left. g_{tT}\right|_{r=d_2}  & = & 0 \ .
\eea

\end{appendix}

\end{document}